\documentclass[a4paper,fleqn,usenatbib]{mnras}
\usepackage{newtxtext,newtxmath}
\usepackage[T1]{fontenc}
\usepackage{ae,aecompl}
\usepackage{graphicx}	
\usepackage{amsmath}	
\usepackage{amssymb}	
\usepackage{CJKutf8}

\title{Short and long term near-infrared spectroscopic variability of eruptive protostars from VVV}
\author[Z. Guo et al.]{Zhen Guo\begin{CJK*}{UTF8}{bsmi}
(郭震)\end{CJK*}$^{1}$\thanks{E-mail: z.guo4@herts.ac.uk},
P. W. Lucas$^{1}$,
C. Contreras Pe{\~n}a$^{2}$,
R. G. Kurtev$^{3,4}$,
\newauthor L. C. Smith$^{5}$, J. Borissova$^{3,4}$, J. Alonso-Garc{\'i}a$^{6,4}$, D. Minniti$^{4,7,8}$,
\newauthor A. Caratti o Garatti$^{9}$ and D. Froebrich$^{10}$
\\
$^{1}$Centre for Astrophysics Research, University of Hertfordshire, Hatfield AL10 9AB, UK\\
$^{2}$Department of Physics and Astronomy, University of Exeter, Stocker Road, Exeter, Devon EX4 4SB, UK\\
$^{3}$Instituto de F{\'i}sica y Astronom{\'i}a, Universidad de Valpara{\'i}so, ave. Gran Breta{\~n}a, 1111, Casilla 5030, Valpara{\'i}so, Chile\\
$^{4}$Millennium Institute of Astrophysics, Av. Vicuna Mackenna 4860, 782-0436, Macul, Santiago, Chile\\
$^{5}$Institute of Astronomy, University of Cambridge, Madingley Road, Cambridge, CB3 0HA, UK\\
$^{6}$Centro de Astronomía (CITEVA), Universidad de Antofagasta, Av. Angamos 601, Antofagasta, Chile\\
$^{7}$Departamento de Ciencias Fisicas, Universidad Andres Bello, Republica 220, Santiago,
 Chile\\
$^{8}$Vatican Observatory, V00120 Vatican City State, Italy\\
$^{9}$Dublin Institute for Advanced Studies, School of Cosmic Physics, Astronomy and Astrophysics Section, 31 Fitzwilliam Place, Dublin 2, Ireland\\
$^{10}$Centre for Astrophysics and Planetary Science, University of Kent, Canterbury CT2 7NH, UK
}

\date{Accepted 2019-12-02. Received 2019-11-29; in original form 2019-10-11}
\pubyear{2019}

\begin{document}
\label{firstpage}
\pagerange{\pageref{firstpage}--\pageref{lastpage}}
\maketitle

\begin{abstract}
Numerous eruptive variable young stellar objects (YSOs), mostly Class I systems, were recently detected by the near-infrared Vista Variables in the Via Lactea (VVV) survey. We present an exploratory near-infrared spectroscopic variability study of 14 eruptive YSOs. The variations were sampled over 1-day and 1 to 2-year intervals and analysed in combination with VVV light curves. CO overtone absorption features are observed on 3 objects with FUor-like spectra: all show deeper absorption when they are brighter. This implies stronger emission from the circumstellar disc with a steeper vertical temperature gradient when the accretion rate is higher. This confirms the nature of fast VVV FUor-like events, in line with the accepted picture for classical FUors. The absence of Br$\gamma$ emission in a FUor-like object declining to pre-outburst brightness suggests that reconstruction of the stellar magnetic field is a slow process. Within the 1-day timescale, 60\% of H$_2$-emitting YSOs show significant but modest variation, and 2/6 sources have large variations in Br$\gamma$. Over year-long timescales, H$_2$ flux variations remain modest despite up to 1.8 mag variation in $K_s$. This indicates that emission from the molecular outflow usually arises further from the protostar and is unaffected by relatively large changes in accretion rate on year-long timescales. Two objects show signs of on/off magnetospheric accretion traced by Br$\gamma$ emission. In addition, a 60\% inter-night brightening of the H$_2$ outflow is detected in one YSO.
\end{abstract}

\begin{keywords}
stars: pre-main sequence -- stars: protostar -- stars: variables: T Tauri -- infrared: stars
\end{keywords}

\section{Introduction}

Accretion is a fundamental part of the star and planet formation process that remains poorly understood \citep[See][]{Hartmann2016}. Historically, steady accretion scenarios were proposed to describe the pre-main-sequence (PMS) stage \citep[e.g.] []{Larson1969, Shu1977, Terebey1984}.  However, for low mass Class I YSOs, the stellar luminosity predicted by steady accretion models is an order of magnitude fainter than is observed \citep{Kenyon1990, Evans2009}. It has been proposed that episodic accretion can solve this  ``luminosity problem'' if it is common in early stellar evolution. Numerous theoretical mechanisms for this phenomenon have been put forward \citep[e.g.][] {Vorobyov2005, Zhu2012, DAngelo2010, Dunham2012, Audard2014}. E.g. \citet{Vorobyov2005, Vorobyov2015} suggest that disc fragments are formed by gravitational instabilities and then migrate inward to be accreted onto the star, or possibly survive as massive substellar companions \citep{Zhu2012}.  Variation in the accretion rate over different timescales has been revealed by photometric monitoring, showing 0.5 dex changes on timescales from hours to weeks and up to 3 dex changes on multi-year timescales \citep[e.g.][] {Joy1945, Grankin2007, Venuti2015, Guo2018a}. 

For most of the past 40 years, eruptive YSOs undergoing episodic accretion were placed in two categories: the EX Lupi events (EXors, with durations of 200 to 400 days) \citep[e.g.][] {Herbig1989, Herbig2008} and decades-long FU Orionis events \citep[FUors, see e.g.][]{Herbig1977eruptive, Hartmann1996} associated with higher accretion rates. EXors in outburst display emission lines and CO emission bands are seen in the near-infrared spectrum \citep{Lorenzetti2009, Aspin2010, Banzatti2015}. By contrast, in the much longer FUor events there is a lack of emission lines and H$_2$O and CO absorption bands are seen in near-infrared spectra \citep [e.g.][] {Reipurth2010, Connelley2018}.

However, in recent years some eruptions with mixed photometric and spectroscopic properties were observed, e.g. V1647~Ori \citep{Aspin2009}, V346 Nor \citep{Kraus2016} and V1318 Cyg S \citep{Magakian2019}. With the advent of the near-infrared VVV survey \citep{Minniti2010}, a large number of eruptive YSOs have been detected \citep[][as Paper I and Paper II]{Contreras2017, Contreras2017b}. Almost all of these are optically obscured Class I YSOs or flat spectrum systems, often with outbursts of intermediate duration and typical amplitudes of 1 to 3 mag in $K_s$. { A new category, ``MNors", was tentatively proposed in Paper II, a tentative category of eruptive YSOs, as they have large $K_s$ amplitudes from VVV light curves and the dominant timescale of variation was typically a few years. V1647~Ori is the putative prototype of ``MNors", but the diversity of these embedded sources indicates that much further study is required.} A recent study employing Gaia DR2 photometry \citep{Gaia2018dr2catalog} and SuperCOSMOS \citep{Hambly2001} surveys similarly concludes that {\it long-lasting} eruptive events are an order of magnitude more common amongst Class I YSOs than Class II systems \citep{Contreras2019}.  The variety of optically detected events continues to increase. For example, the well-studied source V2492~Cyg \citep{Covey2011} shows repetitive brightness changes attributed to both episodic accretion and extinction \citep{kospal2013,giannini2018}. The otherwise FUor-like outburst of PTF14jg shows an anomalously hot spectrum \citep{Hillenbrand2019}, and the burst in MIR was observed 500 days before the optical burst on a FUor type object Gaia17bpi \citep{Hillenbrand2018}.

The near-infrared VISTA Variables in the Via Lactea (VVV) survey monitors $10^9$ sources close to the Galactic plane \citep{Minniti2010, Saito2012}.  $K_s$ photometry has been taken at several dozen epochs since 2010 for a 560~deg$^2$ area of the Galactic disc and bulge at longitudes $-65^{\circ}< l <10^{\circ}$. The survey has been extended to a larger area since 2016 while still monitoring the original fields a few times each year. In Paper I the authors identified 816 high-amplitude variables with $\Delta K_s > 1$ mag as measured in 2010--2012 observing seasons. Among these variable sources, about 50\% were classified as YSOs, most are Class I objects. Paper II examined near-infrared spectra of 37 VVV systems that were taken in 2013--2014, prior to classification of the light curves. Among them, 28 targets were confirmed as YSOs by spectroscopic signatures, such as H I emission lines, molecular hydrogen emission, and H$_2$O and CO features, the remainder being dusty Long Period Variables (typically displaying smooth sinusoidal light curves) and cataclysmic variables. Most of the YSOs (19 / 28) were classified as eruptive variables, and 17 of them were provisionally classified as MNors due to a mixture of FUor and EXor spectroscopic characteristics or intermediate outburst timescale.

In this work, we focus on the near-infrared spectroscopic variability of 14 eruptive YSOs identified in Paper II, with measurements at 1 day and 1 to 2-year intervals.  Variability of four main spectroscopic characteristics is investigated: the continuum, Br$\gamma$ (2.17 $\mu$m), H$_2$ (2.12 $\mu$m), and the first two overtone CO band heads beyond 2.29 $\mu$m. Other features such as broad H$_2$O absorption bands are also variable in some sources. These spectroscopic features are related, directly or indirectly, to the accretion state of the YSO and the associated wind or outflow, so we bring together the results to gain insights of the changing physical conditions around eruptive Class I objects. We build on the previous spectroscopic variability study by \citet{Connelley2014} of a sample of Class I YSOs without large photometric variability between observations.

The paper is organised as follows. The details of the observation and data reduction procedures are described in Section \ref{sec:obs}. Classifications of observed photometric variability,  short to long term spectroscopic variability and emission line profiles are presented in Section \ref{sec:result}. Section \ref{sec:discussion} contains the discussions of physical mechanisms and discussions on individual sources. Finally, our results are summarised in Section \ref{sec:conclusion}. 

\begin{table*} 
\caption{Basic information for all observed objects}
\renewcommand\arraystretch{1.}
\begin{tabular}{l c c c c r c c c c l}
\hline
\hline
Name & RA & Dec & $K_{\rm s}$$^a$ & $\Delta$$K_{\rm s}$$^a$ & $\alpha_{\rm class}$$^a$ & Distance$^b$ & Light curve & Physical  & v$_{\rm LSR}$$^b$ & $t_{\rm exp}$ \\
& J2000 & J2000 & mag & mag & & kpc & class$^a$ & classification$^b$ & km s$^{-1}$ & x 4 (s) \\
\hline
v20  &  12:28:27.97 &  -62:57:13.97 &  11.70  &  1.71 &  0.60 & 2.5 & Eruptive &  MNor  & -   & 158.5 \\
v25  &  12:35:14.37 &  -62:47:15.63 &  12.34  &  1.68 &  0.22 & 2.2 &  Eruptive &  AGB & 21.1 $\pm$ 5.3   & 158.5  \\
v42  &  13:09:34.64 &  -62:49:32.52 &  11.94  &  2.16 &  0.98 & 3.7 &  LPV-Mira &  AGB & -49.0 $\pm$ 5.5   & 158.5 \\
v118 &  14:51:20.97 &  -60:00:27.40 &  13.01  &  4.24 &  0.16 & 2.2 &  Eruptive &  EXor(MNor?) & -   & 158.5 \\
v270 &  16:23:27.14 &  -49:44:43.96 &  16.14  &  3.81 &  1.76 & 2.3  &  Eruptive  & MNor & -87.6 $\pm$ 12.1  & 158.5 \\
v322 &  16:46:24.57 &  -45:59:21.04 &  15.25  &  2.63 &  0.91 & 3.1  &  Eruptive &  MNor &  -59.8  $\pm$  5.5    & 253.6 \\
v374 &  16:58:33.99 &  -42:49:55.25 &  11.98  &  2.41 &  0.91 & 2.9  &  Eruptive &  MNor & -   & 158.5 \\
v473 &  13:10:57.49 &  -62:35:22.34 &  14.53  &  1.50 &  1.82 & 3.7 &  LPV-YSO &  MNor & -   & 253.6 \\
v631 &  15:45:18.36 &  -54:10:36.87 &  13.41   &  2.63 &  -0.14 & 2.3  &  Eruptive &  MNor & -  & 158.5 \\
v662 &  16:10:26.82 &  -51:22:34.13 &  15.30  &  1.83 &  0.81 & 3.1  &  Eruptive &  MNor & -  & 253.6 \\
v665 &  16:09:57.70 &  -50:48:09.42 &  14.09  &  1.63 &  0.95 & 4.3  &  Eruptive &  MNor &  -73.1  $\pm$  5.3   & 158.5 \\
v699 &  16:23:44.34 &  -48:54:55.29 &  16.19  &  2.29 &  2.78 & 4.2  &  Eruptive &  MNor &  -93.5  $\pm$  5.3  & 253.6  \\
v717 &  16:36:05.56 &  -46:40:40.61 &  14.37  &  2.47 &  0.81 & 6.1 -- 10.2  &  LPV-YSO &  MNor &  -126.3 $\pm$ 5.3  & 253.6 \\
v721 &  16:39:48.77 &  -45:48:47.96 &  13.98  &  1.86 &  0.87 & 4.2  &  Eruptive &  FUor &  -89.7 $\pm$ 5.8 &  158.5 \\
v800 &  17:12:46.04 &  -38:25:24.63 &  12.89 &  1.65 &  1.44 & 1.4  &  Eruptive &  MNor  & -  & 158.5 \\
v815 &  14:26:04.95 &  -60:41:16.81 &  14.94  &  1.71 &  1.58 & 3.1  &  Eruptive &  MNor & -44$^{c}$  & 253.6 \\
\hline
\hline
\end{tabular}
\label{tab:info}
\flushleft{$a$: Photometric information and classifications are adopted from Paper I. LPV stands for long period variable. $\alpha_{\rm class}$ is defined as the slope of stellar SED.\\
$b$: Information adopted from Paper II. Most distances are estimated by nearby ($<300$~arcsec) star forming regions. The distances of v20 and v118 are derived by fitting SED models \citep[][error bars in Table D.1 of Paper II]{Robitaille2006, Robitaille2007}. Spectroscopic classifications are the final classifications of Paper II. The $v_{\rm LSR}$ is measured through fitting CO band heads around 2.29 $\rm\mu$m.\\
$c$: $v_{\rm LSR}$ of v815 is estimated by a surrounding HII region [WHR97] 14222-6026 \citep{Walsh1997}.}
\end{table*}

\section{observation and data reduction}
\label{sec:obs}
\subsection{Target information}
For this study, we selected 16 targets for which the discovery spectra (taken in 2013 or 2014) were published in Paper II. They comprise 14 YSOs and 2 asymptotic giant branch (AGB) stars with good quality of discovery spectra. They were expected to be bright enough to obtain comparable observations in 2015 and they appeared to be strong eruptive variable YSO candidates at that time, based on the 2010--2014 light curves. (The AGB stars were identified as such in Paper II after careful consideration; their periods are unusually long and the light curves are imperfectly periodic due to variable extinction on long timescales). Information on the targets, including their variable star designation from Paper I, coordinates, distances, previous light curve classifications, physical classifications, mean $K_s$ magnitudes and amplitude of variability are listed in Table~\ref{tab:info}. The YSOs mostly had ``eruptive" light curve classifications, except 2 YSOs had the ``LPV-YSO" classification due to detection of highly amplitude long period variation in the 2010-2014 light curves. LPV-YSO light curves were distinguished from the ``LPV-Mira" category of the 2 AGB stars by having significant short-timescale scatter about the periodic fit.
The masses of these sources were estimated to lie in the range of 1 to 5~M$_{\odot}$ in Paper II by SED fitting, though individual masses were highly uncertain. Following the classification criteria \citep{Lada1987, Greene1994}, all sources in this work are Class I or flat-spectrum objects with $\alpha > 0.3$ or $ 0.3 \ge \alpha \ge -0.3 $, respectively.  Here $\alpha$ is the slope of the SED from 2 to 22~$\mu$m calculated in Paper I. For convenience we refer to each target with a single ``v", e.g. ``v20" rather than ``VVVv20".

\subsection{FIRE spectra on Magellan telescope, 2015 epochs}

The spectra of 2015 epochs were obtained on two consecutive nights (27 - 28th, April 2015) by the Folded-port Infra-Red Echellette (FIRE) spectrograph on the 6.5 m Magellan Telescope at Las Campanas Observatory, Chile. In the Echelle mode, the FIRE spectrograph provides mid-resolution ({ $R = 6000$ or $\Delta V \sim 50$ km s$^{-1}$}) spectra covering from 0.8 to 2.5 micron. The size of the slit is 0.6 arcsec and the spatial pixel scale is 0.18 arcsec on the detector. All targets were observed once each night. For every visit, four exposures were taken in an ABBA nodding pattern. In some cases, targets were acquired directly using acquisition images taken with the fixed $J$ filter. However, in many cases, blind offsetting was necessary because the target was fainter than $J$ = 20 mag. Individual exposure times are shown in Table~\ref{tab:info}.

\subsection{Data reduction and calibration}

The Echelle mode spectra from the FIRE spectrograph was reduced with FIREHOSE V2.0 pipeline based on the IDL platform \citep{Gagne2015} following similar methodologies described in Paper II. First, locations of each order on the detector are determined by the pipeline from raw images of both flat and scientific exposures. The flat field frames are then generated by combining dispersed twilight sky images taken each night. After tracing each order, the spectra are extracted by the ``optimal mode" of the pipeline via fitting Gaussian profiles. { Spatially extended H$_2$ emission is only detected in one special case, v815,} which is seen up to 0.6 arcsec away from the continuum and the emission was included as the FWHM of the extraction profile is 0.72 arcsec (see further discussions in \S \ref{sec:wind}). The telluric corrections are performed with an interactive interface, using A0-type stars observed immediately before or after the scientific exposures. 

The wavelength calibration is performed by the pipeline. For scientific exposures beyond 2.1 $\mu$m, we apply an extra step of wavelength calibration by OH emission lines from the sky background. A third-order polynomial function is fitted in each spectral order to match the measured central wavelengths with the laboratory wavelengths \citep{Lord1992}. The final wavelength calibration accuracy is 0.02 to 0.2 \AA, throughout the spectral coverage (0.8 to 2.45 $\mu$m). Correction to the heliocentric reference frame was then performed.

The pipeline-produced spectral orders were combined using a custom-written IDL program. The small overlap regions between adjacent orders are calculated as variance-weighted averages. In a few objects, the continuum slopes of telluric calibrated spectra near the overlap of the two longest wavelength orders showed unphysical curvature, which might be caused by mis-tracing the spectra of telluric standards in the dispersed images. These curved joints were artificially fixed by fitting a linear gradient across the overlap region. 

The individual spectra of each target were then averaged to produce the final spectrum. In this step, noise spikes in individual spectra due to e.g. cosmic rays or imperfect telluric correction were detected by visual inspection and the affected wavelengths were removed from the average. A good absolute flux calibration is not possible due to the need for blind offset acquisition in many cases and sometimes poor seeing conditions, so we only present spectra of relative flux in this work.

\subsection{Line profile fitting and equivalent width measurements}

Gaussian profiles are fitted to emission and absorption lines by applying the {\it gaussfit} function in IDL. Heliocentric radial velocities were calculated from the difference between the fitted line centres ($\lambda_0$) and laboratory wavelengths. Equivalent widths, ${\rm EW} =\int^{\lambda_0+3\sigma}_{\lambda_0-3\sigma}F_{\rm line}$($\lambda$)$\,{\rm d}\lambda$/$F_{\rm con}$, are used in this work to study stellar variability. This method is preferred given that the absolute flux calibration was not possible. The line flux, $F_{\rm line}$, is integrated within the 3-$\sigma$ width of the Gaussian profile around the line centre, while $F_{\rm con}$ is defined as the continuum flux at the line centre generated from the fitting procedure. The error bar is calculated following the method established by \citet{Vollmann2006}. The signal-to-noise ratio of the spectra is estimated by the continuum flux level and the standard deviation of the fitting residual. The measurements of radial velocities with respected to the local standard of rest and equivalent widths of Br$\gamma$ and H$_2$ (1.96 $\mu$m, 2.03 $\mu$m, and 2.12 $\mu$m ) lines are listed in Table~\ref{tab:EW}. { Double or multi-component line profiles are detected on a few objects. Based on the spectral resolution ($\Delta V \sim 50$ km s$^{-1}$), double-Gaussian profiles are fitted to three objects (v118, v270, v815) whose separations between peaks are greater than 100 km s$^{-1}$.
} In this work, the double-Gaussian functions are defined as
\begin{equation}
F(\lambda) = F_{\rm con} + A_1\exp\biggl[- \frac{\,(\lambda - \lambda_1)^2}{2\,\sigma_1^2}\biggr] + A_2\exp\biggl[- \frac{\,(\lambda - \lambda_2)^2}{2\,\sigma_2^2}\biggr] ,
\label{eq:dbg}
\end{equation}
where $[A_{1,2}, \lambda_{1,2}, \sigma_{1,2}]$ are Gaussian parameters. Initial guesses of these fitting parameters are given by two individual Gaussian fits near the local maxima, then are applied as inputs to the {\it mpfitfun} IDL procedure to fit the double-Gaussian profile. The fitting results will be shown in \S\ref{sec:lineprofile}.

The measurement of the equivalent width of CO overtone features beyond 2.29 $\mu$m is conducted by fitting the $K$-band continuum using 3rd order polynomial functions. Two causes of uncertainty in our calculation need to be stressed. First, the continuum level is only fitted short ward of 2.29 $\mu$m because it is not well measured beyond 2.4 $\mu$m. Second, the joint between the last two spectral orders is located at 2.295 to 2.305 $\mu$m. The loss of signal around the joint may affect the shape of the first CO overtone bandhead. Therefore we combine the equivalent width of the first and second band heads to reduce the uncertainty. 

\subsection{VVV Photometry}
\label{sec:photometry}
The 2010--2015 $K_s$ light curves (obtained by aperture photometry) of the 16 targets were provided in Paper II. However, point spread function (PSF) fitting photometry offers improved precision and reliability. For instance, PSF-based light curves were previously used to detect variable YSOs and other variable stars in 2 VVV tiles in \citet{Medina2018}. The 2010-2018 light curves were derived using an updated version of DoPHOT  \citep{Schechter1993, Garcia2012}. The light curves were drawn from a preliminary version of a full-time series and astrometric VVV catalogue (L. Smith et al., in prep).

\begin{table*} 
\centering
\caption{Radial velocity and equivalent width measurements of spectroscopic features on YSOs.}
\begin{tabular}{l | l | c c | c c | c c | c c | c }
\hline
\hline
Object & Epoch &\multicolumn{2}{|c|}{Br$\gamma$} &  \multicolumn{2}{|c|}{H$_2$ 1-0 S(1) 2.12 $\mu$m} &   \multicolumn{2}{|c|}{H$_2$ 1-0 S(2) 2.03 $\mu$m}&  \multicolumn{2}{|c|}{H$_2$ 1-0 S(3) 1.96 $\mu$m} & CO \\
\hline
&&$v_{r}$ & EW  (\AA)  &$v_{r}$ & EW (\AA)  &$v_{r}$& EW (\AA) &$v_{r}$  & EW (\AA) & EW (\AA) \\
\hline
V20
      &2014  & -54.9&    -3.6 $\pm$ 0.2& -99.8&   -9.7 $\pm$ 0.1& -86.9&    -4.6 $\pm$ 0.6&-128.0&   -10.1 $\pm$ 0.2 &-9.3 $\pm$ 4.5\\
      &2015.1& -40.0&    -5.1 $\pm$ 0.1 & -81.7&    -6.8 $\pm$ 0.1&-101.4&    -2.6 $\pm$ 0.1&-100.8&    -9.1 $\pm$ 0.3 & -8.0 $\pm$ 4.0\\
      &2015.2& -35.5&    -5.3 $\pm$ 0.2 & -82.3&    -5.9 $\pm$ 0.1& -97.4&    -2.3 $\pm$ 0.1& -95.1&    -7.8 $\pm$ 0.5 & -14.6 $\pm$ 2.9\\
\hline
V118$^*$&2013  &-121.8&    -0.9 $\pm$ 0.5& - & - & - & - & - & -& - \\
      &2015.1& -37.6&    -1.9 $\pm$ 0.4& - & - & - & - & - & - & -\\
      &2015.2& -70.5&    -3.6 $\pm$ 0.8& - & - & - & - & - & -& - \\
\hline
V270$^*$
      &2014  & 58.5 &    -1.1 $\pm$ 0.4 &-13.7&    -0.7 $\pm$ 0.1& - & - & - & - & -60.2 $\pm$ 3.2 \\
      &2015.1& 26.8&    -1.4 $\pm$ 0.5 &-46.7&    -0.6 $\pm$ 0.1&-53.7&    -0.5 $\pm$ 0.1& - & -& -89.9 $\pm$ 4.0\\      
      &2015.2&-72.0&    -1.1 $\pm$ 0.3&-46.2&    -0.7 $\pm$ 0.1& - & - & - & -& -89.9 $\pm$ 3.6 \\
\hline

V322&2013  & - & -& -25.7&    -1.5 $\pm$ 0.4& - & - & -35.4 &    -3.0 $\pm$ 0.5 & 89.7 $\pm$ 5.8\\

      &2014  & - & - & -11.9&    -5.3 $\pm$ 0.6& - & -  &-42.1&    -4.4 $\pm$ 0.7 & 46.6 $\pm$ 14.0\\
      &2015.1& - & - & -35.0&    -5.3 $\pm$ 0.1 &-70.6&    -2.2 $\pm$ 0.9&-47.2&    -9.3 $\pm$ 0.2 & 97.6 $\pm$ 14.0\\
      &2015.2& - & - & -39.7&    -5.3 $\pm$ 0.1 &-55.4&    -3.0 $\pm$ 0.9&-57.8 &    -9.1 $\pm$ 0.2 & 80.1 $\pm$ 11.1\\
\hline
V374
      &2014  & -97.2&    -2.1 $\pm$ 0.2 & -86.8&    -1.9 $\pm$ 0.1& -74.4&    -0.7 $\pm$ 0.1&-100.7&    -2.3 $\pm$ 0.6& -\\
      &2015.1&  56.0&     0.7 $\pm$ 0.1&  -78.2&    -1.7 $\pm$ 0.1& -86.7&    -0.7 $\pm$ 0.1& -95.7&    -3.0 $\pm$ 0.4& -\\
      &2015.2&  61.4&     0.6 $\pm$ 0.1&  -78.6&    -1.6 $\pm$ 0.1& -98.8&    -0.7 $\pm$ 0.1& -97.1&    -1.4 $\pm$ 0.1& -\\
\hline
V473&2013  & - & - &-139.8&   -17.0 $\pm$ 0.1&-144.4&    -4.8 $\pm$ 0.4&-150.4&   -11.5 $\pm$ 0.4& -\\
      &2015.1& - & - &-123.1&   -45.2 $\pm$ 4.7&-136.4&   -19.7 $\pm$ 0.3&-123.9&   -34.5 $\pm$ 6.0& -\\
      &2015.2& - & - &-126.9&   -36.5 $\pm$ 3.2&-142.7&   -16.3 $\pm$ 0.1&-139.1&   -46.6 $\pm$ 7.4& -\\
\hline
V631
      &2014  & -49.4&    -2.0 $\pm$ 0.2& - & - & - & - & - & - & -29.9 $\pm$ 5.5\\
      &2015.1& -47.6&    -2.0 $\pm$ 0.2& - & - & - & - & - & -  & -21.5 $\pm$ 6.8 \\
      &2015.2& -51.5&    -1.3 $\pm$ 0.1& - & - & - & - & - & -  & -3.42 $\pm$ 6.7 \\
\hline
V662
      &2014  &-112.0&    -3.6 $\pm$ 0.4 & - & - & - & - & - & - & -\\
      &2015.1& - & - & - & - & - & - & - & -& - \\
      &2015.2& - & - & - & - & - & - & - & -& - \\
\hline
V665
      &2014  & -23.0&    -2.4 $\pm$ 0.1&-99.5&    -1.3 $\pm$ 0.4& - & - & - & -  & -134.5 $\pm$ 5.4\\
      &2015.1&-28.0&    -2.5 $\pm$ 0.3&-102.0&    -2.5 $\pm$ 0.5& - & - & - & - & -83.2 $\pm$ 6.7 \\
      &2015.2&-33.2&    -2.6 $\pm$ 0.6&-103.3&    -2.8 $\pm$ 0.5& - & - & - & - & -110.0 $\pm$ 6.3 \\

\hline
V699&2013  & 14&    -4.1 $\pm$ 0.5 &-66.6&   -16.8 $\pm$ 0.3 & -68.8&    -6.3 $\pm$ 0.2&-68.6&   -20.3 $\pm$ 0.3  & -124.1 $\pm$ 7.3\\
      &2014  & -0.7&    -3.3 $\pm$ 0.3 &-67.2&   -14.6 $\pm$ 0.2 & -69.2&    -4.9 $\pm$ 0.3&-72.5&   -16.2 $\pm$ 0.4  & -75.9 $\pm$ 9.7\\
      &2015.1& -0.1&    -4.8 $\pm$ 0.4 &-78.7&   -13.9 $\pm$ 0.4 & -82.3&    -4.7 $\pm$ 0.3&-45.9&    -9.4 $\pm$ 0.3  & -127.6 $\pm$ 7.5\\
      &2015.2&-9.4&    -4.9 $\pm$ 0.3 &-73.4&   -11.8 $\pm$ 0.2 & -80.4&    -4.8 $\pm$ 0.4&-70.9&   -13.3 $\pm$ 0.3  & -150.1 $\pm$ 6.8\\

\hline
V717&2013 & - & - &-8.9&    -0.7 $\pm$ 0.1& - & - &-34.2&    -1.4 $\pm$ 0.3 & 55.5 $\pm$ 2.1\\
      &2015.1& - & -  &-24.0&    -5.0 $\pm$ 0.2&-36.7&    -2.4 $\pm$ 0.1& 3.8&   -10.3 $\pm$ 0.9 & 36.6 $\pm$ 5.1\\
      &2015.2& - & -  &-22.1&    -4.3 $\pm$ 0.3&-29.0&    -2.0 $\pm$ 0.1& 15.7&    -6.0 $\pm$ 0.3 & 21.0 $\pm$ 4.7\\
 \hline
V721
      &2014  & - & - & - & - & - & - & - & - & 44.1 $\pm$ 2.6 \\
      &2015.1& - & - & - & - & - & - & - & - & 67.3 $\pm$ 3.6\\
      &2015.2 & - & - & - & - & - & - & - & -& 60.1 $\pm$ 3.8 \\
      \hline
V800
      &2014  & - & -  & -53.6&    -1.9 $\pm$ 0.1& -62.9&    -0.8 $\pm$ 0.3& - & -& - \\
      &2015.1& - & -  & -69.9&    -2.1 $\pm$ 0.1& -84.8&    -0.9 $\pm$ 0.2& -89.0&    -1.6 $\pm$ 0.7& -\\
      &2015.2& - & -  & -69.3&    -2.3 $\pm$ 0.1& -83.7&    -0.8 $\pm$ 0.1& -94.4&    -1.8 $\pm$ 0.2& -\\
\hline
V815$^*$&2013  & - & -  &-111.7&   -91.5 $\pm$ 15.8&-120.3&   -4.0 $\pm$ 6.7&-122.7&   -79.6 $\pm$ 20.1& -\\
      &2015.1& - & -  &-56.4&  -102.9 $\pm$ 21.9&-76.7&   -39.3 $\pm$ 50.3&-100.7&   -33.6 $\pm$ 8.8& -\\
      &2015.2& - & -  &-92.7&   -69.6 $\pm$ 21.3&-121.5&   12.4 $\pm$ 8.5&-104.6&   -51.8 $\pm$ 8.7& -\\
\hline
\hline
\end{tabular}
\label{tab:EW}
\flushleft{ Dash lines indicate that the corresponding lines are not detected. The line features of AGB stars are not listed in this Table.}
\flushleft{$v_{r}$: In unit of km s$^{-1}$. For the targets with $v_{\rm LSR}$ measurements, $v_{r}$ is as the relative radial velocity to the $v_{\rm LSR}$. { One should note the spectroscopic resolution is 50 km s$^{-1}$.}}
\flushleft{$^*$: Emission lines with double Gaussian profiles. Radial velocities shown in this table are fitted by single Gaussian profiles. }
\end{table*}

They benefit from a local relative calibration of the time series photometry to minimise scatter due to uncertainty in the absolute calibration. The absolute calibration is based on the VVV photometric catalogue of \citet{Garcia2018} which is in turn derived from the CASU v1.3 VISTA pipeline \citep{Gonzalez-Fernandez2018}. The data have typical photometric errors ranging from 0.03 to 0.1 mag between $K_s$=11 and $K_s$=16 mag. A few objects are brighter than $K_s$=11 mag during the eruption, which might cause saturation. To reject unreliable detections, we applied the $\chi$ parameter from the output of DoPHOT program as a selection criterion ($\chi < 3$ for point sources) and substituted a saturation-corrected version of the CASU v1.3 pipeline aperture photometry where necessary. More details about the VVV photometry are described in Appendix \ref{sec:vvvlight curve} with light curves shown in Figure~\ref{fig:lc_sum}. 

As the seeing condition was bad ($\sim$ 1.3 arcsec) during part of the observations conducted in 2015, a few objects extended out of the slit on the acquisition images, and blind offset acquisition was usually necessary. The reliable absolute flux calibrations of the spectra were not possible, and we were also not able to schedule simultaneous imaging photometry at the time of observation. Given this circumstance, linear interpolation was applied to the VVV light curves to generate synthetic $K_s$ magnitudes for every spectroscopic epoch. A time step is defined here to estimate the uncertainty of this method as the difference of time between the spectroscopic epoch and the nearest photometric observation. Then, the error is generated based on the typical standard deviation throughout the light curve within the time step. This method is insensitive to any inter-night photometric variability between the two nights in 2015. Hence in the following sections, correlations between $K_s$ continuum flux and emission line intensities are only discussed for 1--2 year timescales.
\section{Results}
\label{sec:result}
\begin{table}
\centering
\caption{Spectral Characteristics of 2015 epochs}
\begin{tabular}{l c c c c c c c c l}
\hline
\hline
Name & Br$\gamma$ & Pa$\beta$ & Na I & H$_2$$^*$ & [Fe II]  &  CO & H$_2$O  \\
\hline
v20 & E & - & E & E & - & E & - \\
v25 & - & - & - & - & -  & A & - \\
v42 & - & - & - & - & -  & A & - \\
v118 & E & E & - & - & -  & - & A \\
v270 & E & -  & E & E & -  & E & - \\
v322 & - & -  & - & E & -  & A & A \\
v374 & A & - & - & E & -  & - & - \\
v473 & - & - & - & E & -  & - & - \\
v631 & E & E & E & - & -  & E & A \\
v662 & - & - & - & - & -  & - & - \\
v665 & E & - & E & E & -  & E & - \\
v699 & E & - & E & E & -  & E & - \\
v717 & - & - & - & E & -  & A & - \\
v721 & - & - & - & - & -  & A & A \\
v800 & - & - & - & E & E  & - & - \\
v815 & - & - & - & E & -  & - & - \\
\hline
\hline
\end{tabular}
\flushleft{Emission lines or band features are marked by `E' while absorption features are marked by `A'. $*$: H$_2$ 1 - 0 S(1) line at 2.12 $\mu$m.}
\label{tab:lines}
\end{table}

\subsection{Spectroscopic features of the mass accretion process}

{ 
Mass accretion processes dominate the stellar activities of YSOs during the protostar and pre-main-sequence stage, and are traced by various spectroscopic features. In the optical, emission lines are widely used to measure the accretion luminosity 
\citep[e.g.][]{Alcala2014}, including hydrogen recombination emission (Balmer and Paschen continuum and line series), Helium I and II emission, and metal emission lines (e.g. Ca H \& K lines, Ca II triplet, Na I D-doublet, O I, etc.).
In the near infrared, a few emission features are found as direct or indirect indicators of the mass accretion process, such as Pa$\beta$, Br$\gamma$, Na I doublet (2.21 $\mu$m), and the CO bandhead beyond 2.29 $\mu$m \citep[e.g.][]{Calvet1991, Muzerolle1998}. 

In general, there are two types of mass accretion process observed on Class 0/I YSOs. The first scenario, the so-called magnetospheric accretion model, is a steady accretion process via funnels linking the circumstellar disk and the stellar surface; the process is controlled by the stellar magnetic field. The prototype of the magnetospheric accretion scenario is AA Tau with asymmetric emission line profiles due to absorption by infalling materials \citep{Bouvier2007}. In this work, we adopted Br$\gamma$ as an indicator of the magnetospheric accretion process, while Na I doublet and CO bandhead emission are indirect indicators. In addition, Br$\gamma$ line profiles are investigated in section \ref{sec:lineprofile} to find evidence for infalling material. Besides the accretion column, highly blue-shifted ($>$ 150 km s$^{-1}$) Br$\gamma$ emission is also observed, associated with Herbig-Haro outflows in a few YSOs \citep{Beck2010}. However, the extended Br$\gamma$ flux in outflows is at least an order of magnitude smaller than the flux coincident with the central star. In this study, Br$\gamma$ is chosen as the main tracer of the magnetospheric accretion process.

The alternative accretion scenario is called boundary layer accretion, wherein mass is directly transferred inward through the accretion disc on to the star \citep[reviewed by][]{Kenyon1999}. The mass accretion rate of the boundary layer accretion process is often orders of magnitude higher than the steady magnetospheric accretion scenario, and is classified as an eruptive stage during the PMS evolution. The spectroscopic prototype of boundary layer accretion is FU Ori, whose near-infrared spectral features include a blackbody component arising in the hot inner disk \citep[T $>$ 1000 K,][]{Zhu2007}, strong CO overtone absorption and a triangular-shaped H-bandpass continuum due to water absorption by low gravity matter in the disc \citep{Connelley2018}. In contrast to the magnetospheric accretion process, no Br$\gamma$ emission was detected in FU Ori or other FUors, which makes Br$\gamma$ an ideal indicator to distinguish different accretion scenarios. The CO and water absorption features in FUors are similar to those seen in late-type stars (e.g. very young brown dwarfs and Mira variables), as noted by \citet{Connelley2018}.  The FUor-like objects in this work are distinguished from such objects by their high amplitude aperiodic photometric variability and in some cases H$_2$ emission lines. Luminosity-based arguments can also be made (see Paper II).}

\subsection{General Spectroscopic Properties}

For most of the objects, the spectra obtained in 2015 have similar properties to the spectra in the previous 2013 and 2014 epochs (published in Paper II). The physical classifications from Paper II, given in Table \ref{tab:info}, are unchanged. Spectral features are listed in Table~\ref{tab:lines}, including lines and molecular-bands. 
 
Emission lines are widely detected among the YSO spectra. Among the 14 young stars, 8 sources had Br$\gamma$ emission in the discovery spectra. In the 2015 data, Br$\gamma$ emission is detected in only 6 objects: it disappeared in v662 and turned to absorption in v374. Pa$\beta$ is only detected on v118 and v631, possibly due to the high extinction in $J$-band. The sodium doublet (Na I, 2.21 $\mu$m) shows a strong correlation with Br$\gamma$ emission, an effect that is observed on 5 objects with only one exception, v118.  As tracers of stellar winds or outflows \citep{Beck2008, Greene2010}, highly blue-shifted H$_2$ lines (1-0 S(0), 2.22 $\mu$m; S(1), 2.12 $\mu$m;  S(2), 2.03 $\mu$m) are seen in 10 out of 14 objects. Another outflow tracer, [Fe II] is only detected in v800, in the 2014 and 2015 data. 
  
In the 2015 epochs, CO bands are detected in 8 YSOs beyond 2.3 $\mu$m, among which 5 sources show emission and 3 show absorption. The 2 AGB stars also show CO absorption (by relatively cool CO as in the earlier data, see Paper II) but no other features. CO overtone emission in YSOs is usually attributed to a hot inner gaseous disc \citep{Tatulli2008, Lorenzetti2009}. We find that the CO emission is well correlated with accretion indicators: all 5 CO emitters are associated with Br$\gamma$ and Na I emission. On the other hand, there is no Br$\gamma$ emission on any object with a CO absorption feature. Deep CO absorption is only seen in v721, which has a typical FUor-type spectrum with H$_2$O absorption and no emission lines. Both CO and H$_2$O absorption became deeper when v721 was brighter. Shallower CO absorption is observed on v322 and v717, both of which were classified as MNors in Paper II because they showed FUor-like spectra in 2013 (CO absorption being attributed to the disc rather than the photosphere) but the eruption duration was less than 4 years, much shorter than classical FUors. 

\subsection{Classification by light curve type and sampling}
\label{sec:class}
The aim of this paper is to investigate the short and long term variability of eruptive YSOs to understand the physical conditions around these objects and hence obtain an insight into the accretion process and disc properties. Before starting the analysis of spectroscopic variability, we will subdivide our targets according to the shape of the light curves and the photometric changes that
occurred between spectroscopic observations. In Paper II, most of these objects were classified as MNors. Among these, 2 YSOs,  v473 and v717, had an LPV-YSO classification, denoting signs of a significant period in the 2010-14 light curves. However, visual inspection and a Lomb-Scargle analysis of the new 8 year light curves show that this possible periodic behaviour did not persist so they are reclassified as having aperiodic eruptive light curves.
 
In this work, we first apply an accumulation function analysis  (see \S \ref{sec:structure}) to subdivide the eruptive YSOs into different groups. Figure \ref{fig:structure} shows that 5 sources have relatively large { photometric} variation (as a fraction of total variation) on short timescales of 10 to 20 days. In 3 of these cases, more than half of total variation occurs in $<$ 30-day intervals. There are 5 YSOs categorized in a separate "multiple timescale" group (see below) that has a substantial variation on timescales of weeks as well as years, evident from inspection of the light curves in Figure~\ref{fig:lc_sum}. The variation exceeds 0.5~mag in $<100$~days in these systems, but these changes have no obvious connection to the inter-year variations.
The remaining 9 YSOs have smoother light curves. In terms of the times sampled by the spectra, these comprise 2 objects in a rising stage (brightening), 4 objects remaining close to maximum brightness after reaching their peak (in-outburst), 2 objects decaying toward the quiescent stage (descending), and 1 object, v800, locating at an intermediate plateau when the spectra were taken. Below we briefly describe the photometric and spectroscopic characters of the YSOs in each sub-group. 

{ Brightening:} Two objects, v374 and v721 were rising towards maximum during the spectroscopic observations. v374 continued to brighten subsequently, while v721 reached its maximum very soon after the 2015 spectroscopic epoch. The spectra of v374 have a unique variability in that the Br$\gamma$ emission line changed to absorption between 2014 and 2015, while the H$_2$ (2.12 $\mu$m) line remain constant (see Figure~\ref{fig:v374br} and \S\ref{sec:v374} for more information).  The spectra of v721 show typical characteristics of FUors \citep[see examples in][] {Connelley2018}, including the absence of emission lines, and absorption by CO and water in the $H$- and $K$-bandpasses. The CO and water absorption strengthened between 2014 and 2015 as the system brightened.

{ In-outburst:} v270, v631, v662 and v665 are classified as ``in-outburst", having reached their photometric maximum near the 2014 epoch and remaining bright in 2015. All spectra were taken in the outburst state of the light curves. V270 and v631 have FUor-like light curves with a single outburst of long duration and high amplitude ($\Delta K_s \sim $3.5~mag for v270, $\Delta K_s \sim $2.5~mag for v631). However, their spectra show typical magnetospheric accretion features rather than FUor-like absorption, including Br$\gamma$, Pa$\beta$ and Na I doublet emission, as well as CO bandheads emission. In addition, v631 exhibits water absorption in all epochs, discussed further in \S\ref{sec:631}. v662 and v665 have low amplitude outbursts ($<2$~mag) that appear to fluctuate between years, likely with a shorter total duration. In v662, the Br$\gamma$ and Na I emission lines disappeared from 2014 to 2015, during which time there was a modest ($\sim$0.5~mag) decline in brightness. V665 is an emission line object in all epochs.

{ Descending:} v322 and v717 show decaying light curves from 2013 to 2015 with $\Delta K_s > 1$~mag. Both objects reached photometric maximums around the time of the discovery spectrum in 2013, then decayed 1.45 and 1.78 mag in $K_s$ by 2015, respectively. Their spectroscopic variations on 2-year-long timescales provide an invaluable view of the cooling down process of apparently FUor-like eruptive events. The spectra of v322 and v717 are similar in 2013, with CO absorption bands beyond 2.29~$\mu$m and no emission lines. In the 2015 epochs, low velocity H$_2$ line series are detected on both objects and the CO absorption features still exist but were much shallower than previous epochs. 

{ V800:} v800 shows no obvious variability in its emission lines or $K_s$ magnitude between the 2014 and 2015 epochs. It brightened by 1~mag between 2010 and 2011, remained relatively stable until 2015 and then brightened further from 12.0 to 9.7 mag in 3 years. Hence, the spectra taken in 2014 and 2015 epochs were taken either pre-outburst (if the early faint measurements were due to extinction) or at an intermediate plateau. The spectra have blue-shifted H$_2$ lines with symmetric Gaussian profiles and [Fe II] lines associated with stellar winds or outflows. No tracers of the magnetospheric accretion process are detected, suggesting either that the accretion rate was too low or that spectroscopic features arising close to the protostar were too heavily veiled \citep[e.g.][]{Hodapp1996}.

{ Multiple timescale:} Five YSOs (v20, v118, v473, v699, and v815) are placed in the multiple timescale category defined above. Emission lines are observed on all objects in 2015, among which v118 shows Br$\gamma$ and Pa$\beta$ emission, v473 and v815 have blue-shifted H$_2$ emission, and v20 and v699 have both Br$\gamma$ and H$_2$ lines (see Table~\ref{tab:lines}). V118 has significant inter-night and inter-year changes in Br$\gamma$ flux and v815 has large inter-night changes in H$_2$ flux.

\subsection{Near-infrared spectroscopic variability: previous studies}

The near-infrared spectroscopic variability of Class I YSO has been studied on timescales from 2 days \citep{Greene2010} to years \citep{Doppmann2005, Connelley2014}. As tracers of stellar winds or outflows \citep{Beck2008, Greene2010}, shock excited H$_2$ lines show typical variations in equivalent width of 15\% on timescales of days and years, suggesting that variability of H$_2$ emission is dominated by short timescales. As a mass accretion indicator in low mass YSOs, the variability of Br$\gamma$ is seen in all the Br$\gamma$ emitters observed by \citet{Connelley2014}. A 15\% variation of equivalent width is detected on timescales from a few days to weeks, indicating an unstable accretion process similar to Class II objects observed at visible wavelengths \citep[e.g.][] {Venuti2015}.  \citet{Connelley2014} noted that the timescale of a few days is comparable with the Keplerian timescale of the inner edge of the dust disc in low mass YSOs.
The median variation increased to 30\% for time intervals exceeding 1 year, suggesting that another mechanism is operating for which the corresponding Keplerian radius is $\sim 1$ au. It is worth mentioning that the samples from \citet{Connelley2014} did not have large photometric variability during the observations ($\Delta K < 0.4$ mag in 100 days), though some had FUor-like spectra. Hence the changes in equivalent widths typically related simply to the variability of line fluxes. 

In the following sections, we discuss the spectroscopic variability of 14 eruptive YSOs on timescales of days and then years.

\begin{table} 
\centering
\caption{Long term variations of $K_s$ mag and equivalent widths of lines}
\begin{tabular}{l | c | c | c | c | c  }
\hline
\hline
Object & Epoch & $\Delta K_s ^\star$ &  LC$^\ast$& \multicolumn{2}{|c}{ $EW_{15}$/$EW_{13,14}$}  \\
\hline
& &(mag) & & Br$\gamma$ & H$_2$ 1-0 S(1)  \\
\hline
  V20 & 2014&  -0.33 & M  &1.46 & 0.66\\
 V118 & 2013&   0.45 & M & 3.01    & -\\
 V270 & 2014&   0.33 &  O &1.99    & 0.83\\
 V322 & 2013&   1.45    &D& -    &3.43\\
 V322 & 2014&   0.26    &D& -    &1.04\\
 V374 & 2014&  -0.30 & B &-0.28 & 0.98\\
 V473 & 2013&   0.07    & M &- & 2.41\\
 V631 & 2014&  -0.02 &  O &0.81 & -\\
 V662 & 2014&   0.62 &  M & 0   & -\\
 V665 & 2014&   0.41 & M &1.06    & 2.13 \\
 V699 & 2013&  -0.39 &  M &1.18 &   0.76\\
 V699 & 2014&   0.10 &  M &1.48 &  0.88 \\
 V717 & 2013&   1.78    & D &-    &  7.03 \\
 V721 & 2014&  -0.90 & B  &- & -\\
 V800 & 2014&  -0.20    & - &- & 1.15\\
 V815$^{\bullet}$ & 2013&  -0.27  & M & - & 0.33 \\
\hline
\hline
\end{tabular}
\flushleft{$\star$: $\Delta K_s = m_{K_s,15} - m_{K_s,13,14}$, the difference of $K_s$-band between 2015 and 2013 or 2014 epochs.\\
$\ast$: Category in this paper based on the photometric behaviour. B: Brightening; D: Descending; O: In-outburst; M: Multiple timescale variable.\\
$\bullet$: High velocity component}
\label{tab:Ks_EW}
\end{table}

\begin{figure*} 
\centering
\includegraphics[width=3.in,angle=0]{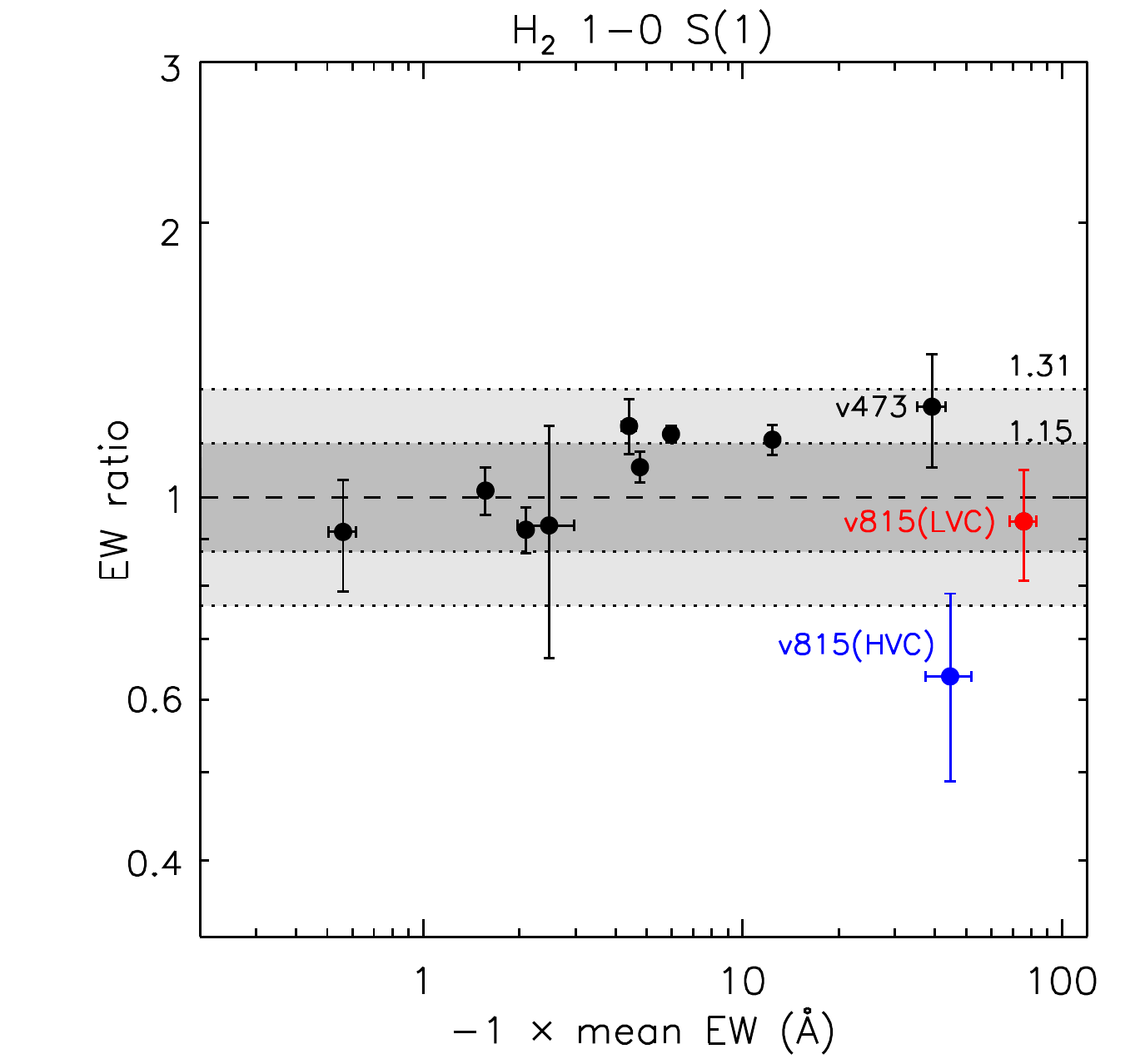}
\includegraphics[width=3.in,angle=0]{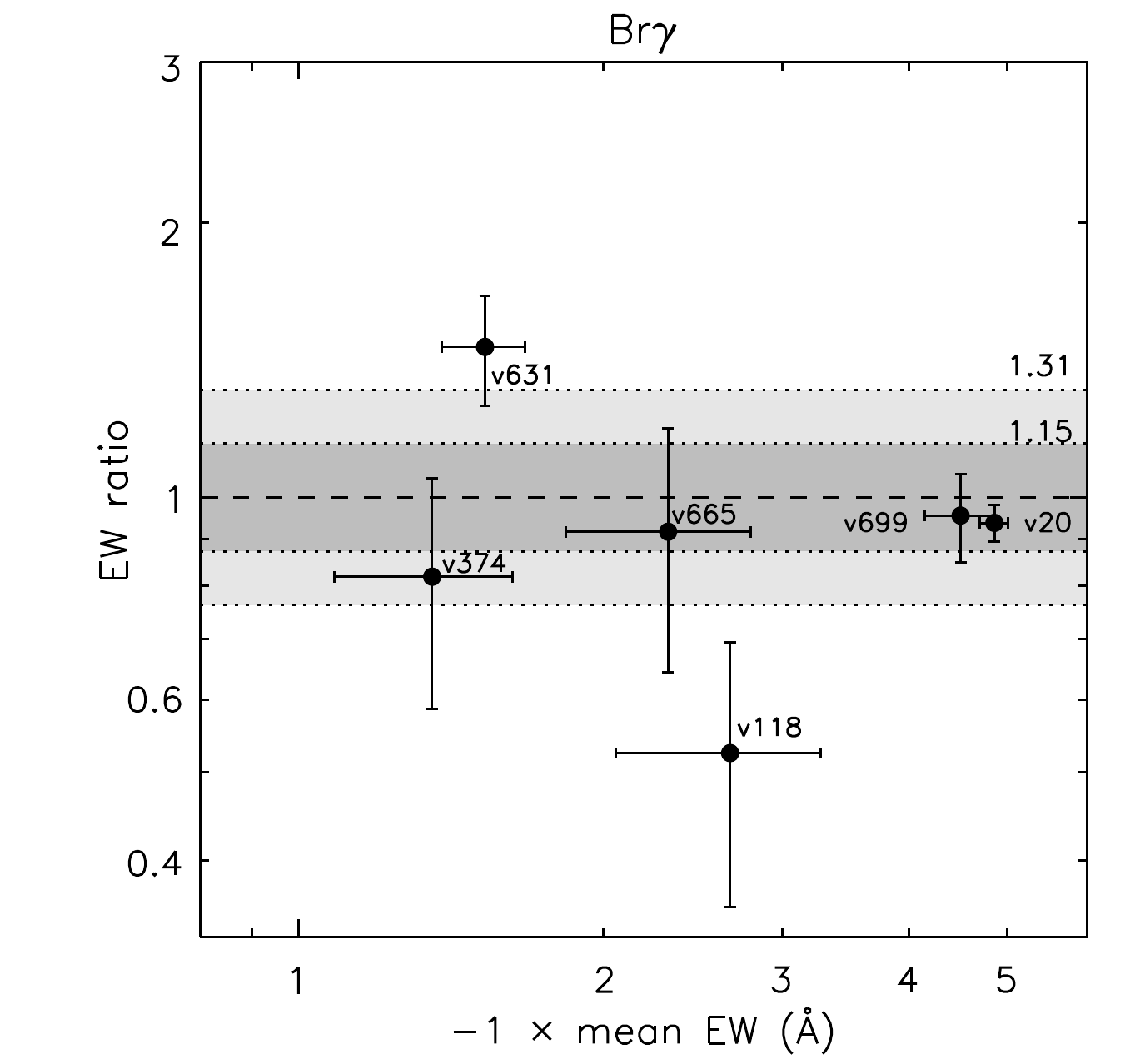}
\caption{Variations of equivalent widths of H$_2$ 1-0 S(1) line and Br$\gamma$ between two consecutive nights in 2015 epochs. In each plot, the ratio of equivalent width between two nights is displayed by the vertical axis in logarithmic scale, while the mean value of the equivalent widths in two epochs is presented on the horizontal axis. One and two times of median average deviations of EW ratios are marked by dotted lines and coloured regions. The high and low blue-shifted components of H$_2$ lines on v815 are shown in blue and red dots, respectively.}
\label{fig:short}
\end{figure*}

\subsection{Short timescale variability of emission lines}

The physical structure of the innermost disc and star-disc interaction of YSOs are revealed by their short timescale variability. The mechanisms of short timescale variability include clumpy accretion process \citep[e.g.][]{Venuti2015}, inner disc extinction \citep[e.g.][] {Bouvier2007}, launching wind or shock in jet \citep[e.g.][] {Connelley2014}, and star spots associated with stellar rotation \citep[e.g.][] {Guo2018b}. Short timescale variations of H$_2$ lines were observed on 3 out of 14 Class I objects from \citet{Greene2010}. Moreover, \citet{Connelley2014} monitored the near-infrared variability of 19 embedded YSOs. Within a 10 day timescale, the median variability of emission lines is $\sim$15\% in equivalent width. 

In Figure~\ref{fig:short}, we present the short timescale ($\sim1$ day) variability of H$_2$ (2.12 $\mu$m) and Br$\gamma$ lines between 2 epochs in 2015. The 1 and 2 times of median average deviation (MAD) of the equivalent width ratios are shown to evaluate the typical variation scales. In general, most objects have fairly modest changes in equivalent width between consecutive nights, though it is clear that measurable changes are quite common for the H$_2$ 1-0 S(1) line.  No clear trend is found between the strength and the variability of emission lines.  Two spectral indicators, H$_2$ and Br$\gamma$, have similar MAD values. Six objects have significant variation. In particular, the short time scale H$_2$ emission variations of two ``multiple timescale" variables, v473 and v815, exceeds 20\%. Separately, large variations in Br$\gamma$ emission are detected on v118 and v631, which are also the only two objects with Pa$\beta$ emission. The short timescale variations of v118 and v631 are probably similar to the variability of Balmer lines observed in CTTSs, which arises as the accretion hot spots and funnels move in and out of the line-of-sight modulated by star and disc rotations \citep{Bouvier2007, Romanova2013, Kurosawa2013}. 
 
\subsection{Long term variability: CO}   

The CO emission/absorption features observed among accreting YSOs reveal the gas structure in the inner part of the circumstellar disc. Previous theoretical work predicts that the CO overtone feature is associated with an accretion disc \citep{Calvet1991}: CO emission is produced at relatively low accretion rates whereas absorption is observed when accretion rates are extremely high. { In addition, photospheric CO absorption is also seen in class II objects \citep{Johns-Krull2001b} and late-type main sequence stars \citep[see Fig. 1 of][]{Greene2002}. However, the CO absorption generated from the inner accretion disc (e.g. FUors) has much stronger variability than the photosphere.}

Excited within a temperature range from $\sim$1500 K to 4500 K, CO overtone emission has a broad velocity profile suggesting a close-in location from 0.05 to 0.3 au on 1 M$_\odot$ stars \citep{Najita2000}. Meanwhile, observation of CO emission in the fundamental bands shows that the inner radius of CO gas is close to or within the co-rotation radius, closer than the inner edge of the dust disc \citep{Najita2003, Najita2007}. Near-infrared surveys discovered a positive correlation between CO overtone and Br$\gamma$ emission over a large spectroscopic sample \citep{Connelley2010}. However, a large scatter is also seen in this correlation which may be a consequence of stellar variability.  Here, we discuss the CO variation on 1--2 year timescales for 3 eruptive sources with CO absorption and on varies timescales for 5 eruptive sources with CO emission.

CO absorption bands are detected in v322, v717, and v721. { The temperature of the CO overtone bands are 2700~K, 1600~K, and 2700~K, respectively, determined by the CO modelling in Paper II. It seems clear that the CO overtone absorption arises in the disc rather than the photosphere, partly because of the strong
variability but also because such cool photospheres would appear much fainter than the measured light curves, at distances over 3 kpc. In this case, we regard these spectra as FUor-like.} Water absorption is seen on v322 and v721, most clearly in the $H$ bandpass. On the other hand, water is not detected in v717, which is faint in $H$.  All 3 YSOs exhibited $|\Delta K_s| > 0.9$ mag changes over a 1--2 year interval, brightening in the case of v721 and fading in the cases of v717 and v322. The variability of CO absorption was much greater on this timescale than the 1-day timescale, which allows us to study the spectroscopic variability during the rising or descending stage of eruptive objects. On all three targets, deeper CO absorption features are seen when the object is brighter. Water absorption behaves the same way in the 2 sources that display it. This is consistent with the expected behaviour of strongly accreting YSOs with optically thick CO gas in the accretion disc and disc-dominated spectra \citep{Calvet1991}. When the disc is heating (cooling), the molecular absorption features become stronger (weaker). This confirmation is valuable because the photometric variations in most classical FUors are very slow after the peak, so such changes have very rarely been seen. For example, CO absorption in the classical FUor V1057 Cyg slowly weakened over the past 30 to 40 years as the outburst faded \citep{Connelley2018}. Separately, the absence of Br$\gamma$ emission at the fainter epoch, $\sim$2 mag below the photometric maximum in v322 and v717, suggests that the magnetic field structure does not recover rapidly to pre-outburst size as the outburst declines. 

\begin{figure} 
\centering
\includegraphics[width=3.in,angle=0]{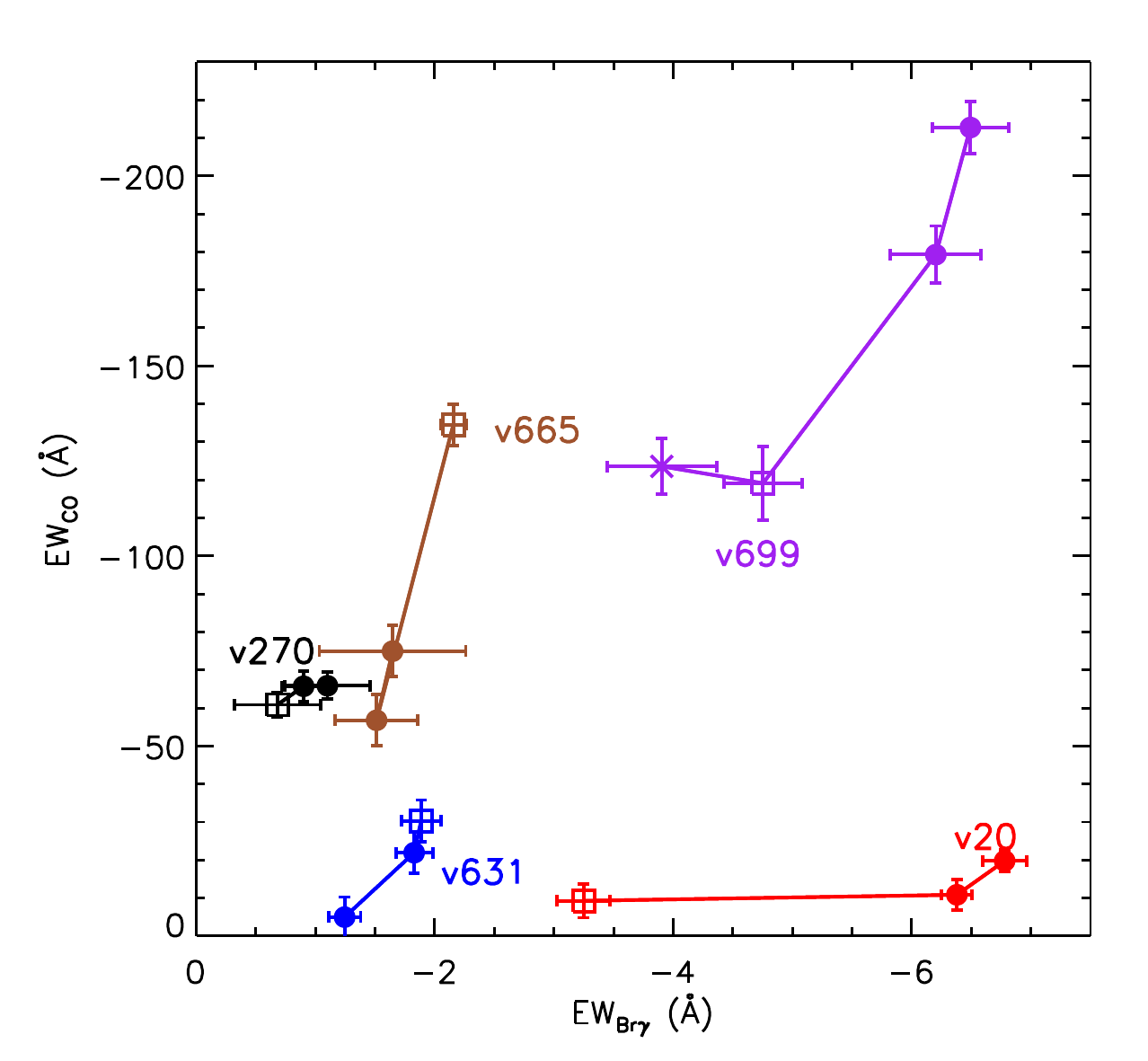}
\caption{Correlations between the equivalent widths of CO band heads and Br$\gamma$ emission. Individual objects are colour coded. observation epochs are marked by different symbols as $\times$ for 2013, $\square$ for 2014, and $\bullet$ for 2015. }
\label{fig:cobr}
\end{figure}

The variability of CO features and Br$\gamma$ emission are compared in Figure~\ref{fig:cobr}. Positive correlations are found between these two spectral features on both short-term and long-term timescales, except for the long term variation of v20 (and a weak correlation in v665). This supports the commonly held view that CO emission is associated with the accretion process. The existence of short timescale variability of CO emission also suggests a close-in location of the warm CO gas. This 1-day-timescale variability is possibly generated by CO gas with an orbital period similar to the stellar rotation, rather than an actual change in the gas density distribution.  The estimated stellar luminosities from SED fitting (Table D.1, Paper II) indicates that CO molecular gas should be able to survive within the co-rotation radius ($\sim$0.05 au).

Even with similar trends, different slopes are seen on the tracks of variations in the CO-Br$\gamma$ space, especially on year-long intervals, which is consistent with the large spread found by other surveys. This large spread corresponds to different inner disc structures among the YSOs. For individual objects, the slope changes might relate to long term effects, such as veiling effects from the stellar continuum, temporary accretion streams and piled up CO gas at the inner disc.  However, one should notice that these 5 CO emitting objects were relatively stable between observations: all of them have $|\Delta K_s| < 0.5$ mag between epochs.

\begin{figure*} 
\centering
\includegraphics[width=3.in,angle=0]{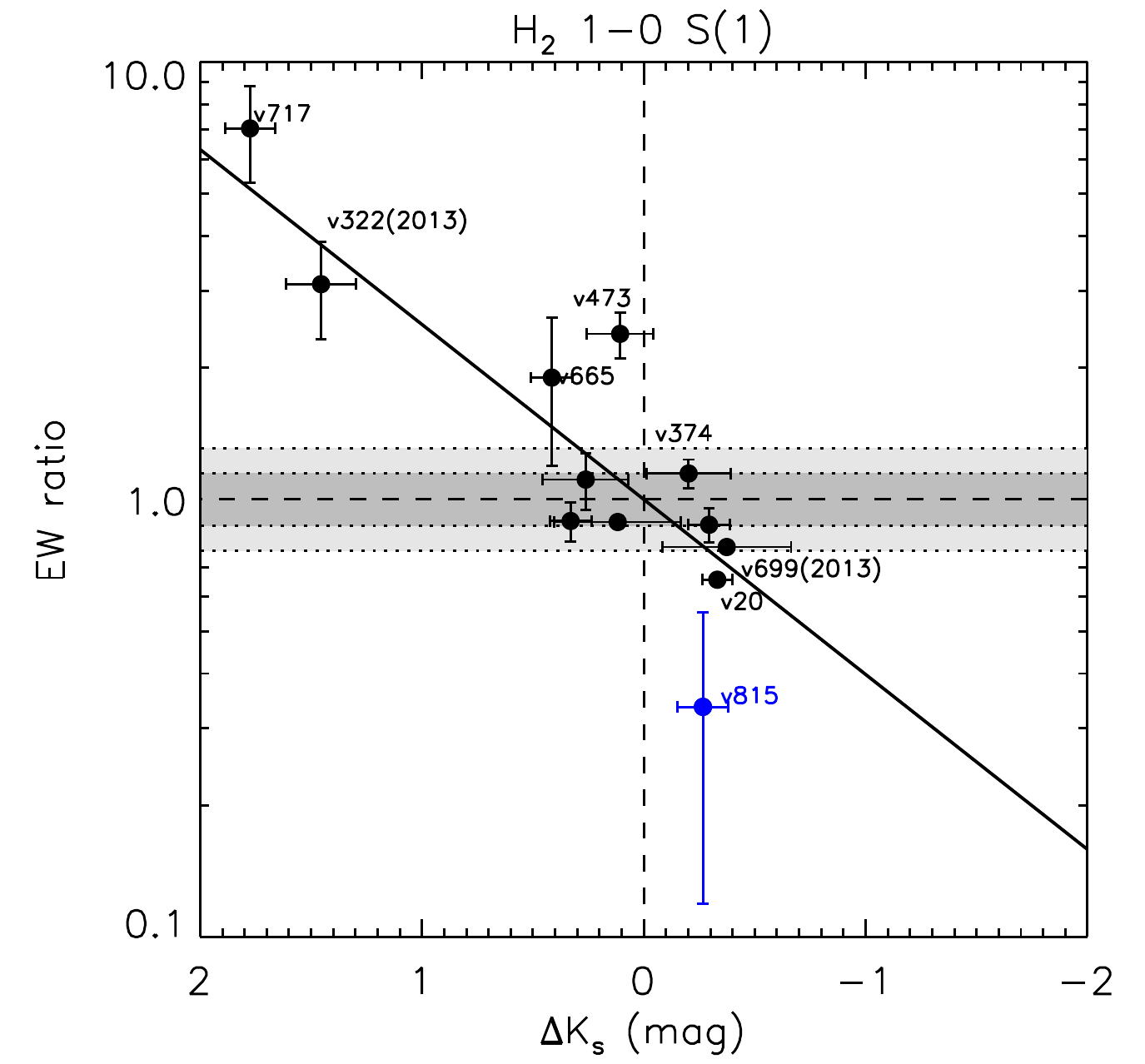}
\includegraphics[width=3.in,angle=0]{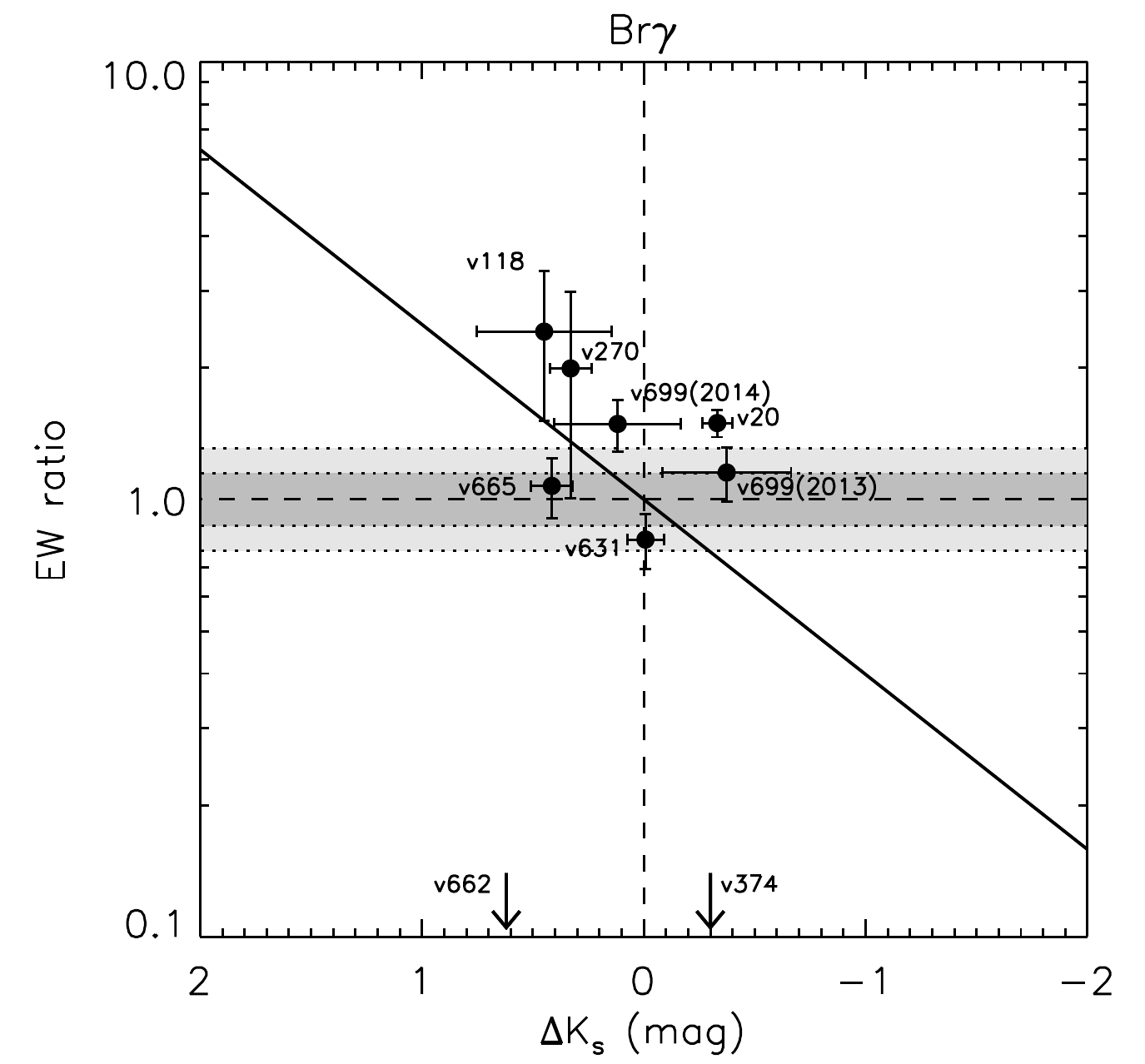}
\caption{Variations of $K_s$ magnitude and equivalent width of H$_2$ 1-0 S(1) ({\it left} panel) and Br$\gamma$ ({\it right} panel) lines on each objects between 2013/2014 and 2015 epochs. The horizontal axis, $\Delta K_s$ is defined as the mean $K_s$ magnitude in 2015 epochs minus $K_s$ magnitude in 2013 or 2014 epochs. The vertical-axis, EW ratio, is the mean equivalent width in 2015 epochs divided by the measurement in 2013 or 2014 epochs. In both panels, highly variable targets are marked out by their names. For a special object, v815, only the high-velocity component of H$_2$ 1-0 S(1) lines are shown in this figure. The thick black solid lines indicate the correlation between equivalent width ratio and stellar luminosity when assuming the line flux is not changing between the two epochs. The downward arrows in the {\it right} panel represent object v374 and v662. Both of them are lack of Br$\gamma$ emission in the 2015 epochs. The grey shadowed regions are 1 and 2 times MAD of short timescale variations (see Figure~\ref{fig:short}).}
\label{fig:magvline}
\end{figure*} 
 
\subsection{Correlation between long term variability of emission lines and $K_s$ magnitude}
\label{sec:longterm}

In this section, we investigate the relationship between the inter-year variability of emission lines and the stellar continuum. In the following, $EW_{15}$ is defined for each target as the mean equivalent width of a line measured in the two 2015 epochs. The variability of emission lines is shown by the ratios of equivalent width measured in different years. The photometric variability is presented by $\Delta K_s$ defined as $m_{K,15} - m_{K,13,14}$ with synthetic $K_s$ magnitudes generated from the VVV light curves. For the $H_2$ emission of v815, we only present the variability of the high velocity component (HVC) as the low velocity component (LVC) remains constant between epochs. 

\subsubsection{H$_2$ 1-0 S(1) line (2.12 $\mu$m)}

 \citet{Connelley2014} found a correlation between the veiling of the spectrum and the equivalent width of H$_2$ lines, indicating that strong winds are shock-excited from the hot inner disc.  In this work, we studied the correlations between the variability of H$_2$ (2.12 $\mu$m) lines and the $K_s$ magnitude, as shown in the left panel of Figure~\ref{fig:magvline}. A positive correlation would place data points in the upper right and lower left quadrants as a larger equivalent width is detected when the star is more luminous in $K_s$-band, and vice versa. However, the result shows that most objects are located in the upper left and bottom right quadrants representing anti-correlations between the stellar $K_s$-band luminosity and the equivalent width of H$_2$ emission. 
 
We consider two scenarios to understand the observations. Most simply, we assume the flux of the H$_2$ line and the $K_s$ magnitude are independent of each other, which is shown by the thick solid lines in Figure~\ref{fig:magvline}. Four objects with significant changes in $K_s$ magnitude, v20, v322 (2013 epoch), v699 (2013 epoch) and v717 are located at or around this line. This suggests that the variable equivalent widths of these objects are simply a consequence of variable continuum levels due to changes in accretion rate or extinction, while the H$_2$ lines arising in winds or outflows remain stable. It is worth noting that v322 and v717 are classified as descending objects, and also have the largest $\Delta K_s$ among all targets. 

A second physical explanation of the anti-correlation is that stellar winds or outflows are terminated by eruptive events. This is suggested by the fact that bona fide FUors with long duration outbursts never display near-infrared H$_2$ emission \citep{Connelley2018}. When an object moves into or out of the eruptive stage, one would expect an increase or decrease of near-infrared luminosity and quenching or resurgence of emission lines. In this case, the equivalent width ratio would have a steeper anti-correlation with $\Delta K_s$ than the previous assumption. Three multiple timescale variables, v473, v665, and v815, have more H$_2$ emission during the faint stage. In addition, one would also expect some delay between the onset of the accretion burst and that of the ejection burst \citep[see e.g.][]{Cesaroni2018}, which require higher cadence spectroscopic surveys on eruptive objects. 

The H$_2$ lines in other objects, including v270, v322 (2014 epoch), v699 (2014 epoch), and v800, have equivalent width ratios around the MAD for short timescales ($\sim$15\%). These objects also have relatively low photometric variability between the two epochs ($|\Delta K_s| < 0.35$ mag). The variation scales of these objects are similar to previous measurements on YSOs with low photometric variability \citep{Connelley2018}.

\subsubsection{Br$\gamma$}

Hydrogen recombination emission lines are generated by the accretion process in low mass YSOs. They are often used to estimate the stellar mass accretion rate \citep[e.g.][]{Calvet2004, Fang2009}. A tight correlation has been found between the luminosity of the Br$\gamma$ line and the accretion luminosity of Class I/II stars \citep{Muzerolle1998c, Natta2006, Manara2013, Alcala2014}. However, the correlation between the equivalent width of Br$\gamma$ and the mass accretion rate shows large scatter since emission lines are veiled by the continuum \citep[e.g. Figure 2 and 4 from][]{Muzerolle1998c}. In classical FUors, the anti-correlation between emission lines and stellar luminosity are widely seen as evidence of changing modes between magnetospheric and boundary layer accretion modes \citep{Hartmann1996}.

The correlations between $\Delta K_s$ and the variability of Br$\gamma$ are shown on the right panel of Figure~\ref{fig:magvline}. The inter-year changes in $K_s$ magnitude are fairly modest for all 8 Br$\gamma$ emitters. Despite this, 6 sources show fairly large changes in equivalent width. Notably, two objects are marked by a downward arrow as their Br$\gamma$ emission in 2015 is either turned into absorption (v374) or not detected (v662). Unlike the 1-0 S(1) H$_2$ line, there is no general correlation between $\Delta K_s$ and the variability of Br$\gamma$ emission.  Two objects, v20 and v699 (2013 epoch) show positive correlations where stronger emission lines coincide with a brighter $K_s$ magnitude. This is consistent with EXor-like behaviour where increasing $K_s$ luminosity corresponds to an enhancement of accretion luminosity. In v662, there is no Br$\gamma$ detection above the 2 $\sigma$ level when it was at a fainter stage in 2015. Anti-correlations are observed on two objects, including v118 and v270. Another object, v665, has similar line profiles and equivalent widths between two campaigns while the $K_s$ luminosity declined 0.41 mag from 2014 to 2015. The spectral gradient between 1.5-2.4~$\mu$m of v665 remains constant, i.e. there was no colour variation. Obscuration by an optically thick disc could explain the simultaneous decrease of the stellar continuum and line flux on v665. 

To summarise, the variability of Br$\gamma$ lines on timescales of 1--2 years was observed in 8 objects. Overall, there is no uniform correlation between the $K_s$ amplitude and the variation of the equivalent width of emission lines, which suggests varies variation mechanisms on different YSOs. Regarding the scale of Br$\gamma$ variability, most targets exceed the MAD of variability on a short timescale, including the absence of Br$\gamma$ emission on two objects. Among highly variables, 2 of them show positive correlations with $K_s$ amplitude while the others have negative correlations.  More specific classifications and physical interpretations will be discussed in following sections.

\begin{figure*}
\centering
\includegraphics[width=6.5in,angle=0]{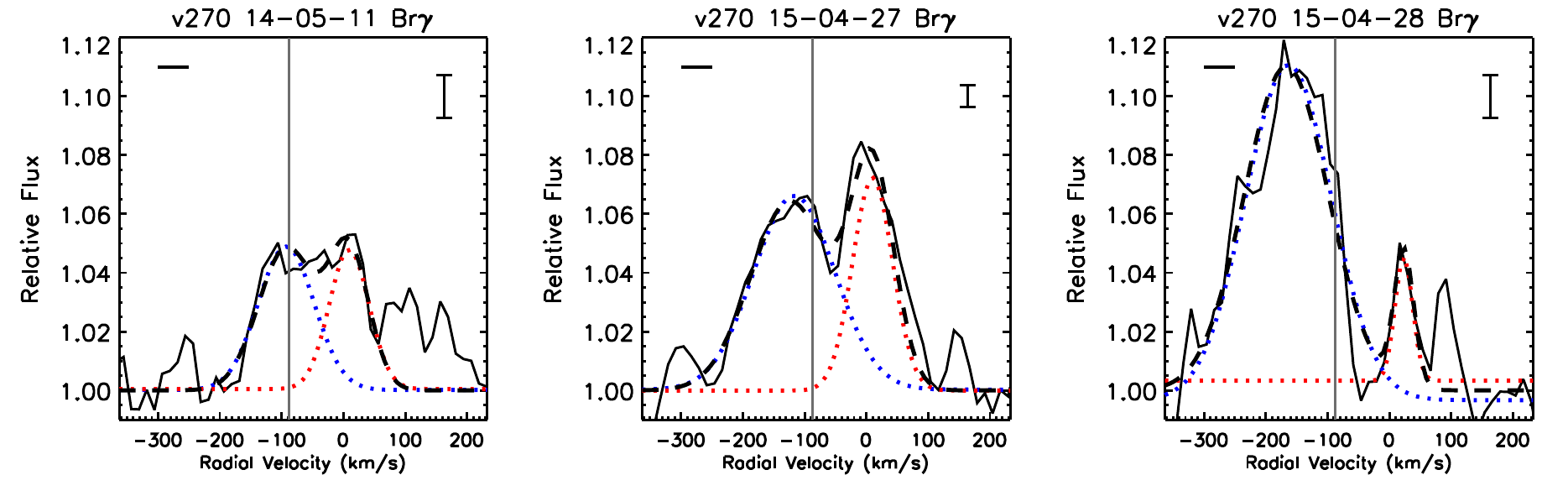}
\caption{Line profiles of Br$\gamma$ ($\lambda_0$ = 2.16612 $\mu$m) on v270 at three epochs. The observation dates are given at the top of each plot. Line profiles are fitted by Double-Gaussian functions shown by thick dashed lines, with blue and red shift components presented by dotted lines. The radial velocities are shown here after removing the system's radial velocity, $v_{\rm LSR}$ = -87.6 km s$^{-1}$, derived from the CO emission lines (Paper II). { The error bars shown in the plots are the standard deviations of the background, and the horizontal lines show the spectral resolutions.}}
\label{fig:v270}
\end{figure*}
\begin{figure*}
\centering
\includegraphics[width=6.5in,angle=0]{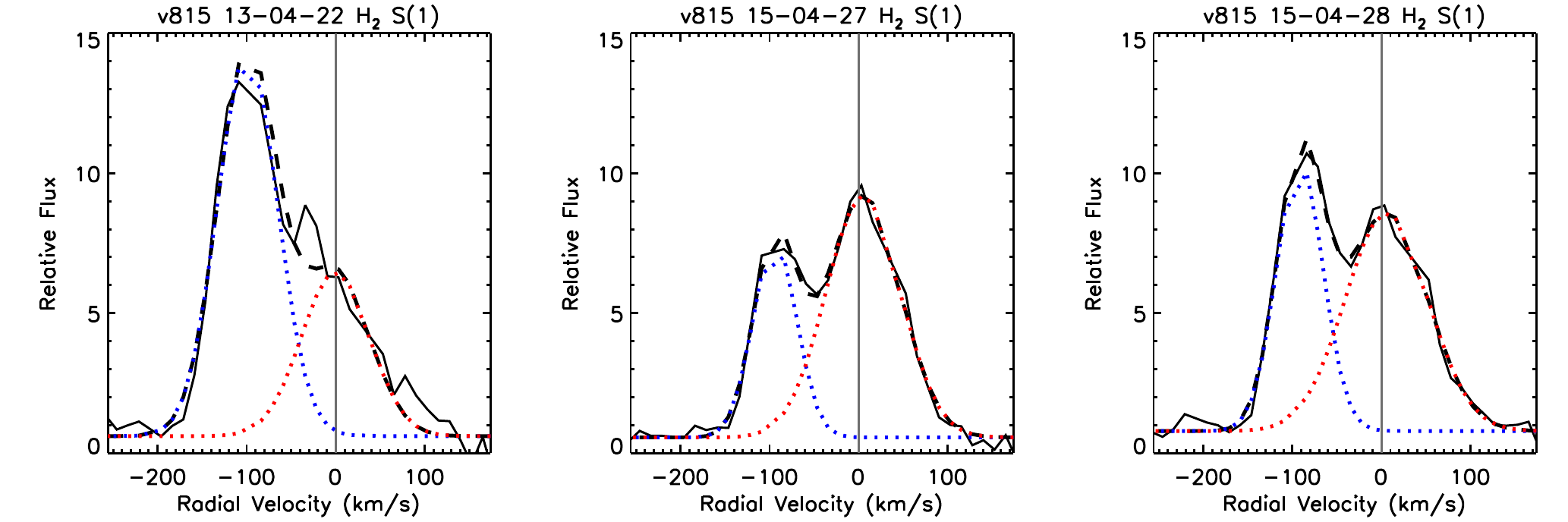}
\caption{Profiles of molecular hydrogen lines ($\lambda_0$ = 2.1218 $\mu$m) on v815 at three epochs (2013, 2015.1 and 2015.2, from left to right). The normalised spectra are shown by black solid lines, with two single Gaussian profiles shown by blue and red dotted curves, respectively. The overall double-Gaussian profiles from fitting procedure are shown by the dashed curves. The radial velocities are shown here after removing the system's radial velocity, $v_{\rm LSR}$ = - 44 km s$^{-1}$. { The separations between two peaks are $100$ km s$^{-1}$ in the velocity space, which are about twice of the spectral resolution.}}
\label{fig:v815tg}
\end{figure*}

\subsection{Variable line profiles}
\label{sec:lineprofile}

The profiles of emission lines are used to investigate the stellar mass loss and accretion flows on many YSOs, especially CTTSs \citep[e.g.][] {Edwards1987, Calvet1992, Muzerolle1998b, Chou2013, Simon2016}.  Double or multiple peak profiles are seen on the emission lines indicating complex circumstellar structures as well as star-disc interactions. For instance, two velocity components are detected in  H$_2$ and [Fe II] emission. Outflows and jets are generally traced by highly blue-shifted components with radial velocity, $v_{r} < -100$ km s$^{-1}$, while low velocity emission is generated by stellar winds \citep{Kurosawa2006, Fang2018}. By contrast, the red-shifted absorption features, or so-called inverse P Cygni profiles, that can be seen in hydrogen recombination lines indicate in-falling material \citep{Edwards1994, Johns1995, Reipurth1996, Alencar2000}.  Therefore, the short term and long term variability of emission line profiles provide clues to reveal the innermost disc structures of YSOs \citep{Scholz2005, Bouvier2007, Stelzer2007}.  In this section, we present the variable line profiles of Br$\gamma$ and H$_2$ 1-0 S(1) emission in v118, v270, and v815, respectively.

\subsubsection{Br$\gamma$}
Variable red-shifted absorption on the line profiles of Br$\gamma$ is predicted by theoretical models under the magnetospheric accretion scenario \citep{Kurosawa2013}. Near-infrared observations of classical T Tauri stars found that 72\% (18/25) of observed Br$\gamma$ emission lines have symmetric broad emission profiles { (FWHM~=~100~-~300~km~s$^{-1}$)} \citep{Folha2001}, 20\% (5/25) have inverse P Cygni profiles and the remainder have an asymmetric emission profile influenced by red-shifted absorption. \citet{Doppmann2005} observed Class I and flat spectrum YSOs and they also measured a red-shifted absorption component in a minority of systems, attributed to in-falling matter.

In this work, among the 6 targets with Br$\gamma$ emission, only 2 objects, v118 and v270, exhibit double-peaked profiles { beyond the spectral resolution ($\Delta V$~=~50~km~s$^{-1}$).} Figure~\ref{fig:v270} shows the double-peaked Br$\gamma$ line profiles of v270, { in which the line fluxes are normalised to continuum flux and the laboratory wavelengths are shifted to the $v_{\rm LSR}$ as the reference wavelengths.} A general change in the line intensity is seen between the three epochs, either resulting from variable mass accretion rate or veiling.  A red-shifted absorption component is often seen as suppression of the red-wing of the stronger blue shifted emission component. Here, short timescale variability is seen by comparing the 2 epochs in 2015: the absorption feature becomes deeper when the general line intensity is higher. This behaviour is consistent with the ``accretion funnel flow'' model originally proposed on AA Tau \citep{Bouvier2007} that the in-falling material along the line of sight is responsible for the red-shifted absorption.  The short timescale of the variability suggests that the inward accretion flow is located close to the star. The enhancement of blue-shifted emission is a result of arising wind. The deep absorption feature in the second epoch of 2015 also infers a close-to-edge-on disc inclination expected from theoretical models \citep{Muzerolle1998, Kurosawa2006}. In addition, another ``multiple timescale" object v118 shows similar low-velocity multiple peaks in Br$\gamma$ along with week-long timescale photometric variability suggesting a cognate physical origin as v270. { Symmetric single Gaussian profiles are detected in the other 4 objects at the spectral resolution (50~km~s$^{-1}$) of the data} Statistically, 67\% (4/6) of the Br$\gamma$ lines measured in this work have symmetric profiles consistent with the 72\% counted from \citet{Folha2001}.

\subsubsection{Molecular Hydrogen}

Among the 10 objects detected with H$_2$ emission, symmetric line profiles are detected in 9 targets, and only v815 shows robust multi-peak H$_2$ profiles. In Figure~\ref{fig:v815tg}, we present the variable line profiles of H$_2$ 1-0 S(1) lines for v815 over three epochs. In the 2013 epoch, the line profile is dominated by a blue-shifted high-velocity component with a low-velocity tail. In the 2015 epochs, asymmetric line profiles are fitted by double-Gaussian functions with results shown in Table~\ref{tab:EW_dbg}.  An overall feature is seen that line profiles are composed of a relatively stable low-velocity component (LVC) and a variable high-velocity component (HVC) { which are well separated at the current spectral resolution.}  In particular, the HVC has mean radial velocities of $v_{\rm LSR}$ = -143 to -135 km s$^{-1}$, while the LVC has mean $v_{\rm LSR}$ = -41 to -38 km s$^{-1}$.   Among all epochs, the $EW_{\rm LVC}$ remains constant, while the $EW_{\rm HVC}$ of S(1) line dropped from $-132.9$~\AA\ to $-34.7$~\AA\, in two years, and then increased about 57\% within a 1-day-timescale. One possible explanation is a sudden increase in the accretion rate and hence the associated outflow. We discuss the variability and physical origin of these H$_2$ lines in Section \ref{sec:wind} by looking at their spatial distributions from raw images. 
\begin{table} 
\centering
\caption{Radial velocity and equivalent width of H$_2$ 1-0 S(1) line of v815}
\begin{tabular}{l | c c | c c | c }
\hline
\hline
& RV$_{\rm HVC}$ & EW$_{\rm HVC}$  & RV$_{\rm LVC}$& EW$_{\rm LVC}$  & $\beta$  \\
\hline
epoch&   (km s$^{-1}$)&  (\AA)& (km s$^{-1}$)  & (\AA)& \\
\hline
2013  & -143 &  -132.9 & -42 & -74.6   & 1.78 \\
2015.1   & -138 &  -34.7   & -39 & -73.3 & 0.47  \\
2015.2    & -136 & -54.5   & -38 & -78.4  & 0.69 \\
\hline
\hline
\end{tabular}
\flushleft{$\beta = EW_{\rm HVC} /  EW_{\rm LVC}$ is defined as the ratio of equivalent widths between HVC and LVC in order to evaluate the variability of line profiles.}
\label{tab:EW_dbg}
\end{table}

\section{Discussion}
\label{sec:discussion}

\begin{figure*}
\centering
\includegraphics[width=6.5in,angle=0]{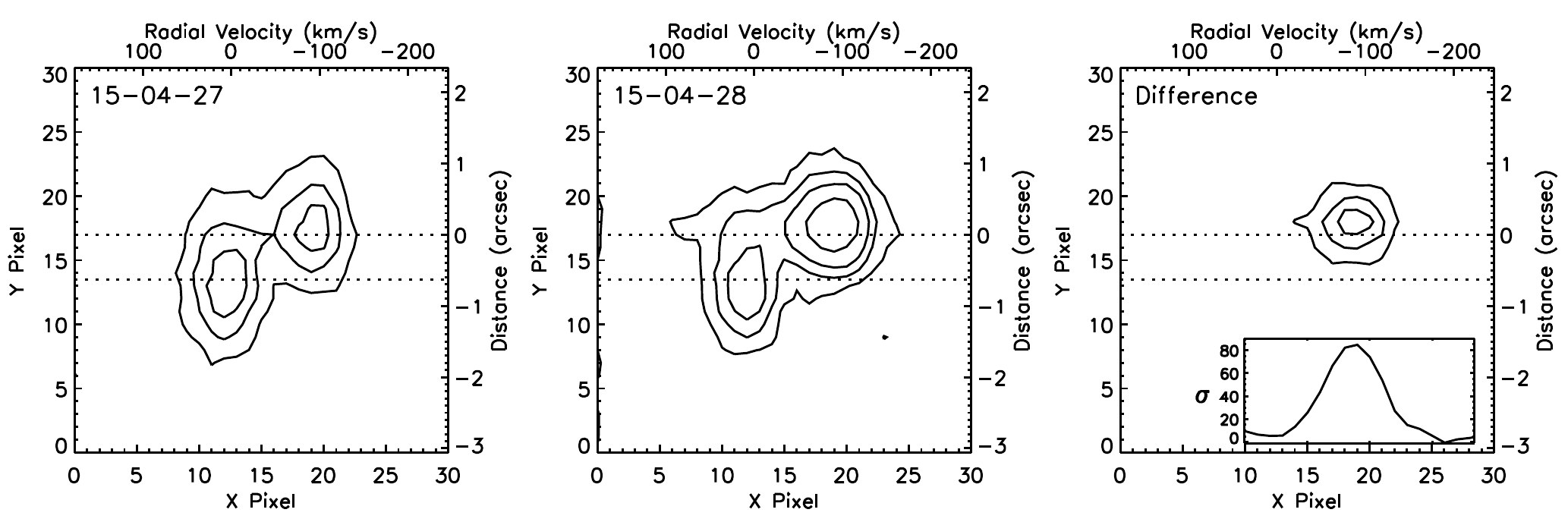}

\caption{Spatial distribution of H$_2$ 1-0 S(1) emission line of v815 on the detector. Contours represent 5, 10, and 15 times of the standard deviation of the background. The X-axis is converted to the radial velocity to the laboratory wavelength by the wavelength solution given by our reduction process. The Y-axis is transformed to the distance from the centre of continuum emission applying the angular resolution as 0.18''/pixel. From left to right, two epochs in 2015 and their difference in flux are displayed in this figure.{ The position angles of the slit are very similar in the two nights.} The integrated flux distribution along the X-axis in unit of uncertainty ($\sigma$) is shown in the $\it Right$ panel.}
\label{fig:v815con}
\end{figure*}

\subsection{Near-infrared spectroscopic variability}

As described in \S\ref{sec:class}, we subdivided the eruptive YSO sample into 5 categories based on their photometric variability and the sampling. It is impractical to compare statistics of spectroscopic variability crossing different sub-groups, so in this section, we discuss the spectroscopic variability in 3 of them: in-outburst, descending, and multiple timescale. 

{ In-outburst:}  The spectroscopic variability of this category is dominated by the short-timescale variations, presumably related to inner disc structures and stellar rotations. In particular, the asymmetric emission profiles of the Br$\gamma$ line in v270 are consistent with the tunnel flow accretion scenario seen among T Tauri stars \citep[e.g.] []{Bouvier2007}. Meanwhile, all emission features on v631 (Pa$\beta$, Br$\gamma$, Na I doublet, and CO overtone) simultaneously changed between two nights in 2015 along with a { $0.1$~mag} changing of $H-K$ colour, suggesting a rapid changing of the mass accretion rate or a veiling effect by a hot inner disc (see discussion in \ref{sec:631}).
 
{ Descending:}   Our analysis in Section \ref{sec:longterm} suggests that the changes in equivalent width of emission lines in v322 and v717 are produced by decaying of continuum levels, indicating H$_2$ emission is independent of the continuum emission from the hot inner disc. We infer that the non-thermal excitation of H$_2$ emission is not affected by inner disc conditions on these two objects, or at least has a much longer reaction timescale.  In addition, no obvious short timescale variability is seen on these two objects. 

{ Multiple timescale:}   This group of objects shows variability on both short and long timescales. Short timescale spectroscopic variability is seen on 3 objects: the variable Br$\gamma$ profile in v118 and the H$_2$ (2.12 $\mu$m) emission in v473 and v815. The former profile changing is similar to H$\alpha$ profiles on T Tauri stars \citep{Bouvier2013}, and is likely correlated with inner disc rotation. Considering the inter-year timescale, all 5 objects in this group have greater variations of EW of emission lines than on the inter-night timescale.  This indicates that the physical changes in this group are actually dominated by the inter-year variation timescale, in agreement with their photometric behaviour.  

There are 5 objects that show both Br$\gamma$ and H$_2$ emission features, including v374 (brightening), v20, v665 and v699 (multiple timescale) and  v270 (in-outburst). Despite the disappearance of Br$\gamma$ in v374, all other 4 objects have an anti-correlation or non-correlation between the two lines, demonstrating separate physical origins. The fluxes of H$_2$ lines are more stable than Br$\gamma$ emission on inter-year timescales, indicating that the location of molecular hydrogen emitting region is distinct from the innermost disc. 

By investigating the variability of eruptive objects in different categories, we find some hints of links between photometric variability and line emission. In sub-groups of long timescale photometric variables, such as brightening and descending, the measured changes in equivalent width are often dominated by the long term variation of the continuum. In the relatively quiescent groups, e.g. v800 and in-outburst, the stellar variability is either unseen or dominated by the inter-night timescale features might be associated with stellar rotation. For multiple timescale variables, even though both short- and long-term spectroscopic variations are detected, the overall spectroscopic variability is still dominated by the long term, in agreement with their $K_s$ light curves. 

\subsection{Individual objects}

In this section, we present photometric and spectroscopic analysis of four individual objects. They were drawn from different sub-categories: v815 (multiple timescale variable), v322 (descending), v374 (brightening), and v631 (in-outburst).

\subsubsection{v815: Launching of outflow on a short timescale}
\label{sec:wind}

Stellar winds and outflows are commonly detected by shock-excited H$_2$ emission among Class I objects \citep{Makin2018}. In long slit spectra, extended H$_2$ emission around YSOs is associated with a disc wind, outflow, HH object, or bow shock \citep{Davis2001, Takami2004, Bally2007, Agra2014}. Typically, two distinct components are found in the emission line profile, an HVC with a blueshift of $\sim 100$~km s$^{-1}$ and an LVC within 5~km s$^{-1}$ of the system velocity. Physically, HVCs relate to jets or outflows, and LVCs are associated with low-speed disc winds or bow shocks. In this work, extended H$_2$ emission is clearly visible in v815 in all epochs in the dispersed images. Surprisingly, the extended emission appears as a LVC at the system velocity, and a { well-separated} HVC with $v = -100$~km s$^{-1}$ relative to the system velocity is coincident with the continuum. The speed of the LVC is in fact likely to be higher (see below) given that we only measure the radial component of velocity. Hereafter we refer to it as the "offset component".

The flux ratio of the H$_2$ 1 - 0 S(1) and 2 - 1 S(1) lines is often used to distinguish excitation mechanisms. According to \citet{Gredel1995}, the flux ratio is 1.9 in the case of UV excitation, 7.7 for shock excited with $T_{\rm gas}$~=~2000 K, and 16.9 for X-ray excitation. For v815, the flux ratios of the integrated H$_2$ line profiles are 6.0$\pm$3.2, 6.4$\pm$3.0, and 9.4$\pm$4.7 at the three epochs. By assuming a typical reddening as A$_V$~= ~20 mag and R$_V$~=~3.1, these flux ratios will be increased by 17\% after de-reddening. This result rules out the possibility of pure UV excitation, similar to a recent measurement of a Class 0 protostar \citep[S68N;][]{Greene2018}. { This method is also applied to other three objects (v20, v473 and v699) which have detectable H$_2$ 2~-~1 S(1) lines. All spectra show flux ratios between 6 and 9.} On the other hand, the excitation mechanism is distinguished by the line width, as X-ray excited H$_2$ lines have low radial velocity ($\Delta v < 10$ km s$^{-1}$) and narrow line-width (FWHM $<$ 10  km s$^{-1}$) \citep[e.g.][]{Bary2003, Bary2008}. For the case of v815, the LVC and HVC have broad line widths (FWHM~$>$~80 km s$^{-1}$), at the the typical line width of thermal excitation and the spectral resolution. Therefore, we confirm that both HVC and LVC are shock-excited.

\begin{figure*} 
\centering
\includegraphics[width=3.in,angle=0]{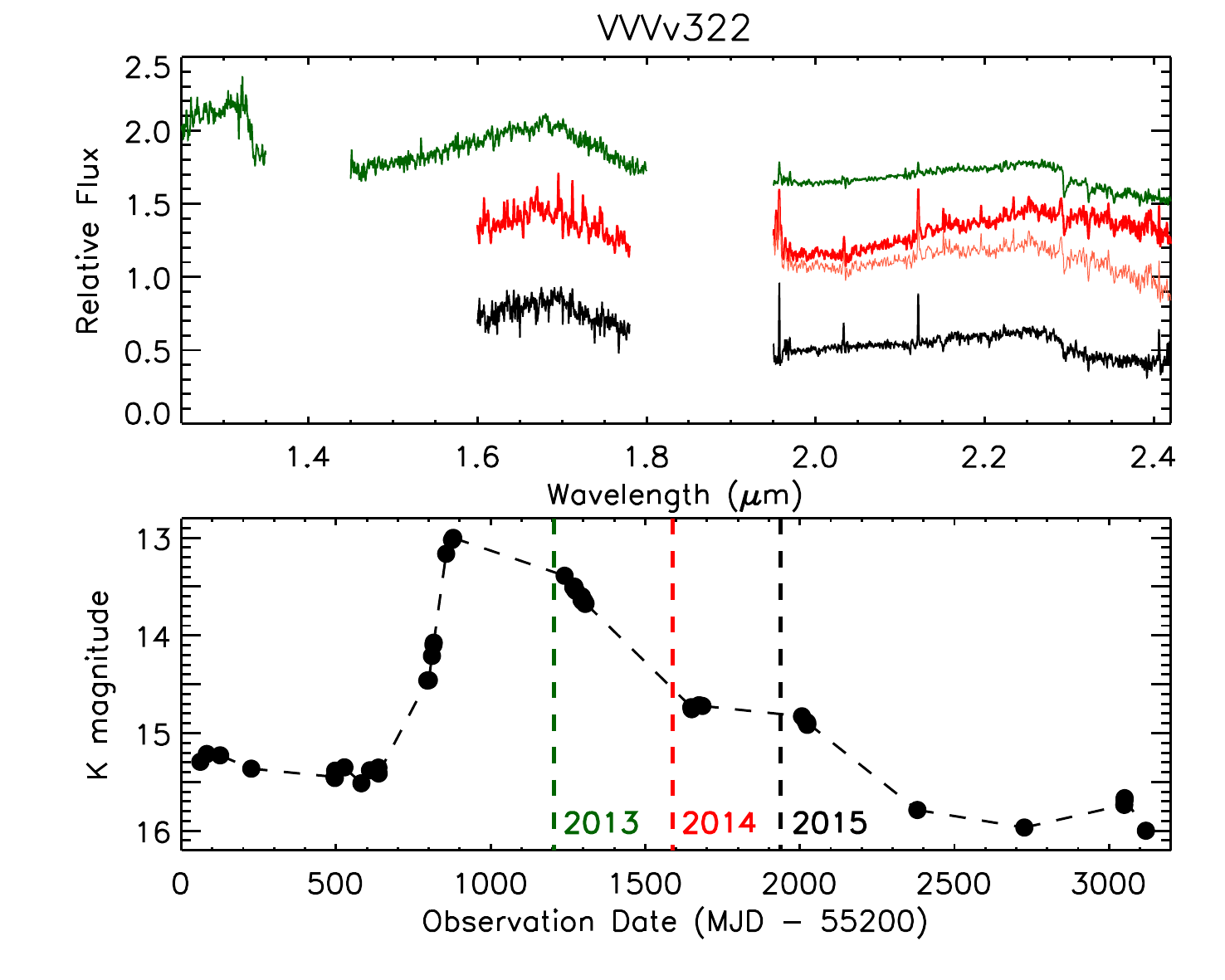}
\includegraphics[width=3.in,angle=0]{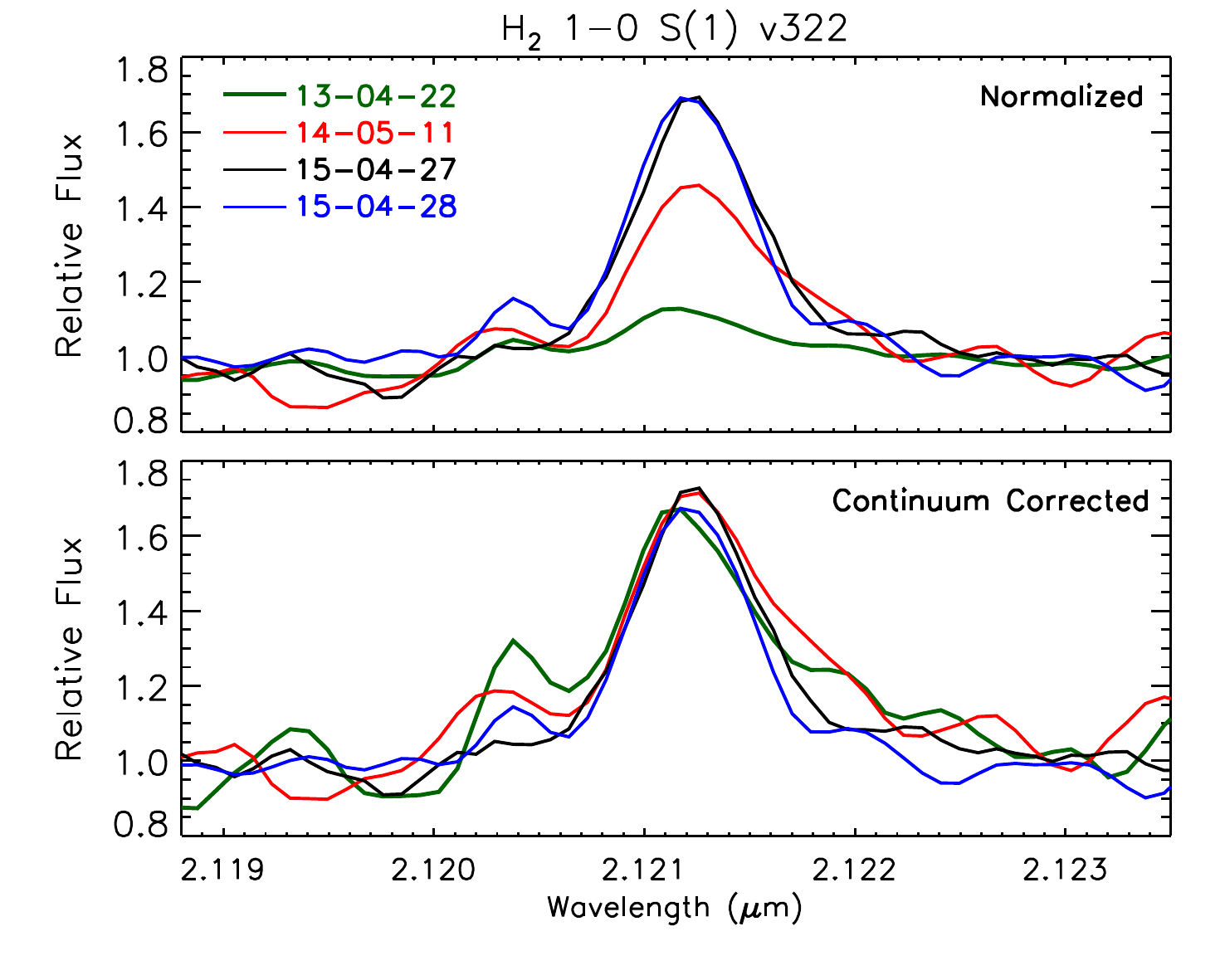}
\caption{De-reddened spectra ({\it Upper Left}) and $K_s$-band light curve ({\it Bottom Left}) of v322 in three epochs. Spectra are colour code by there observation epochs. The thin red line in the {\it Upper Left} panel represents the spectrum from 2014 epoch after removing a 400 K blackbody. The H$_2$ 1-0 S(1) line profiles are shown in the {\it Right} panels. In the {\it Upper Right} panel, spectra are normalised by the continuum levels at the line centre. In the {\it Bottom Right} panel, the variation of continuum emission is removed by $K_s$-band photometry assuming the line flux is independent to the continuum.}
\label{fig:v322s1}
\end{figure*}

The spatial distribution of the H$_2$ (2.12 $\mu$m) emission in v815 is shown in Figure~\ref{fig:v815con}. The HVC is coincident with the stellar spectrum while the offset component has is separated by about 0.6 arcsec. The projected physical separation of the offset component is 1500~au, assuming v815 is associated with the adjacent HII region [WHR97]~14222$-$6026 \citep{Walsh1997} at $d$~=~2.5~kpc. { The position angles (P.A.) of the slit in two epochs are very similar as P.A.s were set to the parallactic angle and the observation times are only shifted by 20 minutes between the two nights.} By taking the difference between two flux calibrated dispersal images in 2015 epochs ($\it Right$ panel, Figure~\ref{fig:v815con}), we found the flux of the HVC increased 62\% between two nights. Meanwhile, there is no variation above the 3$\sigma$ level associated with the offset component. The high velocity excess indicates an enhancement of the stellar outflow that excited the line of sight molecular hydrogen. Such large variability of H$_2$ emission is rarely seen on the 1-day-long timescale. 

The spatial location of the offset component is in contrast with previous observations that suggested LVCs are located close to the central star \citep{Davis2001}. Moreover, bright shock-excited near infrared $H_2$ emission by knots and bow shocks in a jet are typically associated with high velocity matter. This could simply be a geometric effect, if the apparently low velocity, offset component actually has a high velocity in a direction very close to the plane of the sky, whilst the HVC component of the outflow is close to the line-of-sight. \citet{Lucas2000} showed an example of an H$_2$ bow shock from a jet associated with the Class I YSO L1489~IRS, emerging perpendicular to the axis of the main outflow cavity. Other examples of "quadrupolar jets" have been observed in radio observations of CO outflows \citep[e.g.][and references therein]{Hirano1998}. Proposed explanations have included binarity, a precessing jet, or a single outflow with a very wide opening angle. We note that, while v815 might be a more distant source that happens to be projected in the HII region [WHR97]~14222$-$6026, there is no heliocentric distance within the Galactic disc that would cause the radial velocity of the offset component to be large, relative to a system velocity that follows the Galactic rotation curve.

In the case of v815, we observed a 62\% enhancement of molecular hydrogen outflow in a 1-day interval which implies a rapid variation in the mass accretion process. Magnetohydrodynamic simulations predict an unstable accretion process introduced by magnetic Rayleigh-Taylor instability at the inner edge of the circumstellar disc \citep{Kulkarni2008, Romanova2008, Kurosawa2013}. These unstable accretion ``tongues'' have timescales in terms of days and result in stochastic light curves like the photometric variability of v815. An alternative explanation of the fast changes in H$_2$ is through rapidly changing extinction through the line-of-sight, e.g. obscured by a warped inner disc. A highly inclined inner disc is required since there is no obvious day-to-day variation on the continuum slope.

\subsubsection{VVVv322}
\label{sec:322}

The $K_s$ magnitude of v322 rose 2.5 mag from a quiescent state to an outburst within 300 days. The photometric maximum is not observed by the VVV survey and is presumably brighter than 13 mag. After reaching its photometric maximum, v322 underwent a rapid decay, about 1.57 mag per year, lasting for at least 2.5 years. The decay timescale of v322 is shorter than classical FUors \citep{Herbig1989}, but is also longer than the timescale of typical EXors \citep[see the introduction in][] {Herbig2008}. 

Four near-infrared spectra of v322 were observed within 3 years after the burst. The 2013 epoch is soon after the photometric maximum and the remaining epochs are close to the quiescent stage. Following the similar method to \citet{Connelley2018}, the spectra of v322 are de-reddened to match the triangular-shaped $H$-band spectra of FU Ori. Here, we used the reddening curve provided by \citet{Cardelli1989} with $A_V$~=~7.0~mag and $R_V$~=~3.1. The de-reddened and normalised spectra in 3 observation epochs are shown in Figure~\ref{fig:v322s1}. An excess of $K_s$-band flux is seen on the 2014 epoch. This excess is fitted by a 400 K black body spectrum according to the photometric and spectroscopic differences between 2014 and 2015 epochs. As a comparison, this 400 K black body is removed from the 2014 epoch, and the remaining of the spectrum is shown by a thin red line in Figure~\ref{fig:v322s1}.  

Water and CO absorption are detected at all epochs, where the brightest epoch (2013) has the deepest absorption feature.  A weak H$_2$ (2.12 $\mu$m) emission line is detected in the 2013 epoch ($EW$ = $1.5 \pm 0.7$ \AA\,), and much stronger H$_2$ emission lines are seen on the following epochs. In particular, the equivalent width of the H$_2$ line increased by a factor of 3.5 between 2013 to 2015 epochs (listed in Table~\ref{tab:Ks_EW}). Meanwhile, there is no Br$\gamma$ detection in any spectra, suggesting that boundary layer accretion is occurring.

Paper II assumed the remarkable increase in the H$_2$ equivalent width arose from increased emission from a molecular outflow. However, by comparing long term photometric and spectroscopic variations, we infer that the variable equivalent widths of H$_2$ lines is due to the change in the continuum levels instead of the line fluxes (Figure~\ref{fig:magvline}). The normalised spectra near 2.12 $\mu$m are presented in the upper right panel of Figure~\ref{fig:v322s1}. An increase of equivalent width (in absolute terms) is seen from 2013 to 2015. We applied a simple ``de-veiling'' test to investigate whether the variability of equivalent width is due to changes in the continuum or in the line flux. The continuum level of each spectrum was adjusted to the level at the photometric minimum using the synthetic $K_s$ magnitude change while keeping the line flux the same. The ``de-veiled spectra'' are shown on the bottom right panel of Figure~\ref{fig:v322s1}. All line spectra from four epochs share similar line intensities, radial velocities as well as line profiles. This demonstrates that the H$_2$ emission of v322 is independent of the variation of the continuum.

The multi-epoch spectroscopic follow-up of v322 covered both the outburst and relatively quiescent phases of v322, which helps us to reveal the physical picture of this eruptive object. Molecular absorption features and similar spectral shapes to FUors are seen among the de-reddened spectra.  Our results show that the photometric variability of v322 is primarily caused by the variable emission from circumstellar material. The H$_2$ emission in the outflow or wind remains constant and is veiled by the enhanced $K_s$-band continuum. However, Br$\gamma$, the usual indicator of magnetospheric accretion, is not detected even during the relatively quiescent stage in 2015, suggesting that the stellar magnetic field is not quickly reconstructed after the burst. The VVV light curve of v322 demonstrates that this object recently decayed to a level slightly fainter than pre-outburst, i.e. below 16~mag. Therefore, Br$\gamma$ and other emission lines may be expected in the future observations if magnetospheric accretion returns. Future spectroscopy would contribute to the understanding of the reconstruction timescale of magnetospheric accretion after an eruptive event. Similar behaviours of H$_2$  equivalent width, but with smaller $\Delta K_s$, are seen in v20, v374, and v699, all suggesting variable continuum levels instead of line flux. In addition, the $K_s$ excess observed in the 2014 epoch may be explained by a warm inner disc component ($T$ = $400$ K) moving into and out-of the line-of-sight with timescale $\sim 1$~year.

\begin{figure} 
\centering
\includegraphics[width=3.3in,angle=0]{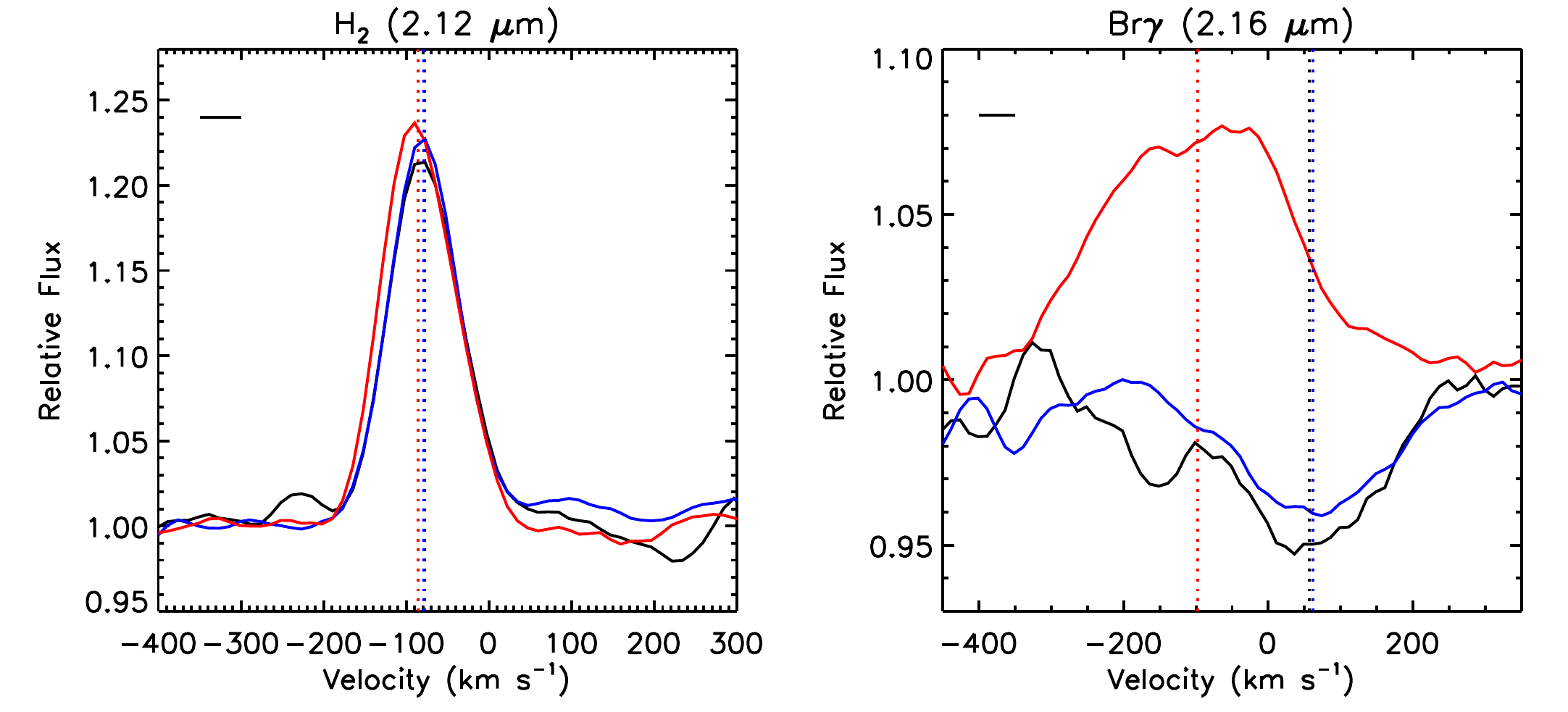}
\caption{Line profiles (H$_2$ (2.12 $\mu$m) and Br$\gamma$ (2.16 $\mu$m)) of v374 in 2014 and 2015 epochs. The spectra are normalised by the continuum. Vertical dashed lines marked the line centres from Gaussian fits. { The spectral resolution (50~km s$^{-1}$) is shown on the plots by horizontal lines}}.
\label{fig:v374br}
\end{figure}

\subsubsection{VVVv374}
\label{sec:v374}

V374 has overall amplitude $\Delta K_s$ = 2.9 mag in the 2010--2018 light curve. At the beginning of the photometric monitoring, v374 dropped 2.1 mag in $K_s$ within 200 days. The fading timescale and amplitude are comparable to EX Lupi in 2012 \citep{Juhasz2012}. A smooth brightening is then seen afterward with a local maximum in 2013. Since the outburst timescale is longer than typical EXors, v374 was identified as an MNor in Paper II.

Three near-infrared spectra are taken in the 2014 and 2015 epochs. From 2014 to 2015, the $K_s$ magnitude brightened by 0.3 mag, and has continued to brighten since then. All spectra show molecular hydrogen emission lines, while Br$\gamma$ and Pa$\beta$ emission are only seen in the relatively faint 2014 epoch. The line profiles of the 2.12~$\mu$m H$_2$ and Br$\gamma$ lines are shown in Figure~\ref{fig:v374br}. The equivalent width and central wavelength of H$_2$ 2.12~$\mu$m line both remain constant between the 2 years. In contrast, Br$\gamma$ line, which has a much broader line width ({ FWHM~=~234~km~s$^{-1}$}), turned from emission in 2014 to absorption in 2015 observations. The central velocity of Br$\gamma$ also shifted from -97.2 km s$^{-1}$ in emission, consistent with the radial velocity of H$_2$, to $+$60~km s$^{-1}$ in absorption. Meanwhile, since there is no robust measurement of the radial velocity of v374, all velocities discussed above are relative to the heliocentric velocity. The spectrum has a steeply rising continuum, suggesting that it is dominated by disc emission. Moreover, the positive velocity of the broad Br$\gamma$ absorption feature rules out a photospheric origin because Galactic plane sources in the VVV disc fields (at 295 < l < 350$^{\circ}$) have negative system velocities. Hence the absorption must arise in an in-falling optically thick accretion stream.

The variability of v374 allows a glimpse of the evolution of the mass accretion process on an inter-year timescale. Further observations would be needed to show whether this is a sustained effect due to a rising accretion rate associated with the brightening stage of the outburst. The relatively small reduction in H$_2$ (2.12 $\mu$m)  equivalent width indicates that emission from the outflow has risen along with the continuum, though not to the same extent.

\subsubsection{VVVv631}
\label{sec:631}
\begin{figure} 
\centering
\includegraphics[width=3.3in,angle=0]{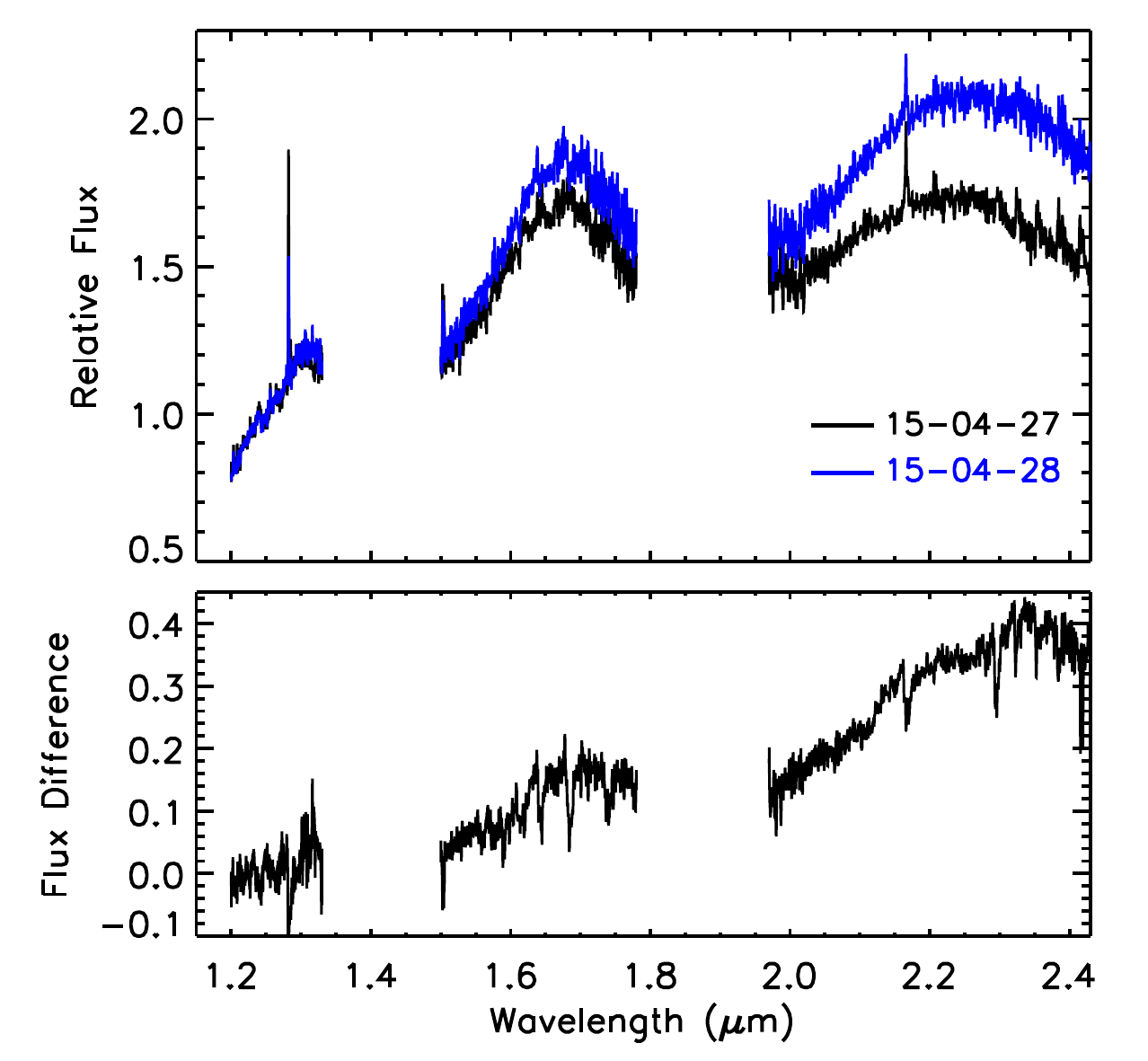}
\caption{{\it Upper panel}: Near-infrared spectra of v631 on two nights (colour coded by date) in 2015. The difference between to normalised spectra is shown in the {\it Lower panel.}}
\label{fig:v631}
\end{figure}

Day to day variation in the shape of the near-infrared continuum is rarely seen in YSOs. In the present dataset, v631 exhibits variability in the $H$ and $K$ continuum between two consecutive nights in 2015. In Paper I, v631 was identified as a flat-spectrum object, a transition stage between Class I and Class II \citep{Grossschedl2019}, based on the infrared SED slope (from 2 to 22 $\mu$m; $\alpha = -0.14$). The $K_s$ magnitude of v631 was $14.11 \pm 0.12$ mag in 1998 measured by 2MASS \citep{Skrutskie2006}, then brightened to 13.3 mag at the beginning of the VVV survey in 2010. It rapidly rose by about 1.5 mag from 2010--2011, then slowly approached to a plateau around 11.0 mag in the following 1000 days. Since then, v631 stayed at this bright stage for at least 4 more years (see Figure~\ref{fig:lc_sum}).

In this work, all spectra of v631 were taken during its bright stage. Emission lines including Br$\gamma$, Pa$\beta$, and Na I doublet, are detected, indicating ongoing magnetospheric accretion. The existence of CO overtone emission (stronger in 2014 than 2015, see Figure~\ref{fig:A3}) suggests a hot inner disc. In addition to emission features, broad water absorption bands are seen across the 1.2--2.4~$\mu$m continuum. This strange combination of CO emission and water absorption has also been observed in other two eruptive YSOs, V1647 Ori and IRAS 06297+1021W \citep{Connelley2018}, but no physical explanation has been put forward. We offer the following interpretation.
The H$_2$O molecule cannot survive at the temperature of the hot surface of the innermost disc that produces CO emission so H$_2$O is never seen in emission. We can assume that the water absorption is generated in a cooler location further out in the system but this then raises the problem of why this absorbing structure does not cause CO to appear in absorption. This can be understood if the absorbing structure is unusually cool: at $T<1500$~K the CO overtone bandheads (and many other lines in the $\Delta v=2$ bands) are weakened by the reduced population in the relevant energy levels. The classical FUor V1057~Cyg provides an example of low temperature CO absorption in a YSO (Hartmann et al.2004), attributed to an ejected dense low temperature shell, and further low temperature CO absorption spectra were presented in \citet{Contreras2014} and Paper II. By contrast, near-infrared water absorption bands become progressively stronger as temperature is reduced, as is seen in brown dwarf spectra.
 
The spectra of v631 observed in the 2015 epochs are shown in Figure~\ref{fig:v631}. The absolute flux calibration was unreliable due to poor seeing on the first night. Here, we normalised the spectra by the median flux between 1.2--1.3~$\mu$m, under an assumption that the $J$-band continuum is constant. A day-to-day variation of the pseudo-continuum slope is then detected, the $H$ and $K$-bandpass continuum having brightened from the first to the second night. { The variation of $H-K_S$ colour in two nights is 0.1 mag, measured by integrating the continuum spectra.} Meanwhile, emission lines and CO band heads are also shallower when the continuum is brighter. The flux differences between the two nights are well-fitted by an extra blackbody emission arisen from a warm inner disc ($T \sim 1000$ K) on the second night. However, it is a weak assumption that $J$-band flux remains constant between two nights. An alternative way of normalisation is by assuming the emission line fluxes are relatively constant, as applied in the previous section. Following this idea, we find $\Delta K_s$=0.99 mag between the two nights. This is unlikely because it is quite large compared with the variation scale in its current bright state ($\sigma = 0.16$ mag) as well as the min-to-max difference in the year 2015 (0.46 mag). 

This short timescale variability of the near-infrared continuum { could be consistent with} an asymmetric structure at the inner edge of the disc. Predicted by MHD simulations, the co-rotating warp structure is generated by star-disc interactions \citep{Romanova2013}. Material from the disc is piled up at one side of the inner disc forming a high density region. In the case of v631, one can expect that the extra 1.5--2.4~$\mu$m emission is contributed by an inner disc warp modulated by the star-disc rotation. However, due to the imprecise flux calibration, we cannot draw an estimate on the height or column density of this warp system. It is still interesting to find such an asymmetric inner disc structure around a flat-spectrum YSO.

\section{Summary}
\label{sec:conclusion}

We have investigated the near-infrared variability of 14 previously identified eruptive protostars via a spectroscopic study with the Magellan telescope, sampling time intervals of 1 day and 1--2 years. Variability in the following features was studied: Br$\gamma$, H$_2$ (2.12 $\mu$m), CO overtone bands, and the $K$-band continuum. From days to years, different variation timescales correspond to different radial locations in the inner disc, often related to the Keplerian orbital timescale. 

The eruptive objects were sorted into groups based on the decade-long photometric behaviour and the portions of the light curves that were sampled. YSOs for which the spectra sampled long term photometric changes were named as ``brightening'' or ``descending'', as appropriate. YSOs which remained at a brighter state were termed ``in-outburst''. Objects with both short- and long-term variability are categorised as ``multiple timescale''. The spectra of the YSO v800 were all taken at a stable, fairly low state so it did not belong to these groups. The results of our spectroscopic monitoring are summarised as follows.

$\bullet$ Major spectroscopic variability is dominated by changes connected with photometric variation on timescales longer than a few days, e.g. an outburst timescale. Spectroscopic variability shows some degree of similarity within each of the groups listed above. Strong short-timescale variations are mostly found in the multiple timescale category, wherein large photometric variations can occur quickly.  Changes in the gradient across the 1.5--2.4~$\mu$m continuum are detected in a few objects. These are well fitted by contributions from a warm black body ($T = 400 - 1000$~K) rather than changes in line-of-sight extinction.

$\bullet$ Most eruptive objects (12/14) have emission lines in their spectra. H$_2$ lines are observed in 10/14 YSOs. This verifies the strong correlation between outflows and eruptive systems found in Paper II, where the incidence was of H$_2$ emission was noted as being higher than in non-eruptive YSOs. { The measurements of the radial velocity and line flux ratio between H$_2$ lines all agree with the shock excitation mechanism of H$_2$ emission.} Two objects without H$_2$ emission show strong hydrogen recombination emission, consistent with the magnetospheric accretion scenario. No emission features were detected on the FUor v721 at any epoch, nor on the FUor-like system v322 in the 2015 observations. In addition, CO overtone bands are observed in 8 YSOs: 3 in absorption and 5 in emission.

$\bullet$ On a 1-day interval, the median average changes in the equivalent width of Br$\gamma$ and H$_2$ are 15\%. The short timescale variation of Br$\gamma$ is consistent with its origin near the centre of the system, as part of the magnetospheric accretion process on to the central star. Variable line profiles are seen on 2 objects, attributed to modulation by orbital motion of matter with a period comparable to the stellar rotation. The inter-night variations in CO emission confirm the location of CO gas close to the protostar and a positive correlation between the {\it short-timescale} variability of CO emission and Br$\gamma$ emission implies that there is a connection between the Br$\gamma$-emitting and CO-emitting regions.
H$_2$ emission is usually relatively stable on the inter-day timescale, in comparison with the inter-year variability. However, one particular YSO, v815, exhibited a 60\% enhancement of its high velocity molecular hydrogen outflow between two consecutive nights, which might be due to a sudden increase in stellar mass accretion rate triggered by disc fragments. Short timescale spectroscopic variability is more commonly detected among ``multiple timescale'' and ``in-outburst'' objects. 

$\bullet$ On the year-to-year timescale, both Br$\gamma$ and H$_2$ lines show statistically greater variability than on short timescales. However, the inter-year photometric changes were fairly small in all of the Br$\gamma$-emitters which limits the conclusions we can draw for this important feature. With this caveat in mind, a range of positive and negative correlations are seen between the variability of Br$\gamma$ and $\Delta K_s$ for different sources, suggesting that the magnetospheric accretion process is not the main variable mechanism for every observed change in eruptive objects. An anti-correlation is found between the equivalent width of H$_2$ lines and $\Delta K_s$ in YSOs with large changes in $K_s$, due to the H$_2$ flux remaining constant between epochs while the continuum flux changes. We conclude that H$_2$ emission is stable among photometric variable YSOs over a 1 to 2 year interval and hence infer that the H$_2$ emission is generated further out from the inner edge of the disc. 

$\bullet$ Three sources with CO overtone absorption show a strong positive correlation between the brightness of the $K_s$ continuum and strength of CO absorption. This indicates that optically thick CO gas is abundant in the accretion disc of these objects, similar to the disc structure of classical FUors \citep{Calvet1991}. The broad water vapour absorption bands that are detected in two of these three sources show a similar correlation between absorption strength and the brightness of the $K_s$ continuum. This implies stronger emission from the circumstellar disc with a steeper disc vertical temperature gradient when the accretion rate is higher. This indicates  that relatively fast VVV FUor-like outburst systems have a physical structure that is in line with the accepted picture for classical FUors.

$\bullet$ Both CO emission and water absorption bands are detected on the YSO v631. Day-to-day variability of CO emission band heads is detected as evidence of its location in the hot inner disc, where the H$_2$O molecule does not survive. We explain the unusual observation of CO emission and water absorption as a consequence of the different temperature dependences of CO absorption and water absorption. If molecular absorption occurs in a structure far out in the system at $T<1500$~K, the CO absorption is weakened, whereas water absorption strengthens at low temperatures. 

$\bullet$ Comparing with the previous near-infrared spectroscopic survey of stable Class I YSOs \citep{Connelley2014}, the VVV eruptive objects exhibit similar spectroscopic variability on day-to-day timescales ($\sim$ 15\%). On 1--2 year timescales, eruptive objects show stronger variability, such as veiling of H$_2$ emission due to a brighter continuum, the disappearance of Br$\gamma$ emission in 2/8 Br$\gamma$-emitting sources and the consistent correlation between the variation of CO absorption and $K_s$ continuum flux. 

This is a first study of the spectroscopic variability of the eruptive YSOs from the VVV survey. The Br$\gamma$ emitting sources had fairly small photometric changes during the observation period and typically only 3 spectra were obtained for each source. Further near-infrared spectroscopic monitoring, ideally with higher cadence, is warranted to investigate variation in the accretion-sensitive Br$\gamma$ line as luminosity changes.

\section*{Acknowledgements}

ZG and PWL acknowledge support by STFC Consolidated Grants ST/R00905/1, ST/M001008/1 and ST/J001333/1 and the STFC PATT-linked grant ST/L001403/1. We thank the staff of the Magellan Telescopes and the European Southern Observatory for their work in operating the facilities used in this project.

We gratefully acknowledge data from the ESO Public Survey
program ID 179.B-2002 taken with the VISTA telescope, and
products from the Cambridge Astronomical Survey Unit (CASU).
This paper includes data gathered with the 6.5 meter Magellan Telescopes located at Las Campanas Observatory, Chile during Chilean programes: CN2014A-16 and CN2015A-78. D.M. and J.A.-G. thank support from the BASAL Center for Astrophysics and Associated Technologies (CATA) through grant AFB170002  and FONDECYT Regular grant No. 1170121. D.M., J.B. and R.K. thank support from the Ministry for the Economy, Development and Tourism, Programa Iniciativa Cientifica Milenio grant IC120009, awarded to the Millennium Institute of Astrophysics (MAS),

A.C.G. acknowledges funding from the European Research Council (ERC) 
under the European Union's Horizon 2020 research and innovation 
programme (grant agreement No.\, 743029).
\label{lastpage}

\bibliographystyle{mnras}
\bibliography{reference}

\appendix
\section{VVV light curves}
By using the VVV $K_s$-band light curve from 2010 - 2014, 816 high variable objects are detected in Paper I with $\Delta K_s$ > 1.0 mag. In this work, we applied the PSF-photometry based light curves extracted by LS with DoPHOT. This new dataset extends the time coverage to 9 years, and provides more robust classification of the photometric variability. Unreliable detections are excluded from the final light curve. The design of the VVV survey allows each source being observed at least twice in one photometric epoch \citep{Minniti2010}. In this case, light curves presented in Figure~\ref{fig:lc_sum} are constructed by averaging $K_s$ magnitude taken within one photometric epoch. The corresponding exposures are ejected from the combined light curve if the magnitude difference within one epoch is greater than 5 times of the photometric uncertainty. The light curves are provided in Table~\ref{tab:lightcurve}. 

\subsection{Notes on individual light curves}

A few objects show different photometric magnitude comparing with the previous published light curve in Paper I. VVVv374 has $K_s$ = 11.12 $\pm$ 0.02 mag from the recent PSF photometry in an epoch around MJD 57227 while the aperture photometry used in Paper I gave 12.032 $\pm$ 0.002 mag. This disagreement results in a divergent classification of the photometric variability as either the eruptive stage of v374 terminated in 3 years or it remains in an rising phase. We manually checked the $K_s$-band imaging data at corresponding epochs from the VISTA Science Archive, and then calculated the relative magnitude of v374 against nearby field stars by applying simple aperture photometry. The measurement agrees with the recent PSF photometric results as v374 stayed at a brightening stage in 2015.  In addition, the first photometric epoch of v118 is rejected by the extremely high $\chi$-value due to saturation. Hence, we adopted the aperture photometric result from Paper I as $K_s$ = 10.6 $\pm$ 0.4 for this particular epoch.

Regular photometric maxima are detected on v42 for every two years, where the $K_s$ magnitudes are brighter than 10 mag and are close to the saturation limit. In these cases, PSF photometry is insufficient to fit the profile and to measure the brightness. For this particular target, we chose to adopt the aperture photometric results from Paper I between 2010 and 2014. Then, for the recent years, the light curve of v42 is provided by the latest PSF photometry.
\label{sec:vvvlight curve}

\subsection{Accumulation function of VVV light curves}
\label{sec:structure}

\begin{figure} 
\centering
\includegraphics[width=3in,angle=0]{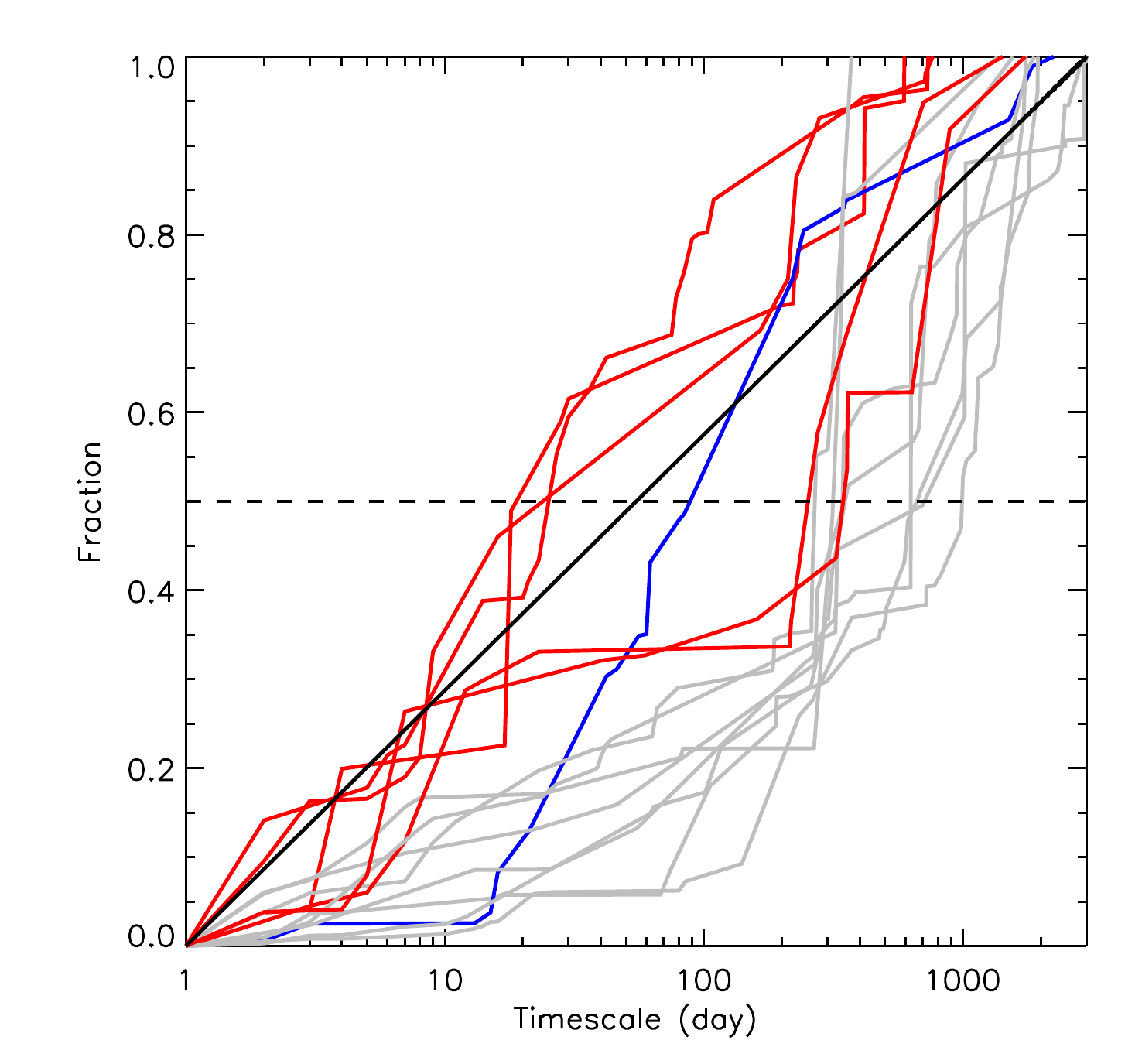}
\caption{Accumulation functions of the $K_s$-band photometric variabilities 14 eruptive YSOs observed in this work. The diagonal black solid line representing a flat increment of photometric variation scale. The horizontal dashed line marks out the 50\% of total variation scale. The red lines representing ``multiple timescale" variables classified in \S~\ref{sec:class}. V322 is marked out by the blue line, while other objects are in grey.}
\label{fig:structure}
\end{figure}

\begin{figure*} 
\centering
\includegraphics[width=6.5in,angle=0]{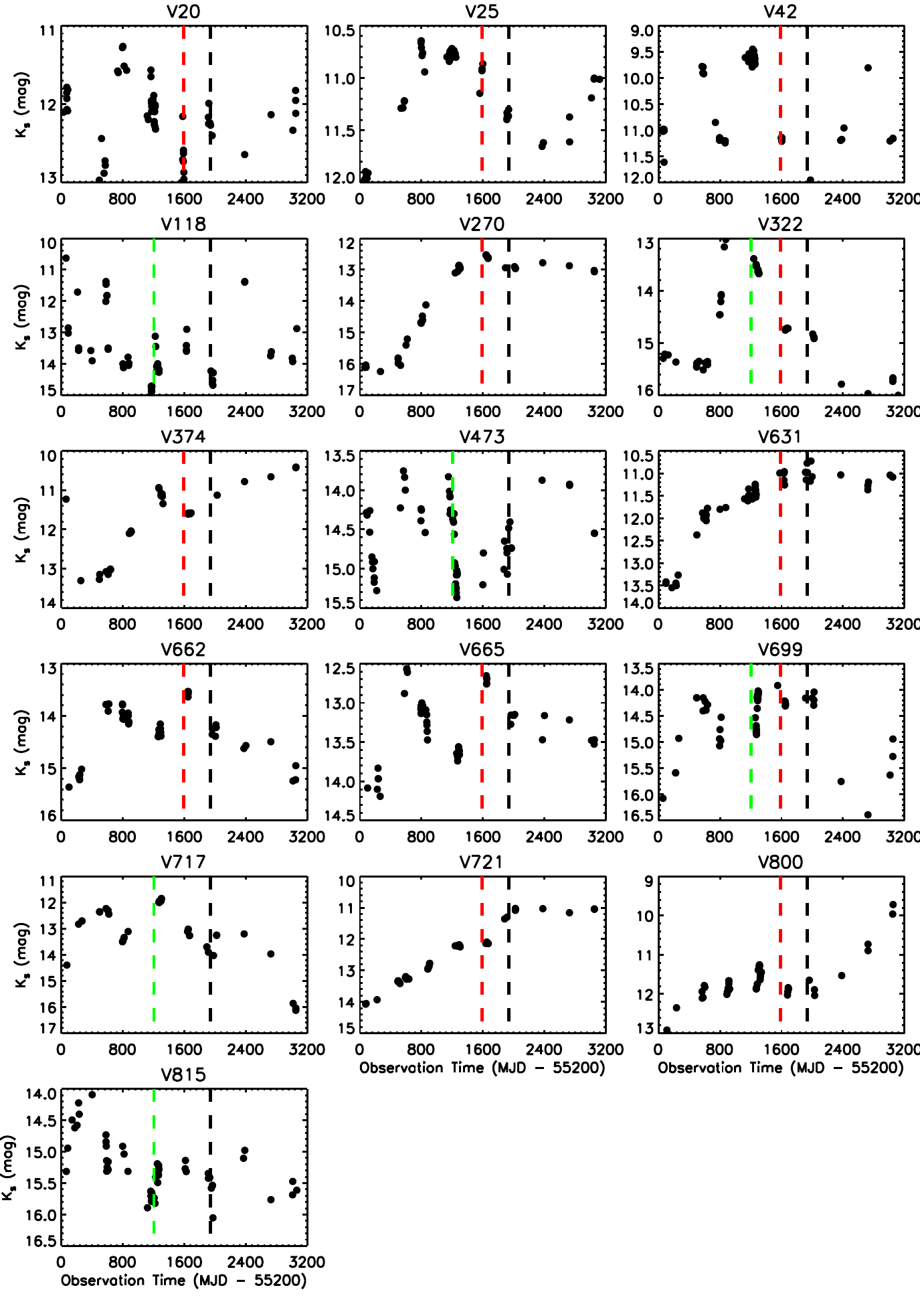}
\caption{$K_s$-band light curves from the Vista Variables in the Via Lactea (VVV) survey. Vertical dashed lines mark out the spectroscopic epochs: 2013 ({\it green}), 2014 ({\it red}), and 2015 ({\it black}).}
\label{fig:lc_sum}
\end{figure*}

To quantify the photometric variability on different timescales, the accumulation function is applied to analysis the VVV light curve. First, the variation timescale ($t$) is set from 1 to 3000 days.
Then, the photometric amplitude ($\Delta K_{s,t_i}$) for each time step ($t_i$) is defined as the maximum $K_s$-band variation within $t_i$. The accumulation functions (shown in Figure~\ref{fig:structure}) between $\Delta K_{s}$ and $t$ are generated on every object to describe the growth of photometric variability through day- to decade-long timescales.

Two groups of variation behaviours are found among eruptive objects. { Five YSOs classified as ``multiple timescale" variables are marked by red lines in  Figure~\ref{fig:structure}, which exhibit rapid photometric variability in timescale that comparable to stellar rotation period ($< 10$~days).} Among them, three YSOs (v20, v118, and v432) show variation reaching 50\% of the photometric amplitude in 30 days. Other objects have low variability within 200 days, but increase fast beyond that. One particular object, v322, exhibited a rapid variation in timescales between 20 to 200 days corresponding to the burst in 2012 (see~\S~\ref{sec:322}). This bipolar distribution is explained by two different variation mechanisms.  The rapid variables are dominated by unstable inner disc structure or clumpy accretion process, while the long-term variables are more like classic eruptive objects like EXors or FUors. 

\begin{table} 
\centering
\caption{VVV light curves of 16 YSOs observed in this paper}
\begin{tabular}{l | c | c }
\hline
\hline
Name &	Time (MJD) &  $K_s$ (mag) \\
\hline
v20  &     55227.2   &    12.0590 \\
v20  &     55269.2    &   11.8133 \\
v20  &     55270.2    &   11.7424 \\
v20  &     55271.2    &   11.8852 \\
v20  &     55272.2    &   12.0250 \\
v20  &     55281.1    &   12.0424 \\
v20  &     55283.2    &   11.7726 \\
v20  &     55693.0    &   12.9246 \\
v20  &     55722.0    &   12.3949 \\
v20  &     55755.0    &   12.8371 \\
... & ... & ... \\
\hline
\hline
\end{tabular}
\flushleft{Note: The light curves were drawn from a preliminary version of a full-time series and astrometric VVV catalogue (L. Smith et al., in prep). The full table will be available as online supplementary material.}
\label{tab:lightcurve}
\end{table}

\begin{figure*} 
\label{fig:A3}
\centering
\includegraphics[width=3.3in,angle=0]{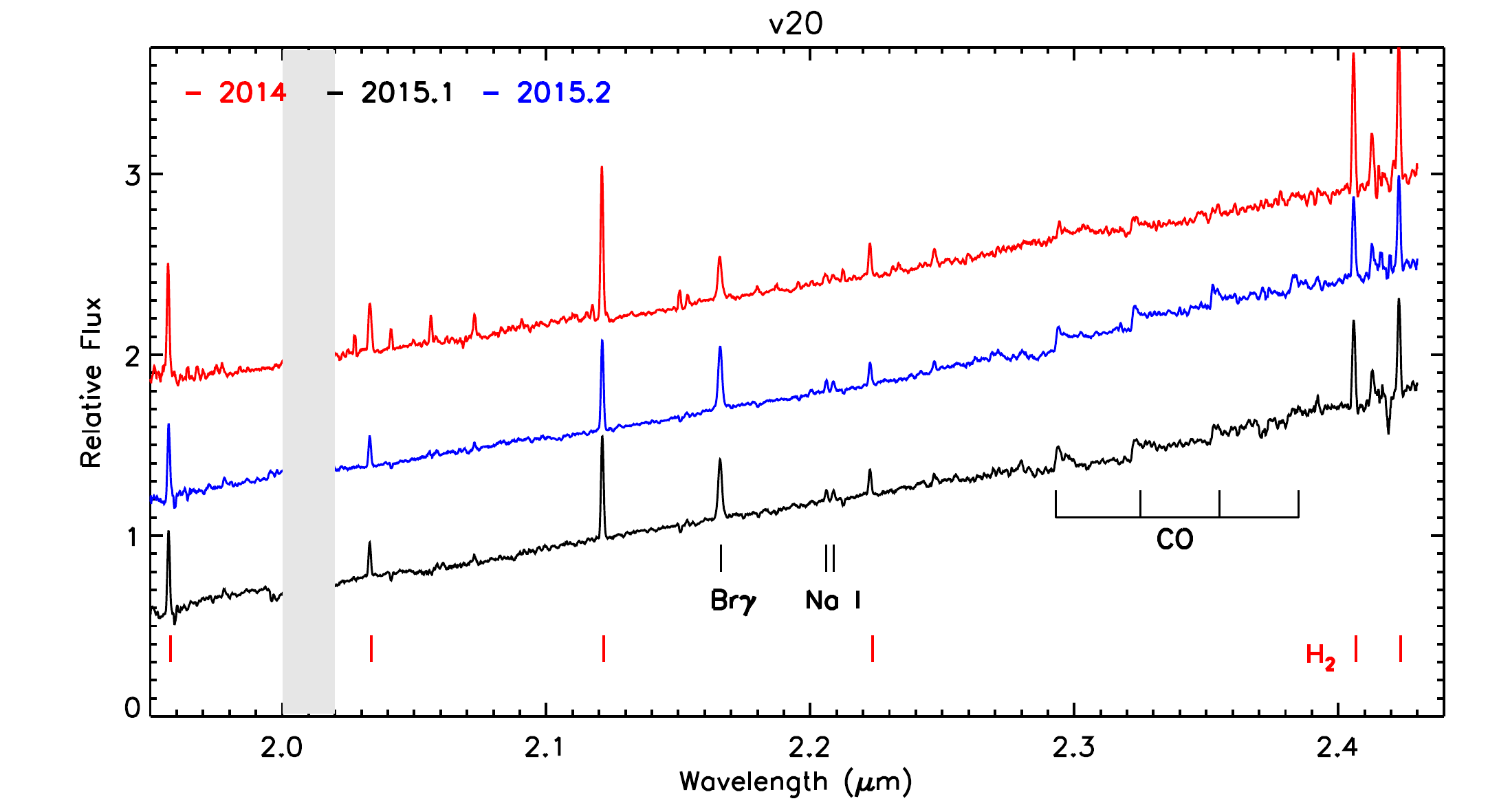}
\includegraphics[width=3.3in,angle=0]{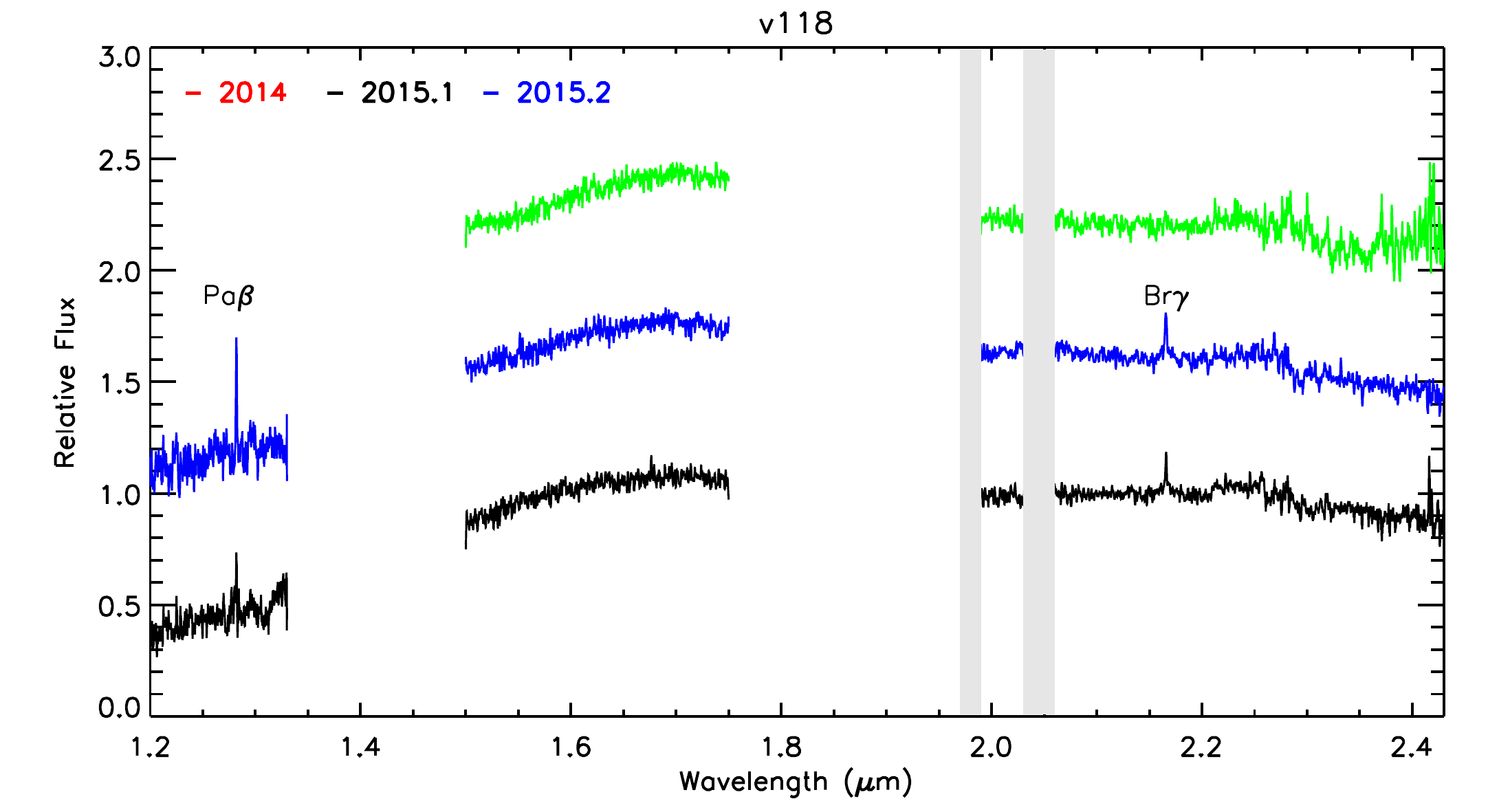}
\includegraphics[width=3.3in,angle=0]{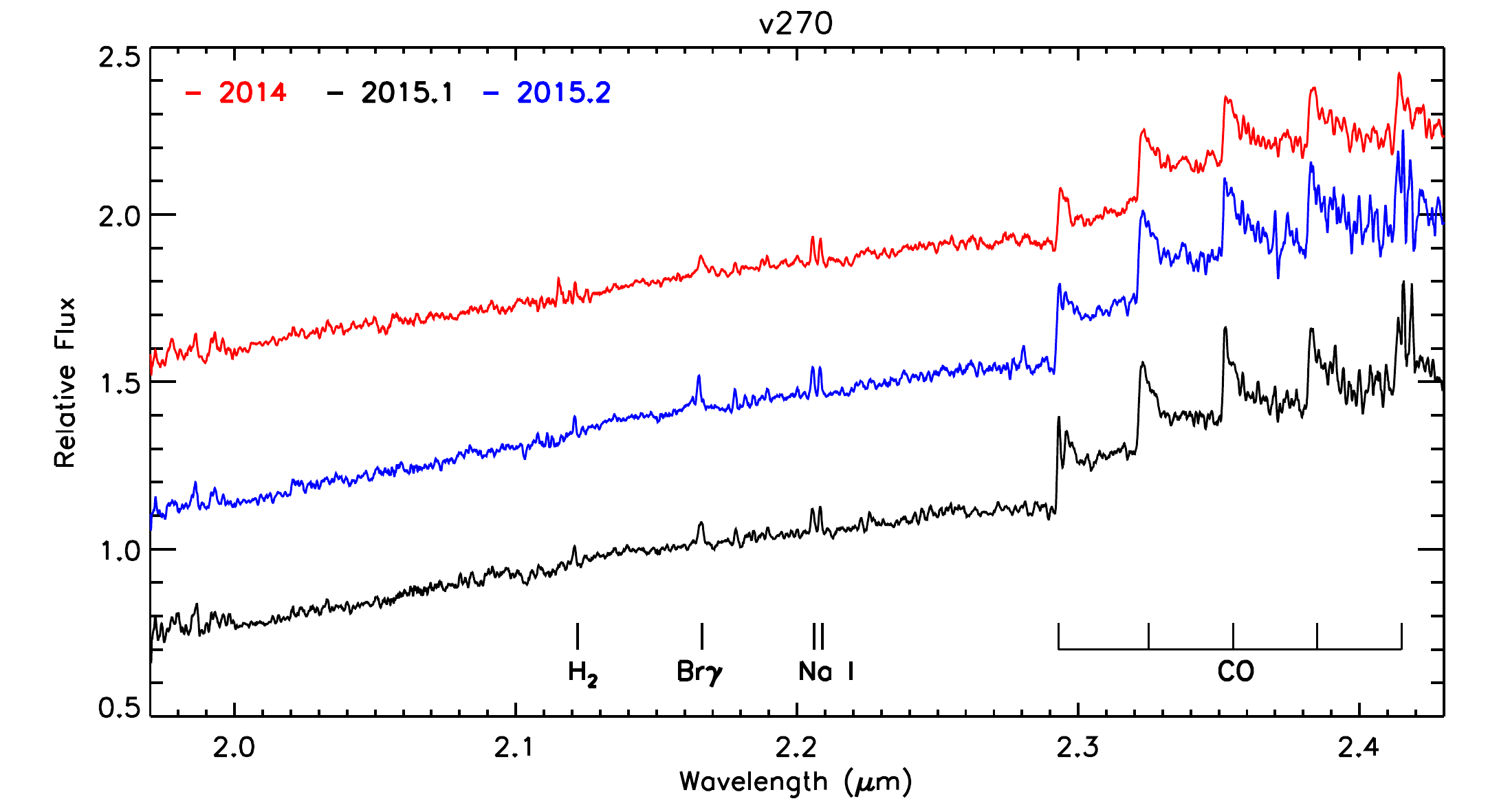}
\includegraphics[width=3.3in,angle=0]{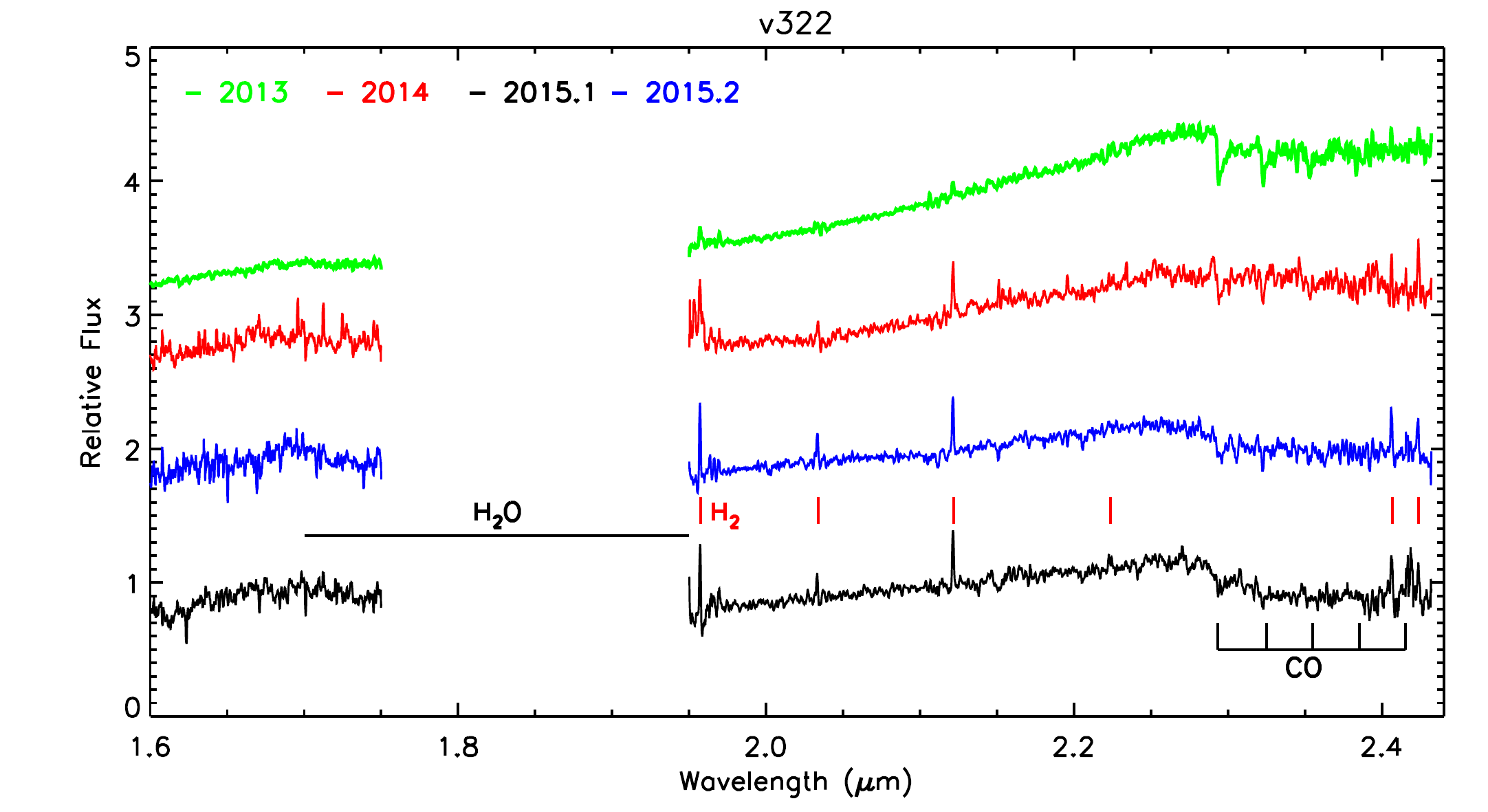}
\includegraphics[width=3.3in,angle=0]{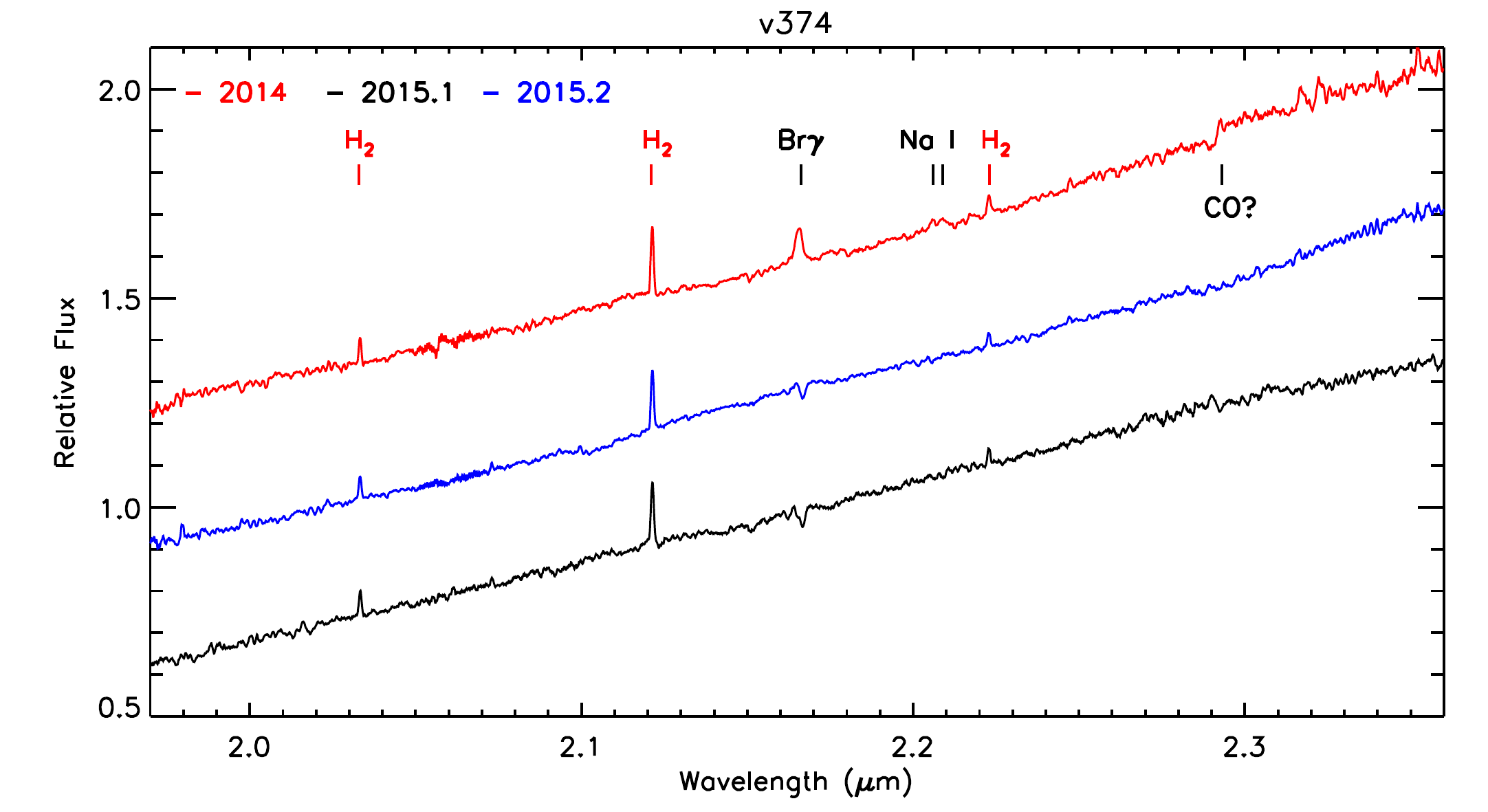}
\includegraphics[width=3.3in,angle=0]{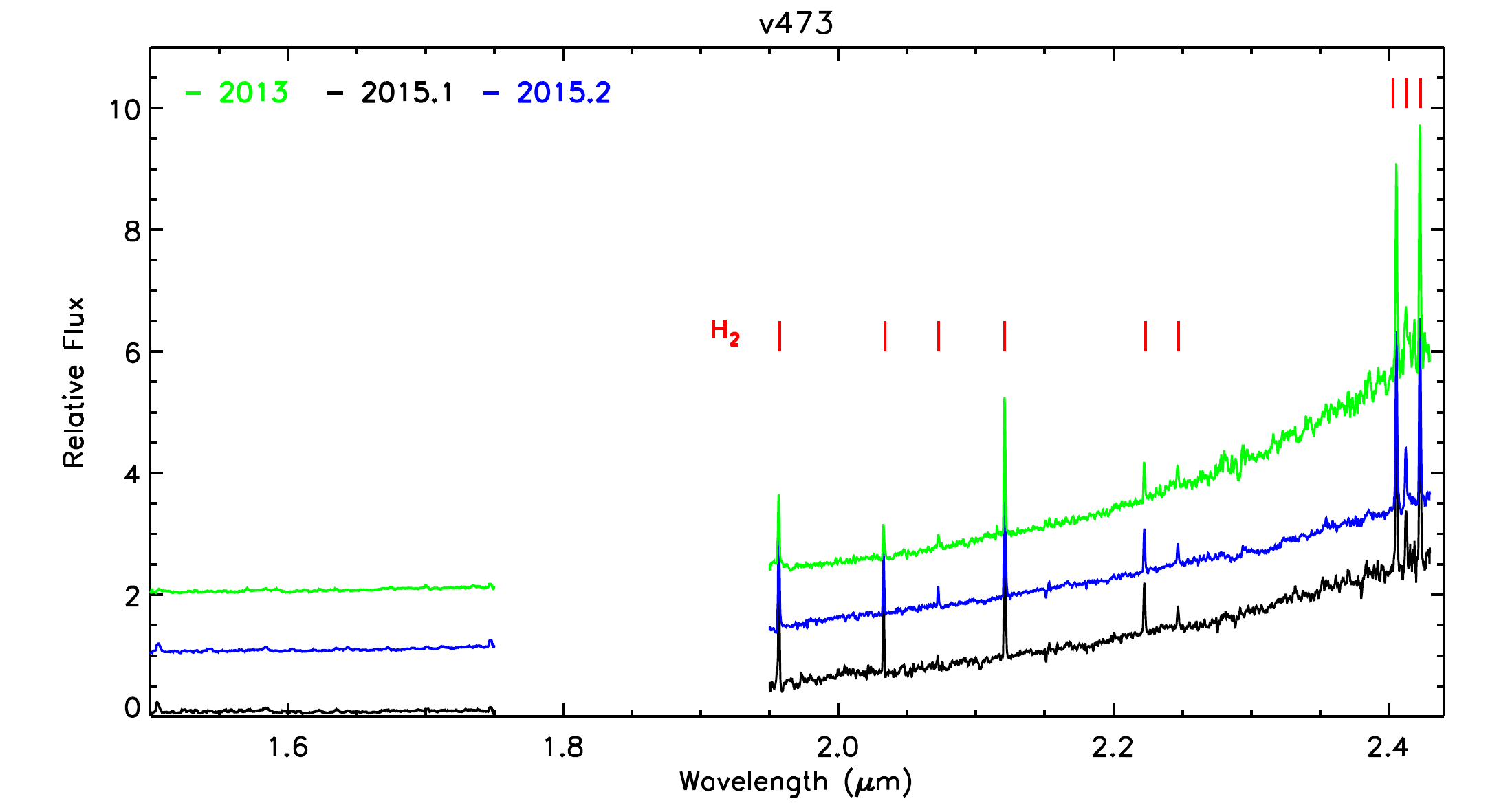}
\includegraphics[width=3.3in,angle=0]{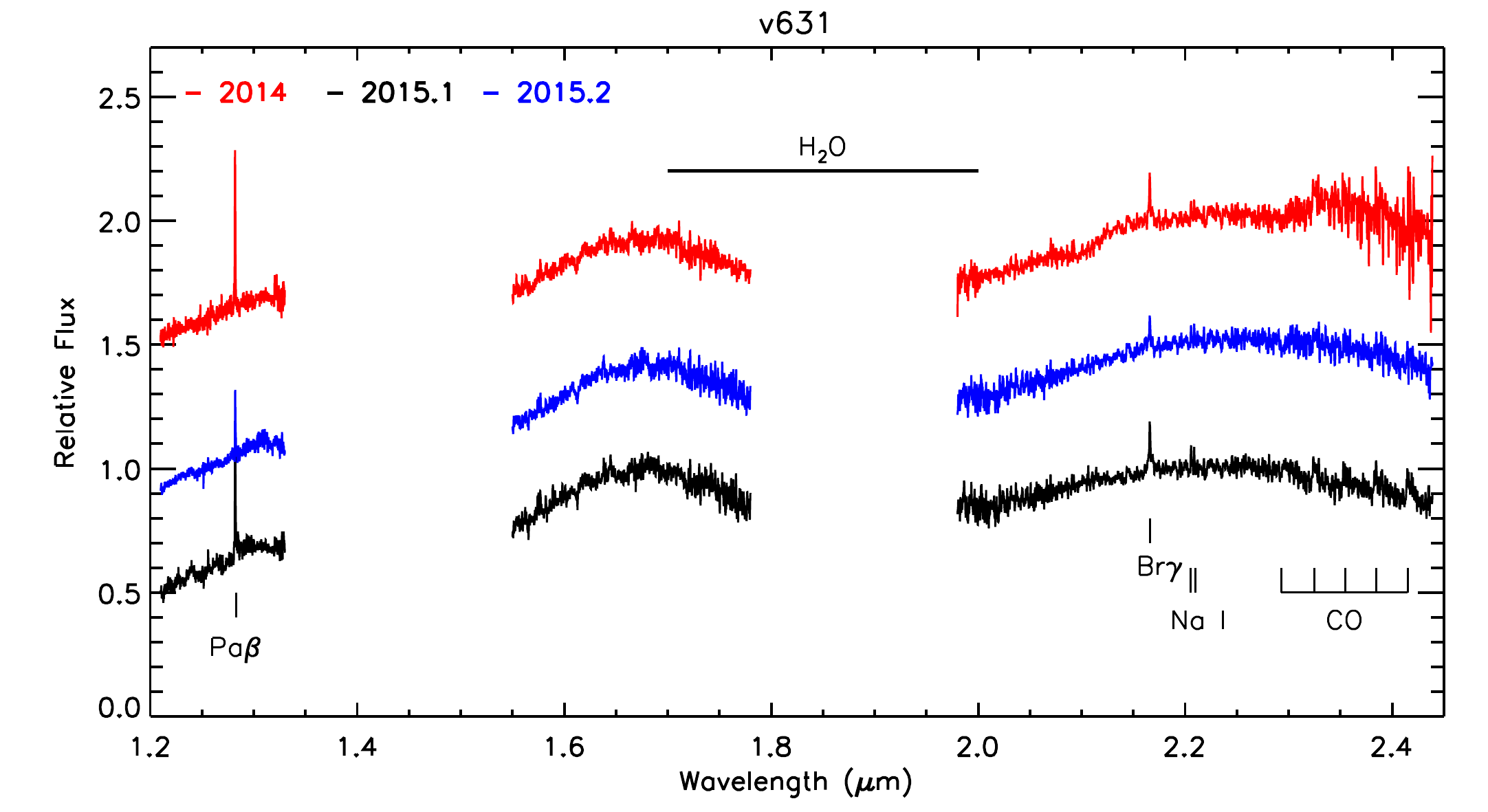}
\includegraphics[width=3.3in,angle=0]{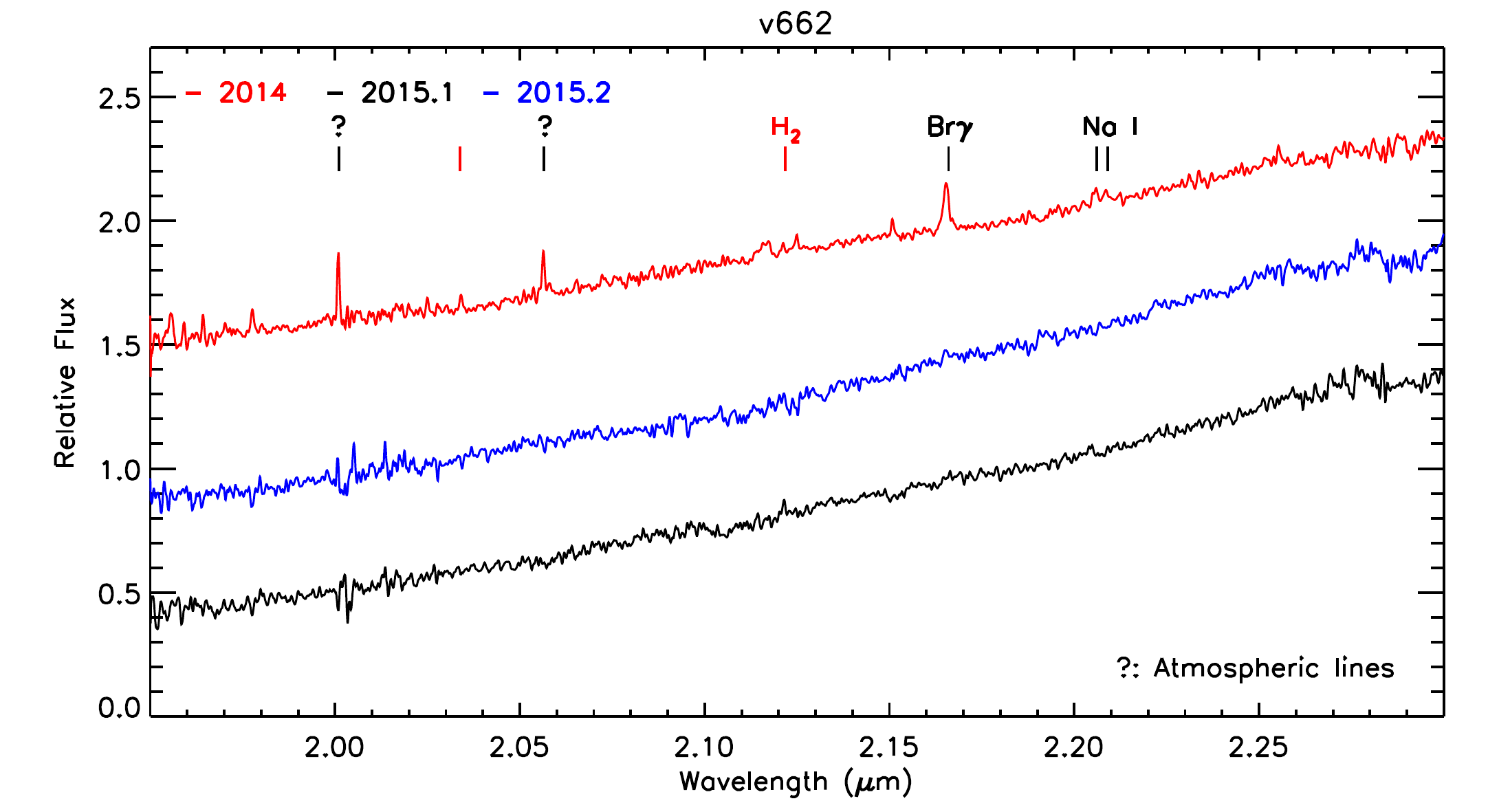}
\caption{Relative spectra of 16 eruptive objects observed in this work. In each plot, spectra is coloured based on their observation epochs as 2013 (green), 2014 (red), 2015-Apr-27 (2015.1; black), and 2015-Apr-28 (2015.2; blue). Spectroscopic features (e.g. molecular hydrogen emission, hydrogen recombination lines, sodium doublets, molecular bands) and atmospheric absorption regions are marked out on some plots as examples.} %
\end{figure*}

\begin{figure*} 
\centering
\includegraphics[width=3.3in,angle=0]{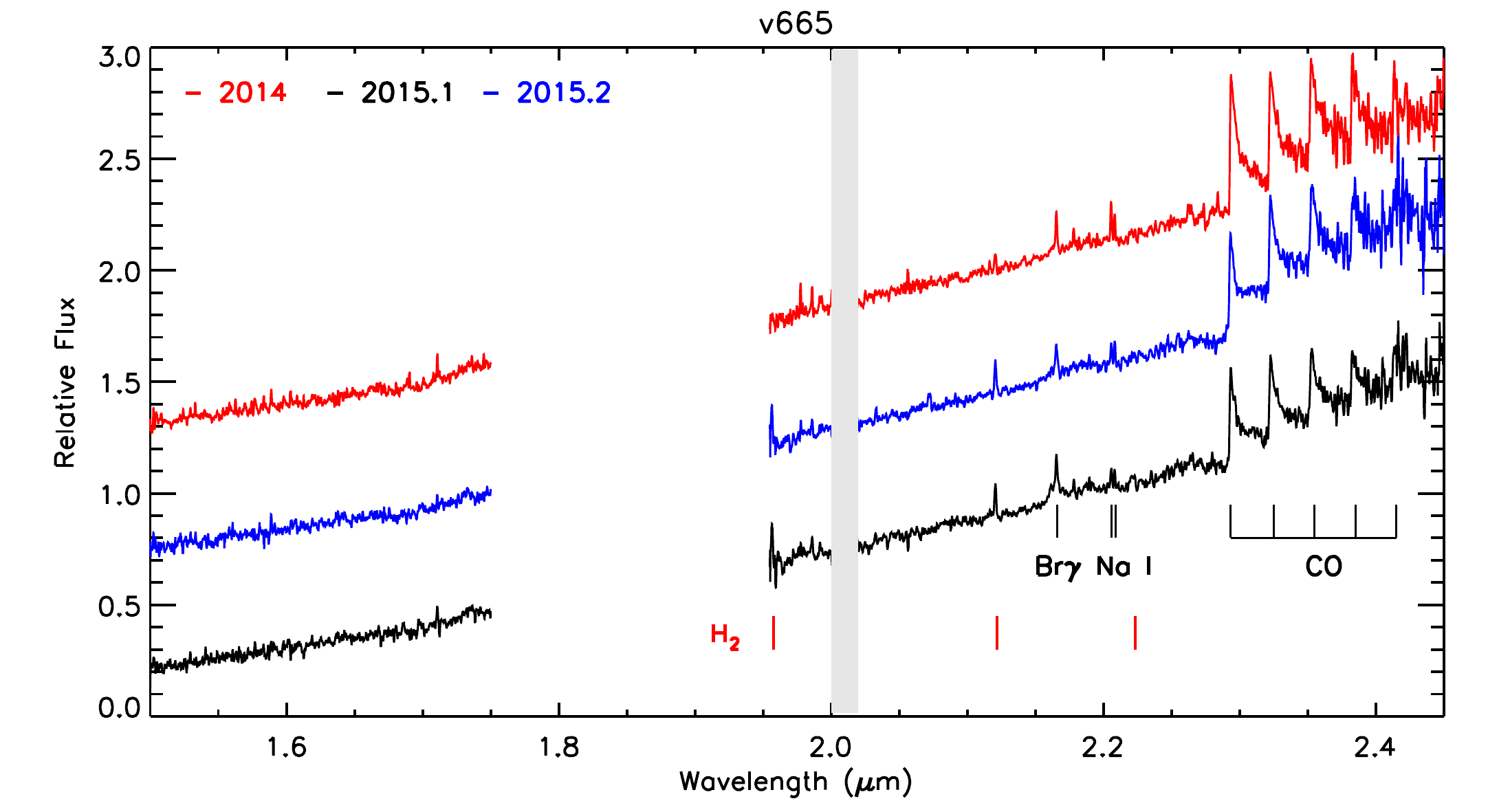}
\includegraphics[width=3.3in,angle=0]{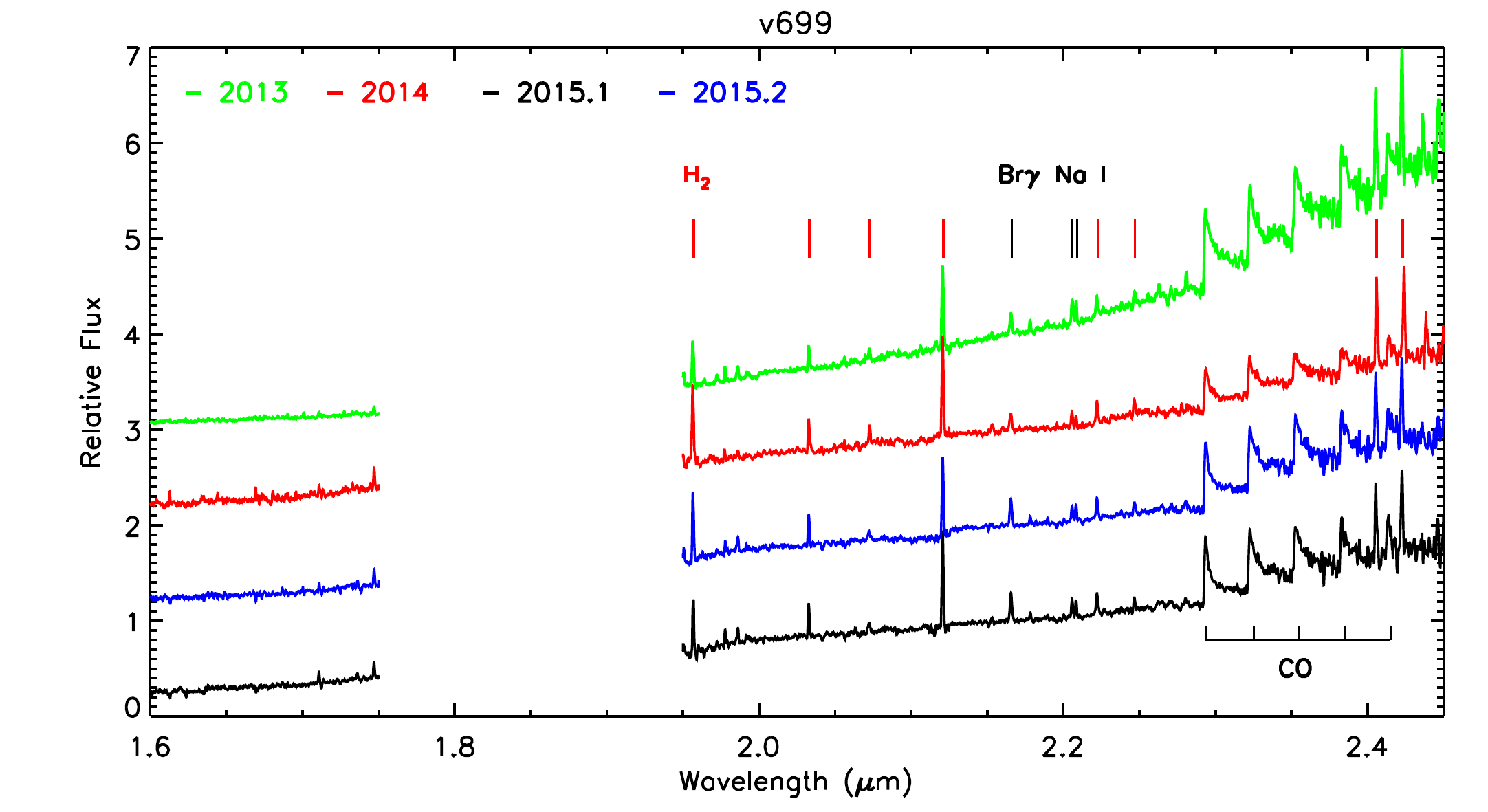}
\includegraphics[width=3.3in,angle=0]{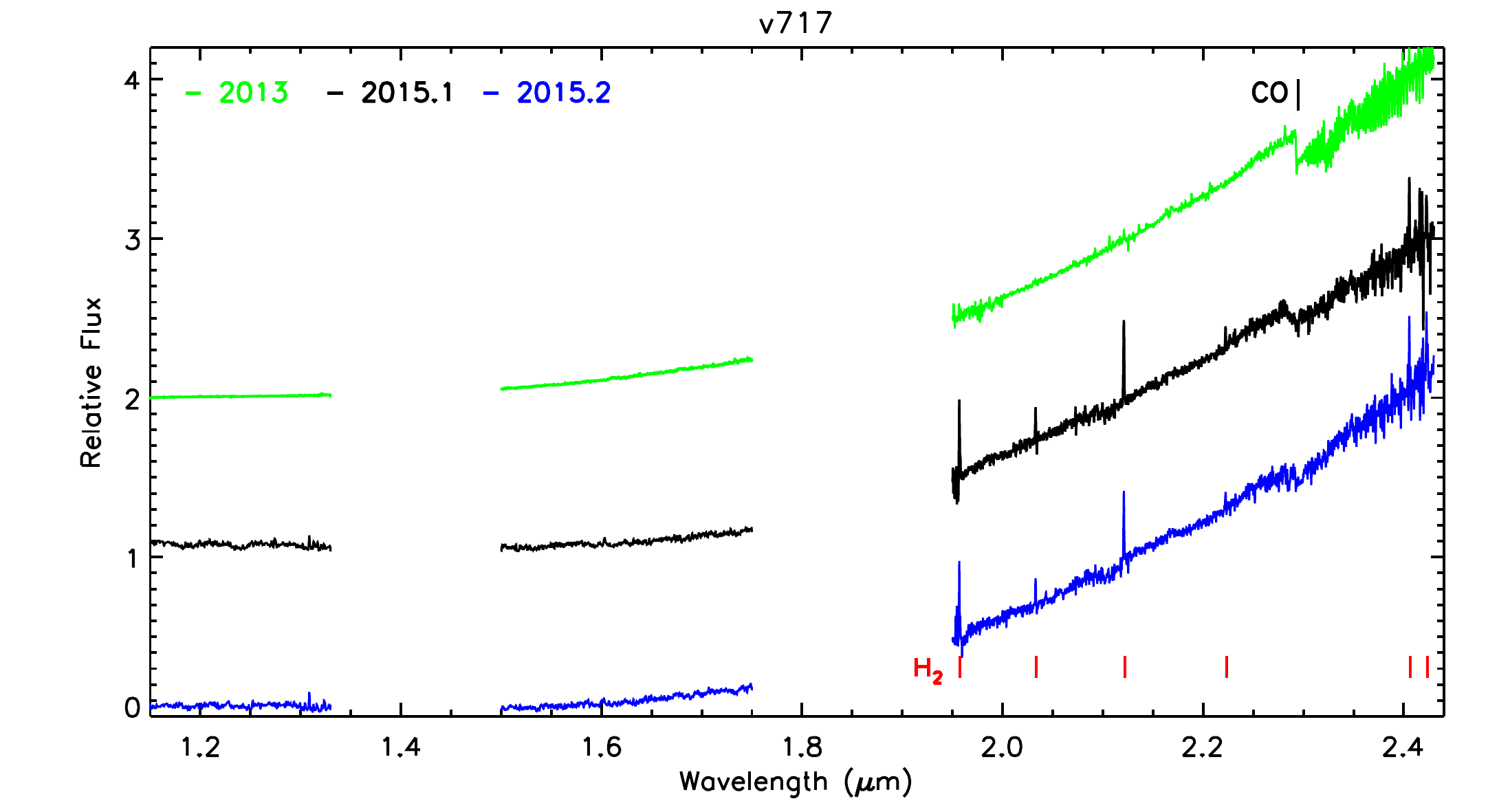}
\includegraphics[width=3.3in,angle=0]{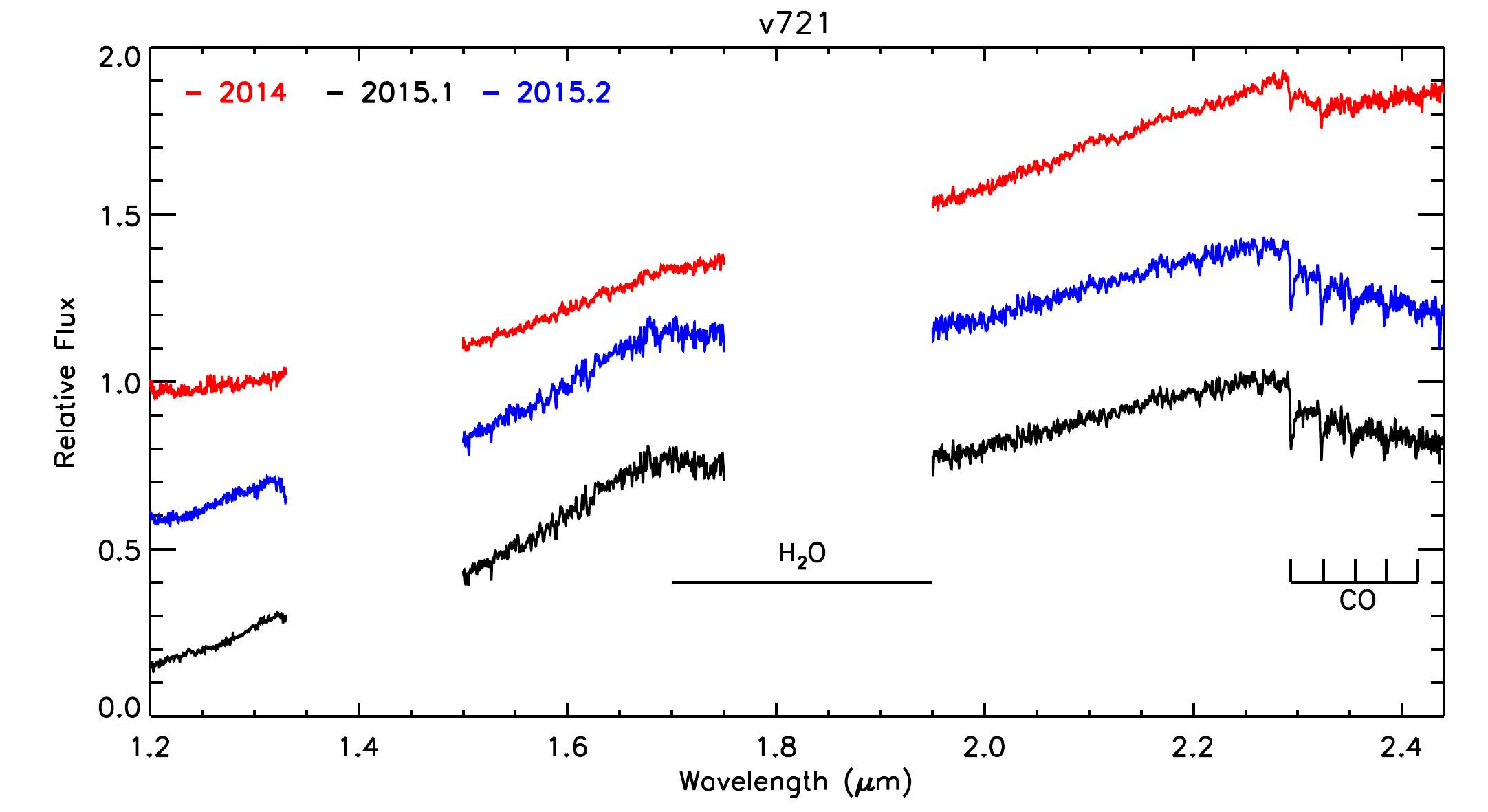}
\includegraphics[width=3.3in,angle=0]{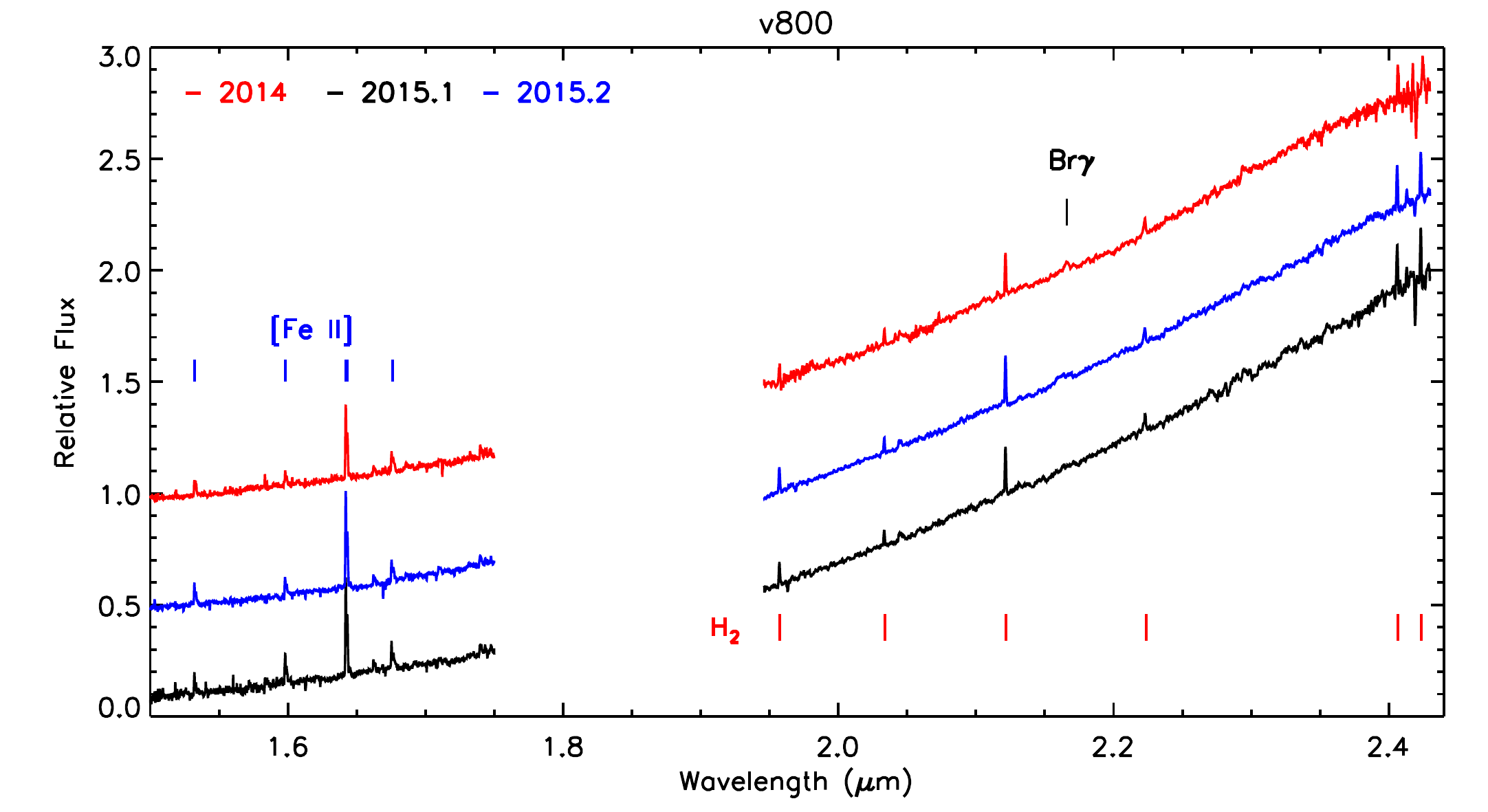}
\includegraphics[width=3.3in,angle=0]{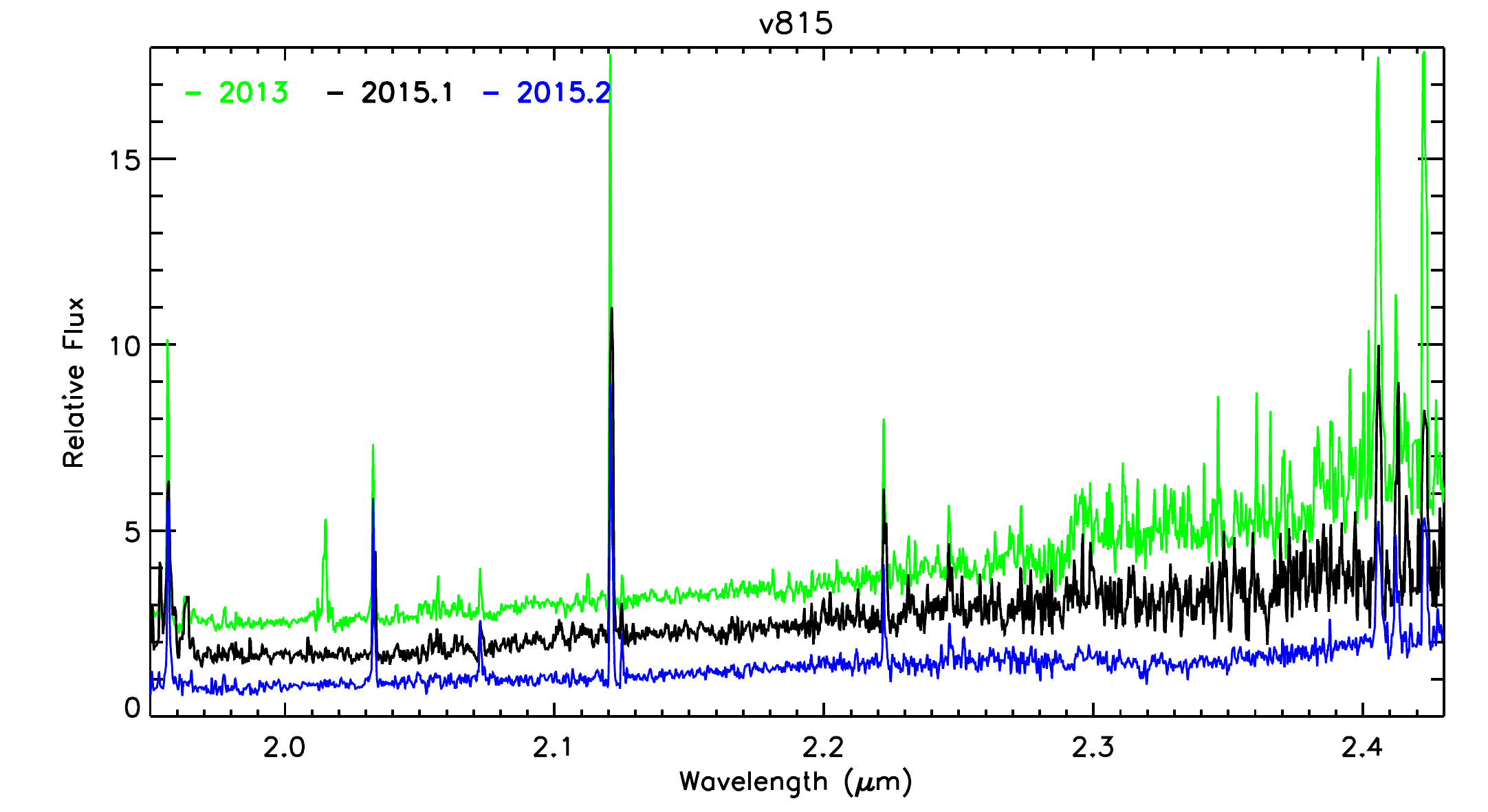}
\includegraphics[width=3.3in,angle=0]{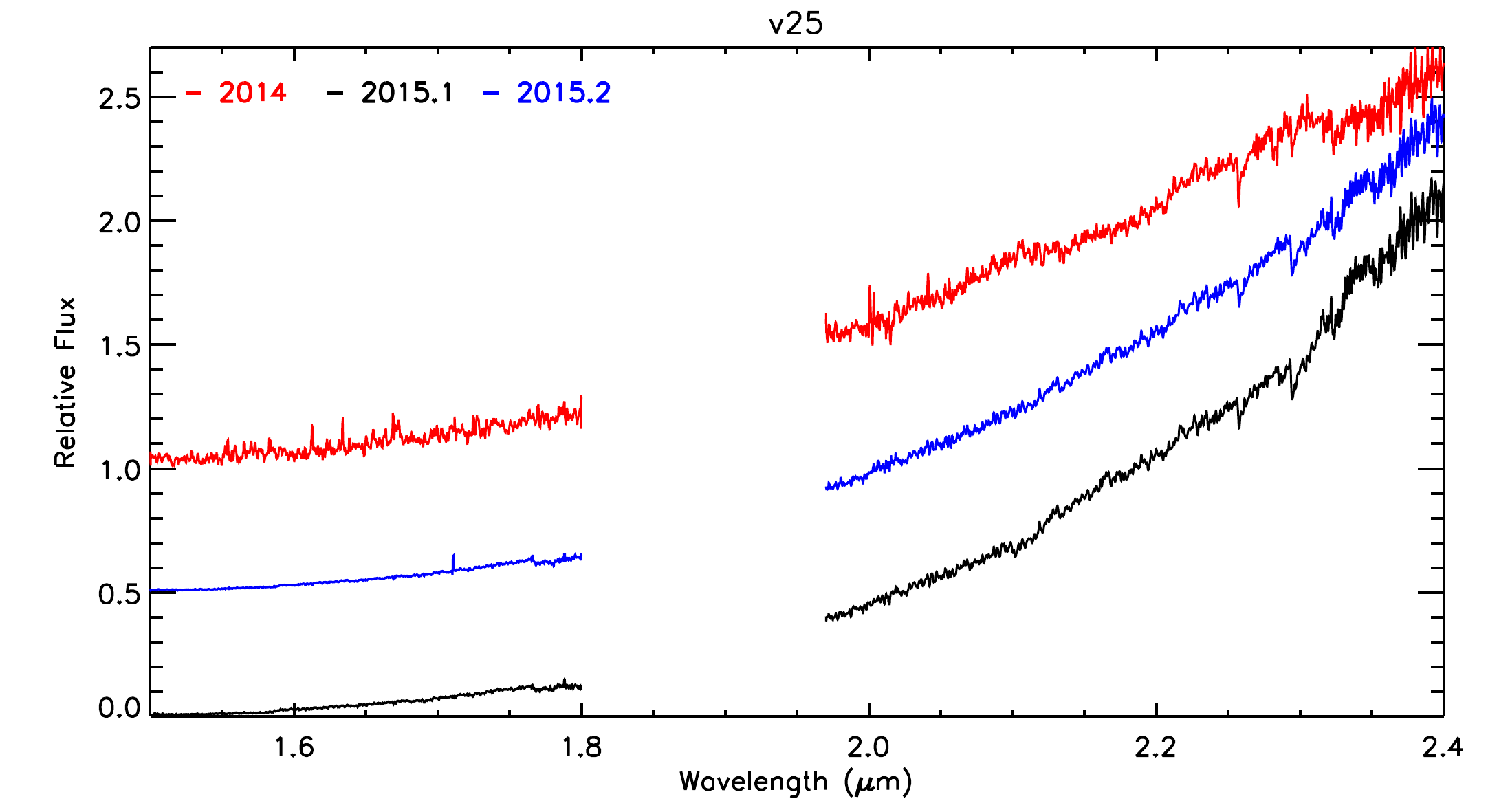}
\includegraphics[width=3.3in,angle=0]{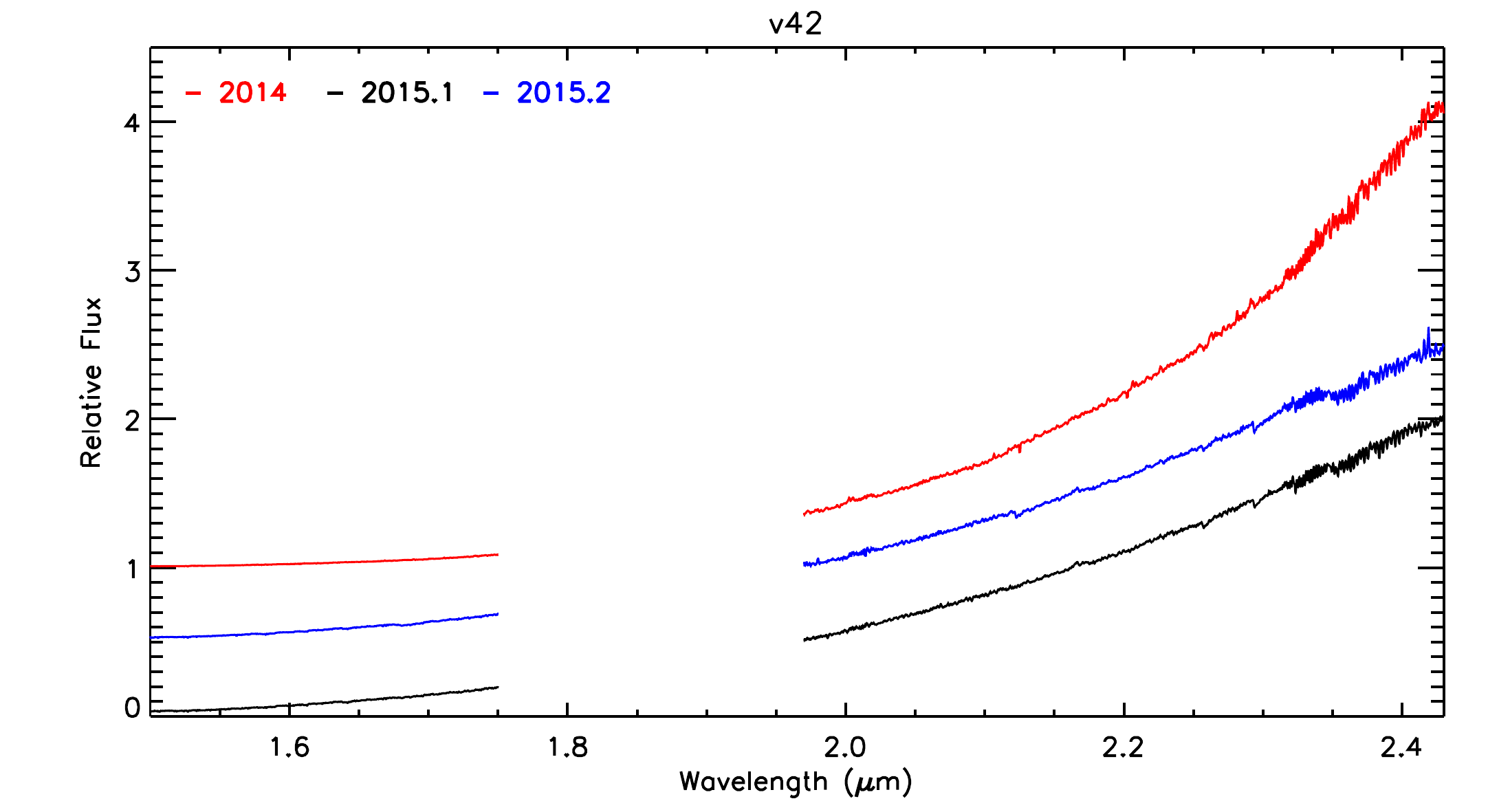}
\caption{continued.}
\end{figure*}

\end{document}